\documentclass[letterpaper,11pt]{article}

\usepackage{amsmath,amssymb}
\usepackage{array}
\usepackage{psfig,float}
\usepackage{amsmath,amsfonts,graphicx,bm}
\usepackage{psfig,float}
\usepackage{amsmath,amsfonts,graphicx,bbm,multicol}

\newcommand{\mP}{{\bar M}_{P}}

\begin{document}

\baselineskip=18pt

\newcommand{\eps}{\epsilon}
\newcommand{\pslash}{\!\not\! p}
\newcommand{\I}{\rm 1\kern-.24em l} 
\newcommand{\Tr}{\mathop{\rm Tr}}



\textwidth 6.5in
\evensidemargin 0in
\oddsidemargin 0in
\textheight 9in

\def\mz{M_Z}
\def\mw{M_W}
\def\ap{A_1}
\def\zp{Z_1}
\def\zx{{Z_X}_1}
\def\zpri{{Z^\prime}}
\def\zpt{\tilde{Z}_1}
\def\zxt{\tilde{Z}_{X1}}
\def\map{M_{A_1}}
\def\mzp{M_{Z_1}}
\def\mzx{M_{{Z_X}_1}}
\def\mzpri{M_{Z^\prime}}
\def\wpL{W_{L_1}}
\def\wpR{W_{R_1}}
\def\wpri{{W^\prime}^\pm}
\def\wpLt{\tilde{W}_{L_1}}
\def\wpRt{\tilde{W}_{R_1}}
\def\mwpL{M_{W_{L_1}}}
\def\mwpR{M_{W_{R_1}}}
\def\mwpri{M_{W^\pm_1}}

\def\beq{\begin{equation}}
\def\eeq{\end{equation}}
\def\bea{\begin{eqnarray}}
\def\eea{\end{eqnarray}}
\def\gev{\rm GeV}
\def\tev{\rm TeV}
\def\fbi{\rm fb^{-1}}
\def\lsim{\mathrel{\raise.3ex\hbox{$<$\kern-.75em\lower1ex\hbox{$\sim$}}}}
\def\gsim{\mathrel{\raise.3ex\hbox{$>$\kern-.75em\lower1ex\hbox{$\sim$}}}}
%
\newcommand{ \slashchar }[1]{\setbox0=\hbox{$#1$}   
   \dimen0=\wd0                                     
   \setbox1=\hbox{/} \dimen1=\wd1                   
   \ifdim\dimen0>\dimen1                            
      \rlap{\hbox to \dimen0{\hfil/\hfil}}          
      #1                                            
   \else                                            
      \rlap{\hbox to \dimen1{\hfil$#1$\hfil}}       
      /                                             
   \fi}                                             %
\def\ptmiss{\slashchar{p}_{T}}
\def\etmiss{\slashchar{E}_{T}}
\providecommand{\tabularnewline}{\\}


\thispagestyle{empty}
\vspace{20pt}
\font\cmss=cmss10 \font\cmsss=cmss10 at 7pt

\begin{flushright}
UMD-PP-07-007\\
BNL-HET-07/12\\
MADPH-07-1496\\
SU-4252-862 
\end{flushright}

\hfill
\vspace{20pt}

\begin{center}
{\Large \textbf
{LHC Signals for Warped Electroweak Neutral Gauge Bosons}}
\end{center}

\vspace{15pt}
\begin{center}
{\large 
Kaustubh Agashe$\, ^{a, b}$, Hooman Davoudiasl$\, ^{c}$,
Shrihari Gopalakrishna$\, ^{c}$,
Tao Han$\, ^{d}$, Gui-Yu Huang$\, ^{d}$, 
Gilad Perez$\, ^{e, f, g}$, Zong-Guo Si$\, ^{h}$, 
Amarjit Soni$\, ^{c}$
} \vspace{20pt}

$^{a}$\textit{Department of Physics, 
Syracuse University, Syracuse, NY 13244, USA}
\\
$^{b}$\textit{Maryland Center for Fundamental Physics,
     Department of Physics,
     University of Maryland,
     College Park, MD 20742, USA}
\\
$^{c}$\textit{Brookhaven National Laboratory,
Upton, NY 11973, USA}
\\
$^{d}$\textit{Department of Physics, University of
Wisconsin, Madison, WI 53706, USA}
\\
$^{e}$\textit{C.~N.~Yang Institute for Theoretical Physics,
State University of New York, Stony Brook, NY 11794-3840, USA}
\\
$^{f}$\textit{Jefferson Laboratory of Physics, Harvard University
Cambridge, Massachusetts 02138, USA}
\\
$^{g}$\textit{Physics Department, Boston University
Boston, Massachusetts 02215, USA}
\\
$^{h}$\textit{Department of Physics, Shandong University,
Jinan Shandong 250100, China}
\end{center}

\vspace{20pt}
\begin{center}
\textbf{Abstract}
\end{center}
\vspace{5pt} {\small \noindent
We study signals at the Large Hadron Collider (LHC) for
Kaluza-Klein (KK) excitations of the electroweak
gauge bosons 
in the framework with the Standard Model (SM) gauge and fermion fields
propagating in a warped extra dimension. 
Such a framework addresses both the Planck-weak and flavor
hierarchy problems of the SM.
Unlike the often studied $Z^{ \prime}$
cases, in this framework, 
there are three neutral gauge bosons due to the underlying 
$SU(2)_L \times SU(2)_R \times U(1)_X$ gauge group in the bulk.
Furthermore, couplings of these KK states
to light quarks and
leptons are suppressed, whereas those to top and bottom 
quarks are enhanced
compared to the SM
gauge couplings. Therefore, the
production of light quark and lepton states is suppressed
relative to other beyond the SM constructions, and the fermionic decays 
of these states are dominated by the top and bottom quarks,
which are, though, overwhelmed by KK gluons dominantly decaying into them. 
However, as we emphasize in this paper, decays of
these states to longitudinal
$W$, $Z$ and Higgs are also
enhanced similarly to the case of top and bottom quarks. 
We show that the $W$, $Z$ and Higgs final states can give significant 
sensitivity at the LHC  to $\sim 2 \; (3)$ TeV KK scale with an integrated luminosity of 
 $\sim 100$ fb$^{-1}$ ($\sim 1$ ab$^{-1}$).  
Since current theoretical framework(s) favor KK masses $\gtrsim  3$~TeV, luminosity
upgrade of LHC is likely to be crucial in observing these states.
}

\vfill\eject
\noindent


\section{Introduction}

The hierarchy between the
Planck scale and the electroweak (EW)
scales has
been one of the deep mysteries of the Standard Model (SM) 
for the past couple of decades. Solutions
to this hierarchy problem invoke new physics
at the weak or TeV scale. Hence, the upcoming
Large Hadron Collider (LHC) with
center of mass energy
of $14$ TeV has the potential to
test such ideas.
In this paper, we focus on one such solution
based on the Randall-Sundrum (RS1)
framework of a warped extra dimension \cite{rs1}. 
Specifically, we consider this framework
with the SM fermion and gauge fields propagating
in the extra dimension (or ``bulk''). 
Such a scenario can also explain the hierarchy
between the SM fermion masses and mixing angles (flavor hierarchy).
Moreover, in this framework, there are Kaluza-Klein (KK)
excitations of the SM gauge and fermionic fields with mass at the TeV scale,
leading to potential signals from
these new states at the LHC. In particular, the 
prospects for detection of the 
KK gluon have been studied recently \cite{kkgluon}, and references
\cite{KKGravitonRefs} studied signals for the
KK graviton in this scenario.

As a next step in this program, 
here we study signals from
KK modes of the EW gauge bosons, focusing
on the neutral ones in this paper. Just like
the case of the KK gluon, 
the fermionic decay modes of the EW KK states
are dominated by the top (and in some cases
bottom) quarks,
in particular, the decays to the ``golden'' leptonic
channels tend
to be suppressed unlike the $Z^{ \prime }$'s
studied extensively
in the literature. However, as we discuss
in this paper,
a new feature for
EW states (with respect to the KK gluon)
is enhanced decays 
(comparable to that into top quarks) of
EW KK states into
{\em longitudinal} $W$, $Z$ and Higgs. 
We therefore focus on the $W$, $Z$ and Higgs final states  
since the decays to top and bottom final states are overwhelmed by decays
of the KK gluon which dominantly decay into them.
In addition, there are multiple 
EW KK states
(namely 3 for neutral and 2
for charged)
which mix with each other, resulting
in interesting phenomenology and decay patterns.
We find that the LHC with $\sim 100$ fb$^{-1}$ 
to $\sim 1$ ab$^{-1}$ luminosity
can be sensitive to masses for
EW states in the $2$ to $3$ TeV range
using the $W$, $Z$ and Higgs final states,
smaller than in the cases of KK gluon
due to the larger cross-section
for the latter.
However, as we will discuss in next section, 
KK masses $\gtrsim 3$~TeV are preferred by
precision electroweak and flavor tests for the simplest existing
models in the literature. So, our results provide a strong motivation for 
LHC upgrade.

The paper is organized as follows. In Sec.~\ref{WrpIntro.SEC}, 
we briefly review the basic setting in the warped extra dimension scenario
focusing on the electroweak gauge bosons, and in Sec.~\ref{summary}
present details on the different neutral states in the theory.
We calculate the widths and branching fractions for their decays 
in Sec.~\ref{zpridec.SEC}.
In Sec.~\ref{signals} we give the main results of our paper.
Here, we consider various signals based on these couplings,
focusing on decays of the neutral modes to $W^+ W^-$, $Z\, h$ and $l^+ l^-$
(even though the latter channel is suppressed, it can be important
due to its cleanness). We defer a study of charged EW states 
to a future publication. 
In appendices~\ref{couplings}~and~\ref{coupl.APP}
we present in detail the couplings of these heavy EW gauge bosons
to the SM fermions and the SM gauge bosons -- in
particular, we present a derivation of couplings of heavy EW gauge bosons to the 
SM gauge bosons, and the corresponding Feynman rules 
of the couplings of the KK gauge bosons to the SM fields.

\section{Warped Extra Dimension: Lay of the Land}
\label{WrpIntro.SEC}
\subsection{ Original RS1}

The framework is based on a slice of AdS$_5$. Owing to the warped geometry,
the relationship between the $5D$ mass scales
(taken to be of order $4D$ reduced Planck scale, $\mP$) and those in an
effective $4D$ description depends on the location in the extra dimension.
The $4D$ (or zero-mode) graviton is localized near the ``UV/Planck''
brane which has a Planckian fundamental scale, whereas
the Higgs sector is localized near the ``IR/TeV'' brane where it
is stable near a warped-down fundamental scale of order TeV. 
The crucial point is that 
this
large hierarchy of scales can be generated via a modest-sized radius of
the $5^{\rm th}$ dimension: $\hbox{TeV} / \mP
\sim e^{ - k \pi r_c }$, where
$k$ is the curvature scale
and $R$ is the proper size of the extra dimension; $k R \approx 11$.
Furthermore, such a size of the extra dimension 
can be stabilized by suitable mechanisms \cite{Goldberger:1999uk}.
Finally, based on the AdS/CFT correspondence
\cite{Maldacena:1997re}, RS1
is conjectured to be dual to $4D$ composite Higgs models
\cite{Arkani-Hamed:2000ds, Contino:2003ve, Agashe:2004rs}.

In the original RS1 model, the entire SM (including the fermions
and gauge bosons) are assumed to be localized on the TeV brane.
The key feature of
this model is that the only
new particles are the KK gravitons
with no SM gauge quantum numbers (color/electroweak charge).\footnote{There
is also the radion, the modulus corresponding to fluctuations
of the size of the extra dimension.} 
These
KK gravitons have a mass $\sim$ TeV
and are localized near the TeV brane so that KK graviton coupling
to the {\em entire} SM is only $\sim$ TeV suppressed. Hence,
KK graviton production
via $q \bar{q}$ or $gg$ fusion at the LHC [or via
$e^+ e^-$ at International Linear Collider (ILC)] followed by
decays to dileptons or diphotons gives striking 
signals \cite{Davoudiasl:1999jd}.

\subsection{ SM in bulk}

However, it was subsequently
realized that to solve
the Planck-weak hierarchy problem
only the SM Higgs boson has
to be localized on/near
the TeV brane -- the rest of the
SM (fermion and gauge fields)
can be allowed
to propagate in the extra dimension \cite{bulkgauge, gn, gp}
since their masses are protected by gauge and chiral symmetries.
Moreover, such a scenario enables a solution
to the following problem of the {\em original} RS1 model.
Namely, the higher-dimensional operators in the $5D$
effective field theory (from physics at the cut-off)
are suppressed only by
the warped-down cut-off $\sim$ TeV
[assuming $O(1)$ coefficients for these operators], 
giving too large contributions to
flavor changing neutral current
(FCNC) processes and observables related to SM electroweak precision
tests (EWPT).
The point is that in this {\em new} scenario 
(with the SM in the bulk) the
SM particles are identified with the zero-modes of the $5D$ fields
and the
profile of a SM fermion in the extra dimension
depends on its $5D$ mass parameter.
We can
then choose to localize 1st and 2nd generation fermions near the Planck brane
so that the
FCNC's from higher-dimensional operators are suppressed
by scales $\gg$ TeV which is the cut-off
at the location of these fermions~\cite{gp, hs}.
Similarly, contributions to EWPT from cut-off
physics are also suppressed.

As a further bonus, we obtain a solution to the flavor puzzle
in the sense that hierarchies in the SM Yukawa couplings arise without
introducing hierarchies
in the fundamental $5D$ theory~\cite{gn, gp, hs}:
the 1st/2nd generation fermions
have small Yukawa couplings to Higgs which is localized near the
TeV brane.
Similarly,
the top quark can be localized near the TeV brane
to account for its large Yukawa coupling.

%
%
In this framework, there are 
KK excitations of SM gauge and fermion fields 
in addition to those of the graviton. These 
states have mass
in the TeV range and are localized near
the TeV brane (just like KK gravitons).
Hence, we obtain 
new possibilities for
collider signals, but at the same time,
there are new contributions to FCNC's and EWPT
which are
{\em calculable} in the $5D$ effective field theory (EFT).
However, due to various symmetries (approximate flavor 
or analog of GIM mechanism of
the SM \cite{gp, hs, aps}
and custodial isospin \cite{Agashe:2003zs}),
we can show that gauge KK masses as small as $\sim 3$ TeV
are consistent with oblique electroweak (EW) data \cite{Agashe:2003zs} 
(we comment on
non-oblique effects such as $Zb \bar{b}$ below) and FCNC's 
\cite{NMFV}.\footnote{
See references \cite{others1, others2}
for other studies of FCNC's in such frameworks.
Note that 
beyond the SM 
operators with $(V-A)\otimes (V+A)$ Lorentz structure mediate
enhanced contributions to
$\Delta S=2$ processes such as $\epsilon_K$~\cite{epsKenh}. Within our framework
these contributions
 are proportional to $m_d m_s$~\cite{aps}. Nevertheless, without
further structure these
 contributions would generically yield a lower bound on the KK gluon
of ${\cal O}$(8 TeV) \cite{private}.}

Let us consider the
top and bottom sector in detail to determine
the couplings to KK states. Due to heaviness of top quark
combined with constraint from shift in $Z b \bar{b}$,
one
possibility is to
localize $t_R$ very close to TeV brane with $(t,b)_L$ having
a profile close to flat \cite{Agashe:2003zs}.
Even with this choice of the profiles, the gauge KK mass scale is constrained
by $Z b \bar{b}$ to
be $\stackrel{>}{\sim} 5$ TeV~\cite{Carena:2006bn,Agashe:2005dk}, 
i.e., a bit higher than that allowed by oblique EW data.  However,
a
%
%
custodial symmetry to suppress $Z b \bar{b}$ \cite{Agashe:2006at}
can relax this constraint on the
gauge KK mass scale and moreover allows even
the other extreme case: $(t,b)_L$ very close to the TeV brane
and $t_R$ close to flat
and also the intermediate possibility with both
$t_R$ and $(t,b)_L$ being near, but not too close to
TeV brane~\cite{Carena:2006bn,Carena:2007ua,Contino:2006qr,Medina:2007hz}.
The bottom-line is that,
with this custodial symmetry for $Z b \bar{b}$ and for certain choices of profiles
for $t_R$ and $(t,b)_L$ in the extra dimension, gauge KK masses as low as
$\sim 3$ TeV can be consistent with $Z b \bar{b}$
as well. 

Clearly, couplings  of
gauge KK modes to light fermions (to top and bottom) 
are suppressed (enhanced) compared
to the SM gauge coupling 
simply based on the overlap of the 
%
%
corresponding 
profiles in the extra dimension (the zero-mode or SM gauge boson
has a flat profile in the
extra dimension).
As a consequence,
production 
of the gauge KK modes tends to be 
suppressed compared to the $Z^{ \prime }/W^{ \prime }$'s
often studied in the literature. Moreover, their fermionic
decay modes are dominated
by top and bottom quarks (which are not easily detectable modes).
In spite of these difficulties, it was shown in references 
\cite{kkgluon} that 
the LHC can be sensitive to 
KK gluon masses up to $\sim 4$ TeV
based on decays to top quarks.

\subsection{ EW gauge states}

However, for EW KK modes, there is  a possibility of sizable decays
to cleaner final states (compared to the KK gluon) as follows.
The crucial point being that by the 
equivalence theorem, {\em longitudinal} $W$ and $Z$
(denoted by $W^\pm_L$ and $Z_L$) are
effectively the {\em un}physical Higgs
(``would-be'' Goldstone bosons) and are therefore localized near
TeV brane (just like the physical Higgs). So, the decay widths 
for EW KK states in the $W_L/Z_L$
channels are the same size as in those of the physical Higgs/top 
quark.\footnote{This feature is expected 
based on the AdS/CFT correspondence
since such a warped extra dimensional
framework is 
dual to $4D$ composite Higgs models: 
after all, EW KK are states conjectured to be
dual to techni-$\rho$'s
and hence it is not surprising that they
are strongly coupled
to techni-$\pi$'s, i.e., longitudinal $W$ and $Z$.}
Clearly, branching ratio of EW KK states
to a pair of $Z/W$'s is sizable; in particular,
$Z_LZ_L$ is not allowed
(it is for KK graviton!), but $WW$, $ZW$, $Z h$ and $W h$ are good 
decay channels. 
As a corollary,
{\em production} of EW KK states via longitudinal $W$ and $Z$ fusion
(weak boson fusion, WBF) can be potentially important. 
Such
%
%
effects were not analyzed before in this
class of models, including
in the recent paper \cite{Ledroit:2007ik}\footnote{Although
reference \cite{Birkedal:2004au} did study
decays of electroweak states into $W/Z$
in Higgsless models,
where light fermions are (almost) decoupled 
from the gauge KK states (unlike in our case) in order to 
suppress the $S$ parameter. Hence,
the  
production of these states has to proceed via WBF.
Whereas, in this paper,
we consider
production of
these states via light quark-anti-quark annihilation 
(which turns out to be the dominant mechanism) as well.
Moreover, the KK
mass scale in the Higgsless models is lower ($\stackrel{<}{\sim} 1$ TeV)
than in the framework studied here.
} which focuses on decays to top and bottom final states. 
However, the signal from electroweak neutral states
in top/bottom final state is 
likely to be swamped by the KK gluon 
which dominantly decays to this 
%
%
final state
with a coupling larger than that for the case of EW KK states.
Our motivation is to study the
heavy electroweak gauge bosons and hence we consider
their decays to 
the top/bottom
final state only in passing
and focus on the $W/Z/$Higgs final state instead.  
To summarize, the relevant coupling to the KK gauge states can be 
described schematically
(see section \ref{couplings} for more details),  neglecting effects 
related to EWSB, via ratio of
RS1-to-SM gauge coupling
\begin{eqnarray}
{g_{\rm RS}^{q\bar q,l\bar l\, Z_{KK}^{(1)} }\over g_{\rm SM}}
&\simeq&
- \xi^{-1}\approx - {1\over5} \nonumber \\
{g_{\rm RS}^{Q_3\bar Q_3 Z_{KK}^{(1)} }\over g_{\rm
    SM}},\ \ 
{g_{\rm RS}^{t_R\bar t_R Z_{KK}^{(1)}}\over g_{\rm
    SM}} 
& \simeq & 
1 \; \hbox{to} \; \xi \; ( \approx 5 ) \nonumber \\
{g_{\rm RS}^{ HH Z_{KK}^{(1)}}\over g_{\rm
    SM}}  
& \simeq & 
\xi \approx 5 \; \; \; \left( H = h, W_L, Z_L \right)
\label{RScouplings}
\end{eqnarray}
where $q=u,d,s,c,b_R$, $l =$ all leptons, $Q^3= (t, b)_L$, $Z_{KK}^{(1)}$ represents the 
 first KK state of the gauge fields (in the KK-basis), 
$g_{\rm RS}^{xyz}, g_{\rm SM}$ stands for the RS KK mode and the three SM (i.e.,
$4D$) gauge couplings respectively, and $\xi \equiv \sqrt{ k \pi r_c }$ (cf Eq.~(\ref{xidefn.EQ})).
Also, 
$H$ includes both the physical Higgs ($h$) and {\em longitudinal}
$W$ and $Z$.
EWSB induces mixing between EW KK states
which we discuss in what follows.


\section{Summary and overview of the electroweak gauge sector}
\label{summary}
Here we give a summary of the various EW gauge bosons present
in the model and refer the reader to the 
appendices for
details of their properties.
The electroweak 
gauge group in the bulk is $SU(2)_L \times SU(2)_R \times U(1)_X$.
So,
we have 
3 electrically neutral towers from the $U(1)_{L, R, X}$
gauge sectors.
The $U(1)_{ R, X }$ towers mix via the boundary condition on
the Planck brane, and the Higgs vev which couples to $U(1)_{L, R}$ mixes these
towers further.
The Higgs is localized near the TeV brane.

We will find it
convenient to rewrite and reorganize the
neutral towers into towers of photon, $Z$ 
(same combinations as in the SM) and $Z_{ X }$ -- which is the
combination of $U(1)_{ R, \; X }$ orthogonal to $U(1)_Y$ -- towers.
Before turning on the Higgs vev, 
zero-modes are present only in the photon and $Z$ towers.
Even {\em after} EWSB, the
photon tower does not mix with the other two
towers nor do the various modes (both zero and KK)
of this tower mix with each other -- the zero-mode photon is then identified
with the SM photon. The $Z$ and $Z_{ X }$ towers do mix 
via the Higgs vev -- specifically, the zero-mode $Z$ mixes with KK
modes from {\em both}{ towers and the KK modes of the two towers mix
with each other as well (cf Eqs.~(\ref{LzpriM.EQ}),~(\ref{sth01.EQ})~and~(\ref{sth01X.EQ})).
The lightest mode of the resulting mixtures is the SM $Z$. 
We will discuss the phenomenology of
only the 1st KK mode in each tower (for simplicity
and also because the effects of heavier KK modes is suppressed)
denoting it by $\ap$, $\zp$ and $\zx$ respectively in the KK basis,
and as $\ap$, $\zpt$ and $\zxt$ for the mass basis eigenstates, collectively referring 
to these mass eigenstates as $\zpri$.

Similarly, there are 
2 charged towers corresponding to
$W_L^{ \pm }$ and $W_R^{ \pm }$ -- only the former tower 
has a zero-mode. Due to Higgs vev, these two towers mix just like
for the neutral sector and the resulting lightest mode 
is the SM $W$ (cf Eqs.~(\ref{LwpriM.EQ})~(\ref{sth0L.EQ})~and~(\ref{sth0R.EQ})).

As explained above, the heavy gauge bosons will decay
mostly to longitudinal $WW$, longitudinal $Z h$, $t\bar t$ and $b\bar b$ since
the couplings to these final states are in fact enhanced relative
to the SM, whereas the couplings 
to leptons and light quarks are suppressed relative
to the SM (see Eq. (\ref{RScouplings})).
As mentioned above, there are various possibilities
for quantum numbers of the top and bottom
quarks and their profiles 
in the extra dimension (for details see Secs.~\ref{fermion}~and~\ref{Zbb}). For
the analysis in this paper, we will choose
$(t,b)_L$ to be a doublet of $SU(2)_R$ with
an approximately flat profile -- the motivation
being to suppress corrections to $Z b_L \bar{b}_L$ and
flavor violation, with $t_R$ being a singlet or triplet
of $SU(2)_R$ and localized near the TeV brane. It is possible
to obtain a good fit to the precision electroweak data
(in particular $T$ parameter can be positive and of the required size)
for such a choice of parameters \cite{Carena:2006bn}.

Having made this choice for top and bottom quarks, we would like to mention that
our 
focus in this paper is on the production
of the heavy electroweak gauge bosons via
quarks in the initial state followed
by decays to $WW$, $Z h$ final states.
The {\em total} production
cross-section of the heavy gauge bosons and the {\em partial}
decay widths to these final states
are not affected significantly by the choice
of top and bottom profiles and representations.  The partial
decay widths to $t\bar t$ and $b\bar b$ and so the total width 
and, in turn, the production cross-section
for {\em specific} final states are of course affected by 
the choice of representation and profile of the
top and bottom quarks, 
but not by more than an $O(1)$ factor.

\begin{figure}[tb]
\begin{center}
\scalebox{0.6}{\includegraphics[angle=0]{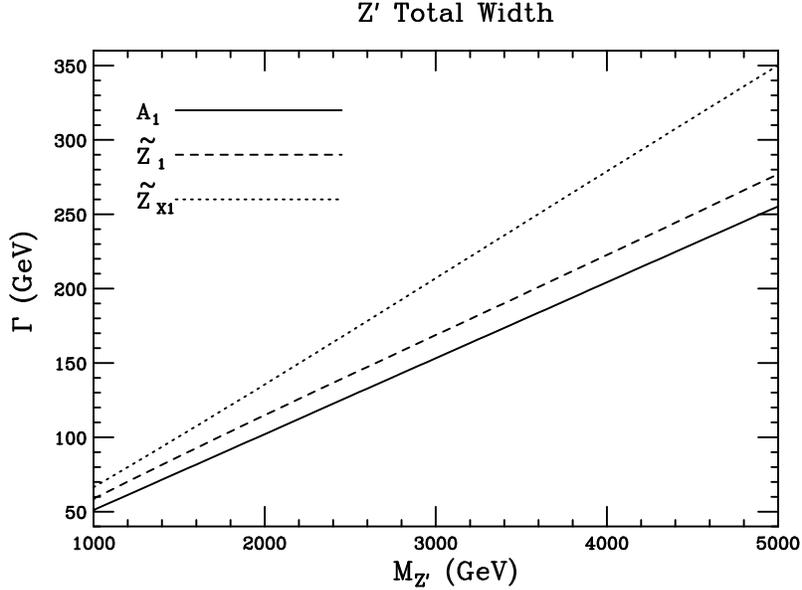}}
\caption{The total width of $\zpri$ as a function of its mass. 
\label{Gam.vs.M.FIG}}
\end{center}
\end{figure}
%

\section{$\zpri$ decays}
\label{zpridec.SEC}
The decay widths for the leading channels of the neutral KK gauge bosons, which
are generically denoted by $\zpri$ unless specified otherwise, are given by
\bea
\Gamma(\ap \rightarrow W_L W_L) &=& \frac{e^2 \kappa^2}{192\pi} \frac{\mzpri^5}{m_W^4} \ ; \ \ \ \ \kappa \propto \sqrt{k\pi r_c} \left(\frac{m_W}{\mwpri}\right)^2 \ , \\
\Gamma(\zpt,\zxt \rightarrow W_L W_L) &=& \frac{g_L^2 c_W^2 \kappa^2}{192\pi} \frac{\mzpri^5}{m_W^4} \ ; \ \ \ \ \kappa \propto \sqrt{k\pi r_c} \left(\frac{m_Z}{(\mzp,\mzx)}\right)^2 \ , \\
\Gamma(\zpt,\zxt \rightarrow Z_L h) &=& \frac{g_Z^2\kappa^2}{192\pi} \mzpri \ ; \ \ \ \ \kappa 
\propto \sqrt{k\pi r_c} \ , \\
\Gamma(\zpri \rightarrow f \bar{f}) &=& \frac{(e^2,g_Z^2)}{12\pi} \left( \kappa_{V}^2 + \kappa_{A}^2 \right)\mzpri \ , 
\label{GammaZpri.EQ}
\eea
where for a quark final state the appropriate color factor (3) should be included 
(which has not been included above), and $\sqrt{k\pi r_c} = \xi$ as 
described in Sec.~\ref{twosmwz.sec}. 
$\kappa$ is the coupling of the $\zpri$ to the 
respective final states relative to that of the corresponding SM coupling. 
The fermion couplings have been defined such that the coefficient of 
$\gamma_\mu$ is $g_Z \kappa_V$ and that of $\gamma_\mu \gamma_5$ is $g_Z \kappa_A$ 
(the $\kappa$ along with the SM factors are given in Table~\ref{fermKappa.TAB} via
$\kappa_V=(\kappa_R+\kappa_L)/2,\  \kappa_A=(\kappa_R-\kappa_L)/2$ ). 
Since $\ap$ is the KK excitation of the photon, the physical Higgs modes are not 
available for it to decay into.
Next to the equations above, the order of magnitude of $\kappa$ is shown without the $(\zp,\zx)$ 
mixing factors. Including these mixing factors, the $\kappa$ are more accurately 
written as
\bea
\kappa_{\ap WW} &=& -2 s_{0L} \ , \\
\kappa_{\zpt WW} &=& s_{01} c_1 - s_{01X} s_1 - 2 c_1 s_{0L} \ , \label{zptK.EQ} \\
\kappa_{\zxt WW} &=& s_{01} s_1 + s_{01X} c_1 - 2 s_1 s_{0L} \ , \\
\kappa_{\zpt Zh} &=& \sqrt{k \pi r_c} (c_1 + \frac{g_R}{g_L} c_W c^\prime s_1) \ , \\
\kappa_{\zxt Zh} &=& \sqrt{k \pi r_c} (s_1 - \frac{g_R}{g_L} c_W c^\prime c_1) \ . \label{zxtK.EQ}
\eea
where $s_{01}$ is the (sine of the) $Z^{(0)} \leftrightarrow \zp$ mixing angle where
$Z^{(0)}$ is the $Z$ zero-mode, $s_{01X}$ that of
$Z^{(0)} \leftrightarrow \zx$, $s_1$ that of $\zp \leftrightarrow \zx$ and $s_{0L}$ that of 
$W^{(0)} \leftrightarrow \wpL$. As explained in Sec.~\ref{summary}, expressions for these mixing 
angles are given in Apps.~\ref{couplings}~and~\ref{coupl.APP}.

 \begin{figure}[tb]
\begin{center}
\scalebox{0.33}{\includegraphics[angle=0]{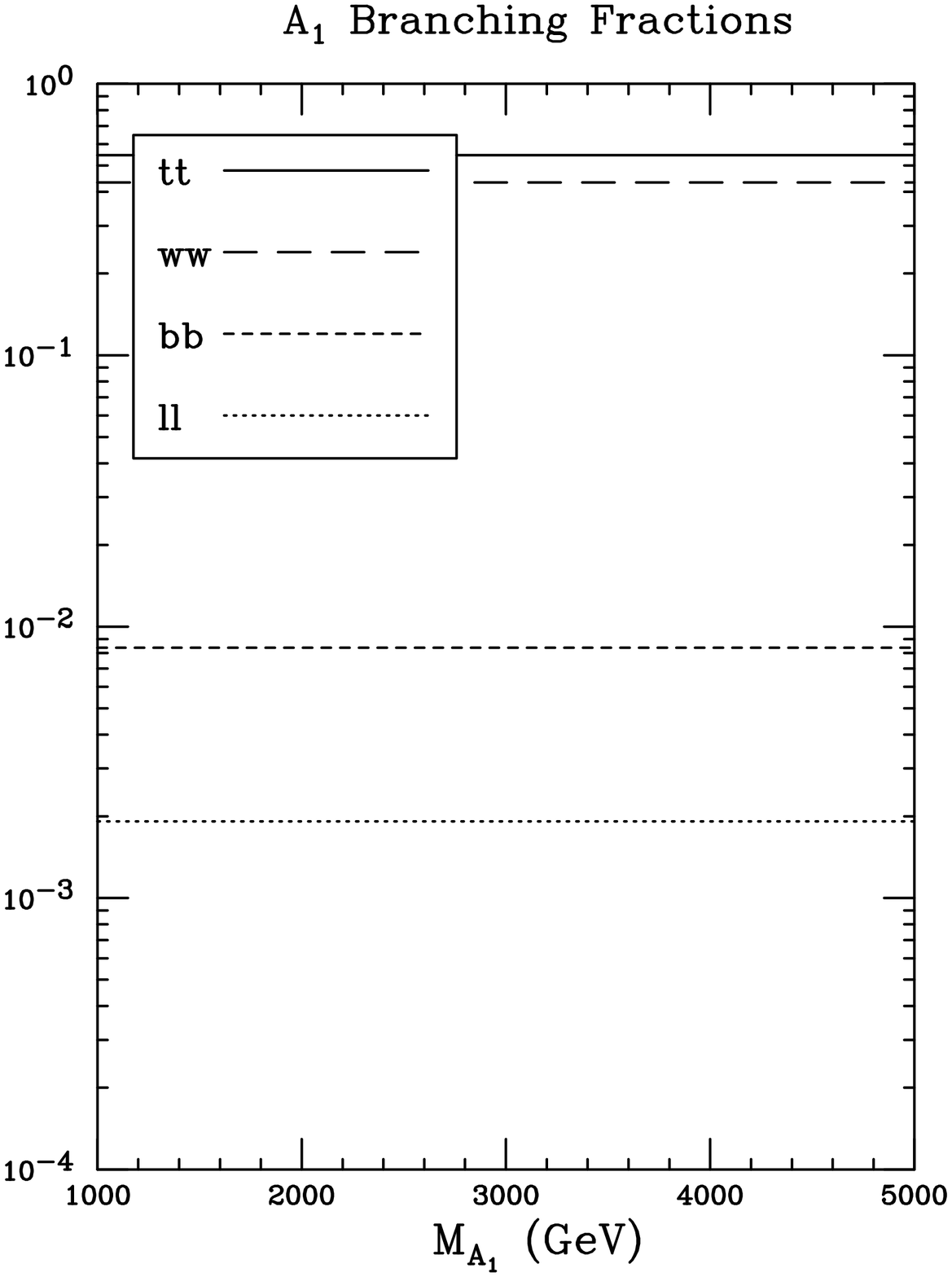}}\scalebox{0.33}{\includegraphics[angle=0]{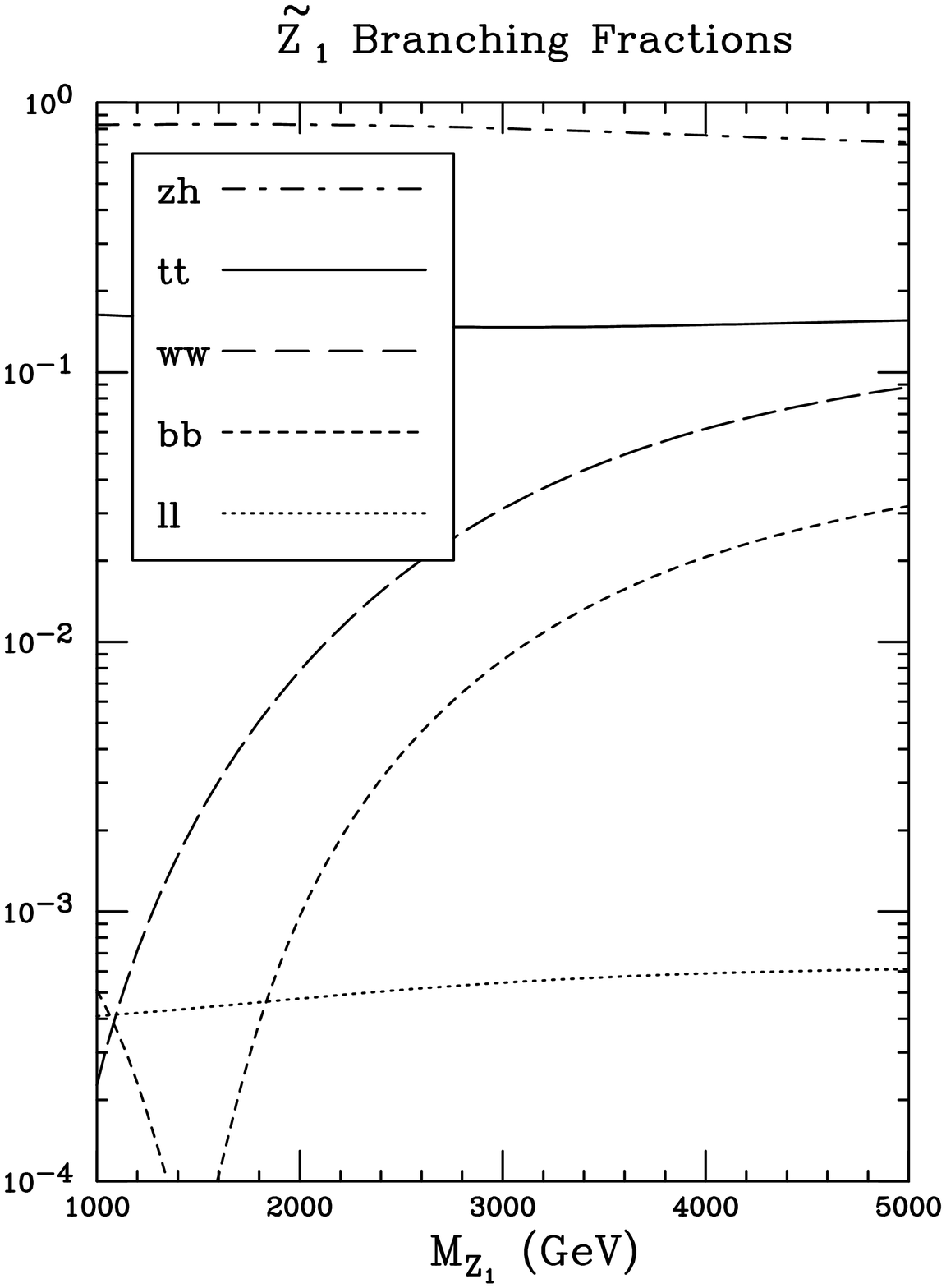}}\scalebox{0.33}{\includegraphics[angle=0]{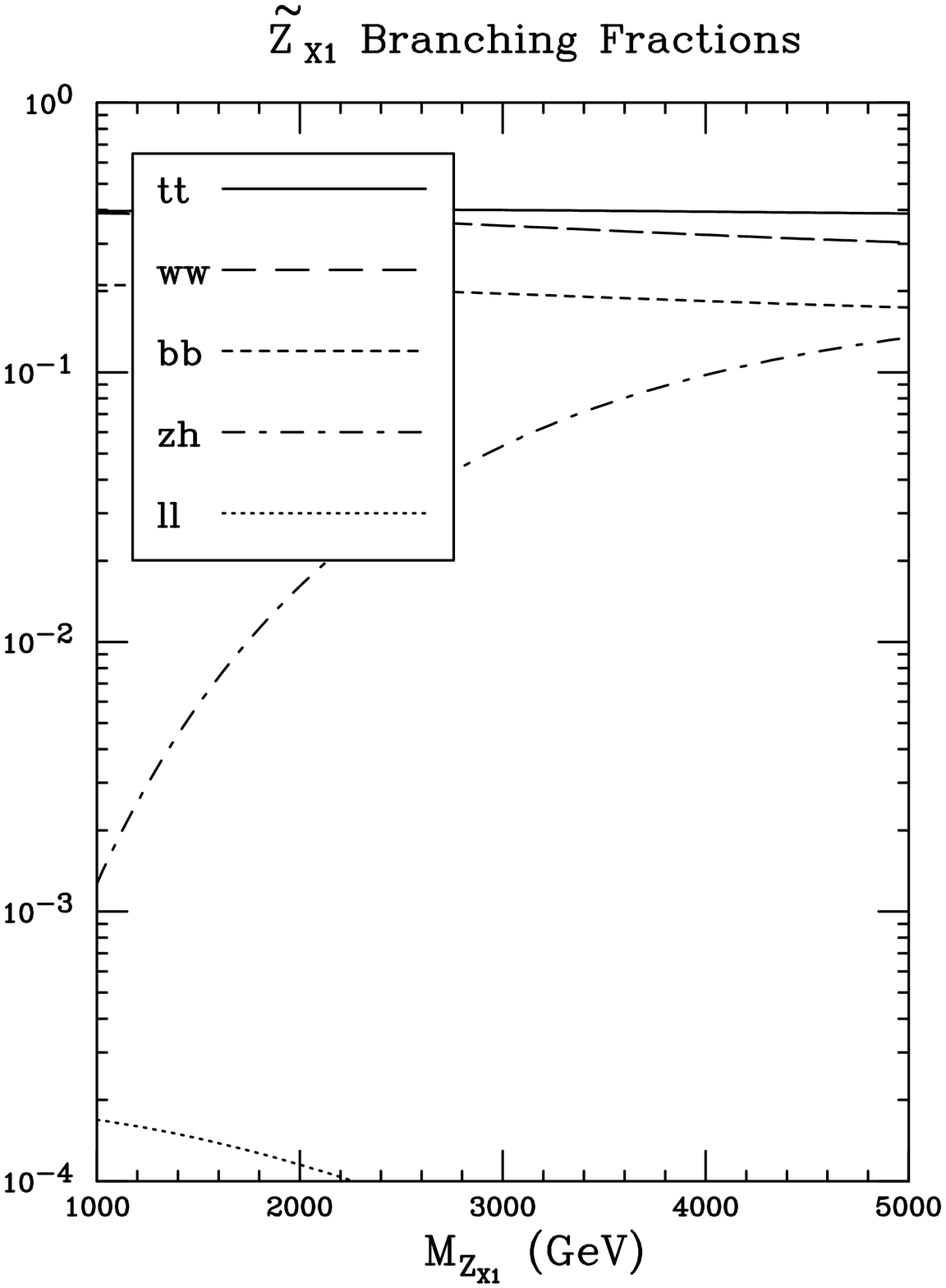}}
\caption{The branching ratios of $\zpri$ into the various modes as a function of its mass for $\ap$ (left), 
$\tilde \zp$ (center) and $\tilde{Z}_{X_1}$ (right). 
\label{BR.vs.M.FIG}}
\end{center}
\end{figure}

In Fig.~\ref{Gam.vs.M.FIG} we show the decay width as a function of $\mzpri$.
In our numerical study we set $g_R = g_L \approx g$, the SM $SU(2)_L$ coupling. 
We base our numerical study of the decay widths and BR's on the analytical calculations
presented above, with some checks performed using the program BRIDGE~\cite{Meade:2007js}.
The widths are all linearly proportional to the mass after properly taking into account the 
mixings in the couplings, 
being about $5\%$ of its mass and thus remain relatively narrow.
In Fig.~\ref{BR.vs.M.FIG} we show the $\zpri$ branching ratios into the various modes
of our current interests.  We see that for $\ap$, all channels have the trivial mass
dependence. There is no $Zh$ channel, and the two leading channels
$t\bar t$ and $WW$ are comparable. 
For $\zpt$, the leading channel is $Zh$ and the next is $t \bar t$.
The suppressed coupling to $WW$ can be understood from the equivalence 
theorem -- for the mass range shown it happens that the 
eaten charged Goldstone boson almost decouples from this state\footnote{Here
the $SU(2)_{L , R }$ couplings are set to be equal, as explained in 
appendix A.}. 
A similar argument, but for the eaten neutral Goldstone boson
explains the suppression of the $Zh$ mode in the case of $\zxt$.
In all cases, the charged lepton mode $\ell\ell$ is very small, ranging in $10^{-3} - 10^{-4}$.
As a representative, 
in Table~\ref{zpri_width.TAB} we show the partial widths and the decay branching ratios for 
$\mzpri = 2$~TeV. 
\begin{table}[h]
\caption{Partial widths and decay branching ratios for $\mzpri=2$~TeV. 
\label{zpri_width.TAB}}
\begin{center}
\begin{tabular}{|c|c|c|c|c|c|c|}
\hline 
&
\multicolumn{2}{c|}{$\ap$}&
\multicolumn{2}{c|}{$\zpt$}&
\multicolumn{2}{c|}{$\zxt$}\tabularnewline
\cline{2-3} 
\hline 
&
$\Gamma$(GeV)&
BR&
$\Gamma$(GeV)&
BR&
$\Gamma$(GeV)&
BR\tabularnewline
\hline 
$\bar{t}t$&
$55.8$&
$0.54$&
$18.3$&
$0.16$&
$55.6$&
$0.41$\tabularnewline
\hline 
$\bar{b}b$&
$0.9$&
$8.7\times10^{-3}$&
$0.12$&
$10^{-3}$&
$28.5$&
$0.21$\tabularnewline
\hline 
$\bar{u}u$&
$0.28$&
$2.7\times10^{-3}$&
$0.2$&
$1.7\times10^{-3}$&
$0.05$&
$4\times10^{-4}$\tabularnewline
\hline 
$\bar{d}d$&
$0.07$&
$6.7\times10^{-4}$&
$0.25$&
$2.2\times10^{-3}$&
$0.07$&
$5.2\times10^{-4}$\tabularnewline
\hline 
$\ell^{+}\ell^{-}$&
$0.21$&
$2\times10^{-3}$&
$0.06$&
$5\times10^{-4}$&
$0.02$&
$1.2\times10^{-4}$\tabularnewline
\hline 
$W_{L}^{+}W_{L}^{-}$&
$45.5$&
$0.44$&
$0.88$&
$7.7\times10^{-3}$&
$50.2$&
$0.37$\tabularnewline
\hline 
$Z_{L}h$&
-&
-&
$94$&
$0.82$&
$2.7$&
$0.02$\tabularnewline
\hline 
Total&
$103.3$&
&
$114.6$&
&
$135.6$&
\tabularnewline
\hline
\end{tabular}
\end{center}
\end{table}

The $\zpt$ and $\zxt$ BR's into some modes show interesting behavior due to the following:
For decay into $WW$ and $Zh$ the dependence of the couplings shown in 
Eqs.~(\ref{zptK.EQ})$-$(\ref{zxtK.EQ}) have
nontrivial dependence as a function of $\mzpri$ (cf App.~\ref{coupl.APP}). 
In particular, in some cases, the various 
mixing angle terms can conspire resulting in an accidentally small coupling
which leads to a small BR, and since the mixing angles depend on $\mzpri$, the BR varies with mass.
Also, for fermionic modes the couplings as shown in 
Eq.~(\ref{GammaZpri.EQ}) and Table~\ref{fermKappa.TAB} 
can lead to nontrivial behavior with $\mzpri$, and in certain cases, 
depending on the $SU(2)_L$ and $SU(2)_R$ charges 
of the particular fermion, can again lead to an accidentally small coupling.
The profiles of the left- and right-handed fermions in the extra dimension also determine
the coupling, and thus whether couplings can be accidentally small
or how they depend with $\mzpri$.

\section{Heavy Gauge Boson Production and Their Search at the LHC}
\label{signals}
We now consider the $\zpri$ production at the LHC.  
We depict the representative Feynman diagrams  in Fig.~\ref{prodchs.FIG} 
for the potentially large  production channels in hadronic collisions
and we show the numerical results in Fig.~\ref{xsmassz.FIG} versus its mass. 
Figure \ref{xsmassz.FIG}(a) shows the production rates for the KK interaction eigenstates
versus the mass parameter
$m_{KK}$ by pulling out the model-dependent overall coupling constant squared
($\lambda^2$) as given in Table~\ref{lambda} of 
Appendix~\ref{coupl.APP} (including the SM couplings  in the curves), 
which reflect the bare-bone features convoluted with the parton distribution functions. 
As one may anticipate, the two leading channels for the $Z_{KK}$ production are
from Drell-Yan (DY) production shown in Fig.~\ref{prodchs.FIG}(a) and the
weak boson fusion (WBF) shown in Fig.~\ref{prodchs.FIG}(b). Although the WBF
process is formally higher order in electroweak couplings, the $t$-channel enhancement
of gauge boson radiation off the quarks and the strong couplings of the longitudinal 
gauge bosons at higher energies could potentially bring this channel comparable or larger 
than that of DY for $m_{KK}>1$ TeV. In spite of the enhanced coupling of $b\bar b$
to $Z_{KK}$, this contribution is still much smaller than that from the light quarks
due to the small $b$-quark parton density at high $x$ values. 

Figure \ref{xsmassz.FIG}(b) includes the full couplings and mixings for the mass eigenstates
and gives the absolute normalization versus the physical mass for a generic $Z'$. 
Although the couplings of $\zpri$ to light fermions are suppressed in the RS model
setting, the main production mechanism is still from the DY as shown in Fig.~\ref{xsmassz.FIG}(b), 
with about $91\%$ from  light valence  quarks and $9\%$ from $b\bar b$ for a 2 TeV mass.
With the enhanced coupling of $\zpri$  to the {\em longitudinal} gauge bosons, one would 
naively expect a large contribution from the WBF
process as implied in Fig.~\ref{xsmassz.FIG}(a).  However, 
%
since the triple $WW\zpri$ vertex is only induced by the EWSB and the coupling strength is 
proportional to  $\xi (m_Z/\mzpri)^2$, the suppression as seen clearly in
Fig.~\ref{xsmassz.FIG}(b).
There are other production channels to contribute. For instance, 
due to the substantial coupling of $Z'$ to the top quark, one may also consider the
process of $\zpri$ radiation off a top quark.  
This is suppressed by a three-body kinematics
and was shown to be much smaller than $b\bar b \to \zpri$ \cite{Djouadi:2007eg}.
Similarly, the process $gg\to \zpri^*$ via heavy quark triangle diagrams 
 must go through an off-shell production and is highly suppressed \cite{Djouadi:2007eg}.
One may also consider the associated production $Z'W$ or $Z'h$, but they are
subleading to the DY process and we will not pursue any detailed studies for those
channels. 

\begin{figure}
\begin{center}
\begin{minipage}{0.3\textwidth}
\includegraphics[width=0.85\textwidth]{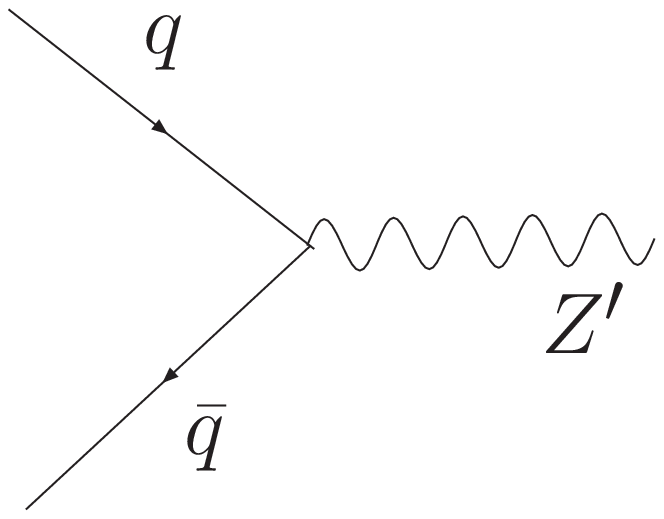} \vspace*{-0.5cm}\begin{center}(a)\end{center}
\end{minipage}
\begin{minipage}{0.4\textwidth}
\includegraphics[width=0.8\textwidth]{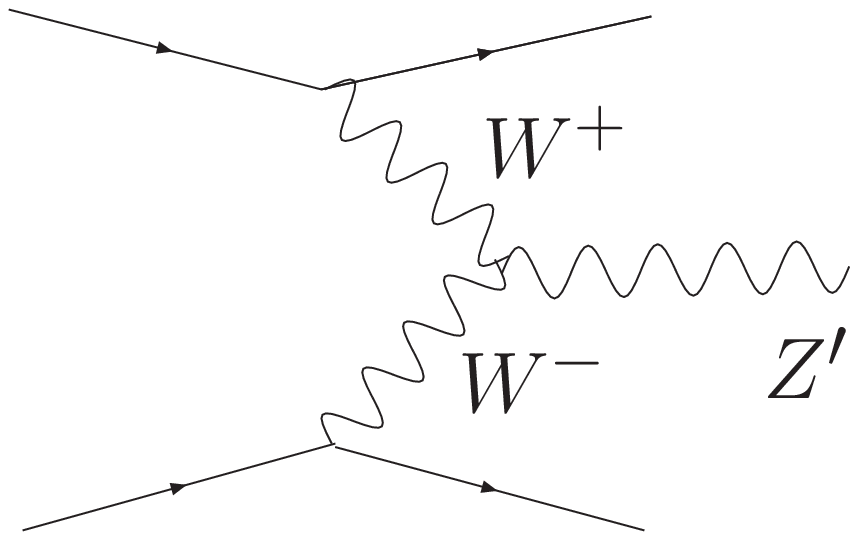} \vspace*{0.0cm}\begin{center}(b)\end{center}
\end{minipage}
%
%
%
%
%
%
%
%
%
%
%
\end{center}
\caption{Representative Feynman diagrams for the $Z^\prime$ production channels.
\label{prodchs.FIG}}
\end{figure}


To further quantify the search sensitivity to the $\zpri$, we will thus concentrate 
on the DY process shown in Fig.~\ref{prodchs.FIG}(a). 
We include the coherent sum of the $\ap$, $\zpt$ and $\zxt$ contributions to a particular 
final state in the following. 
Throughout this section, we set $g_R = g_L \approx g$, the SM $SU(2)_L$ coupling.
We include b-quarks in the 
initial state along with the light quarks. We use the CTEQ6.1M parton distribution
functions \cite{Pumplin:2002vw} for all our numerical calculations. 
We have obtained the results in this section by incorporating
our model into CalcHEP~\cite{CalcHEP} and performed some checks by adding our model into 
Madgraph~\cite{Alwall:2007st}.

\begin{figure}
\begin{center}
\scalebox{0.4}{\includegraphics[angle=0]{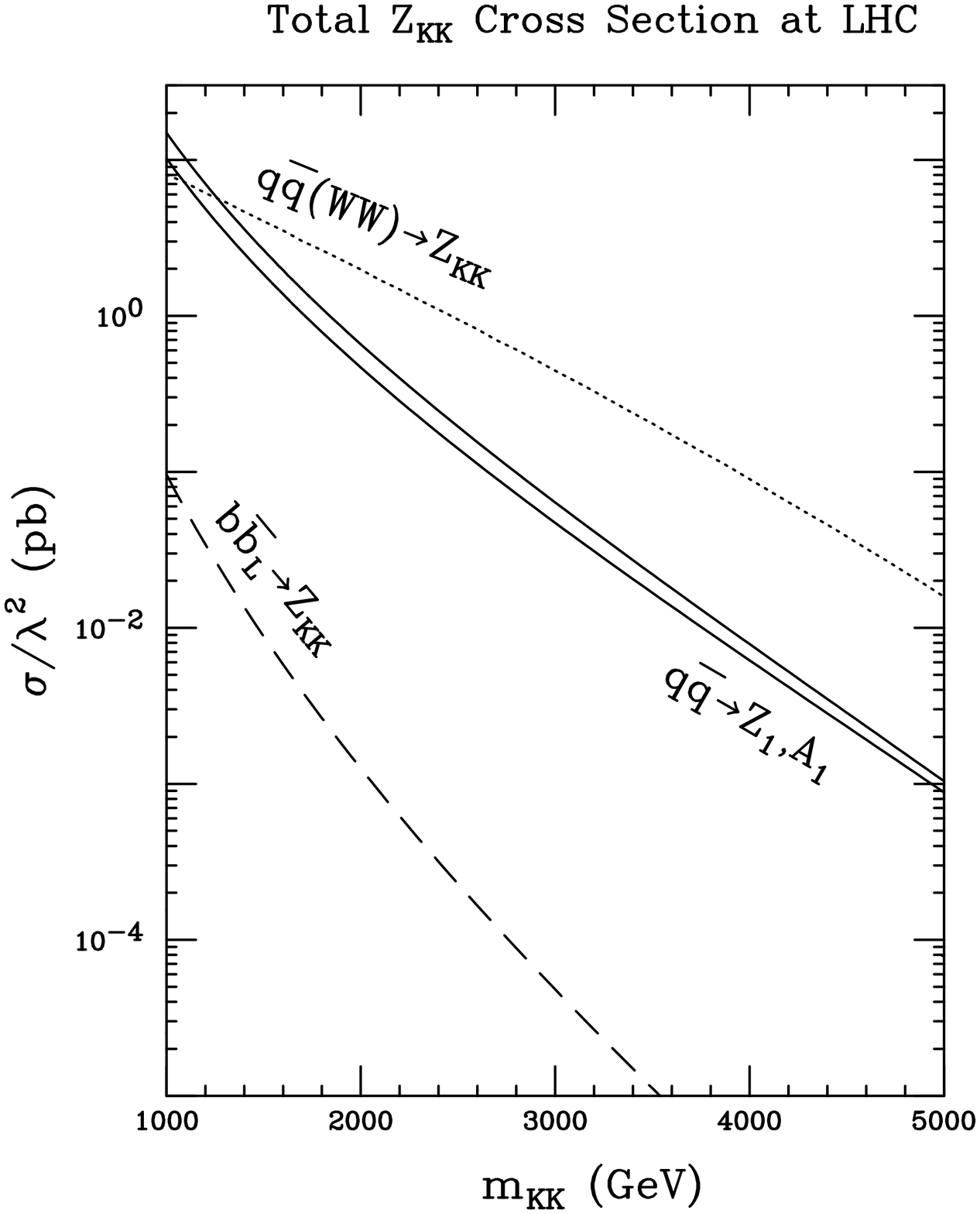}}
\scalebox{0.4}{\includegraphics[angle=0]{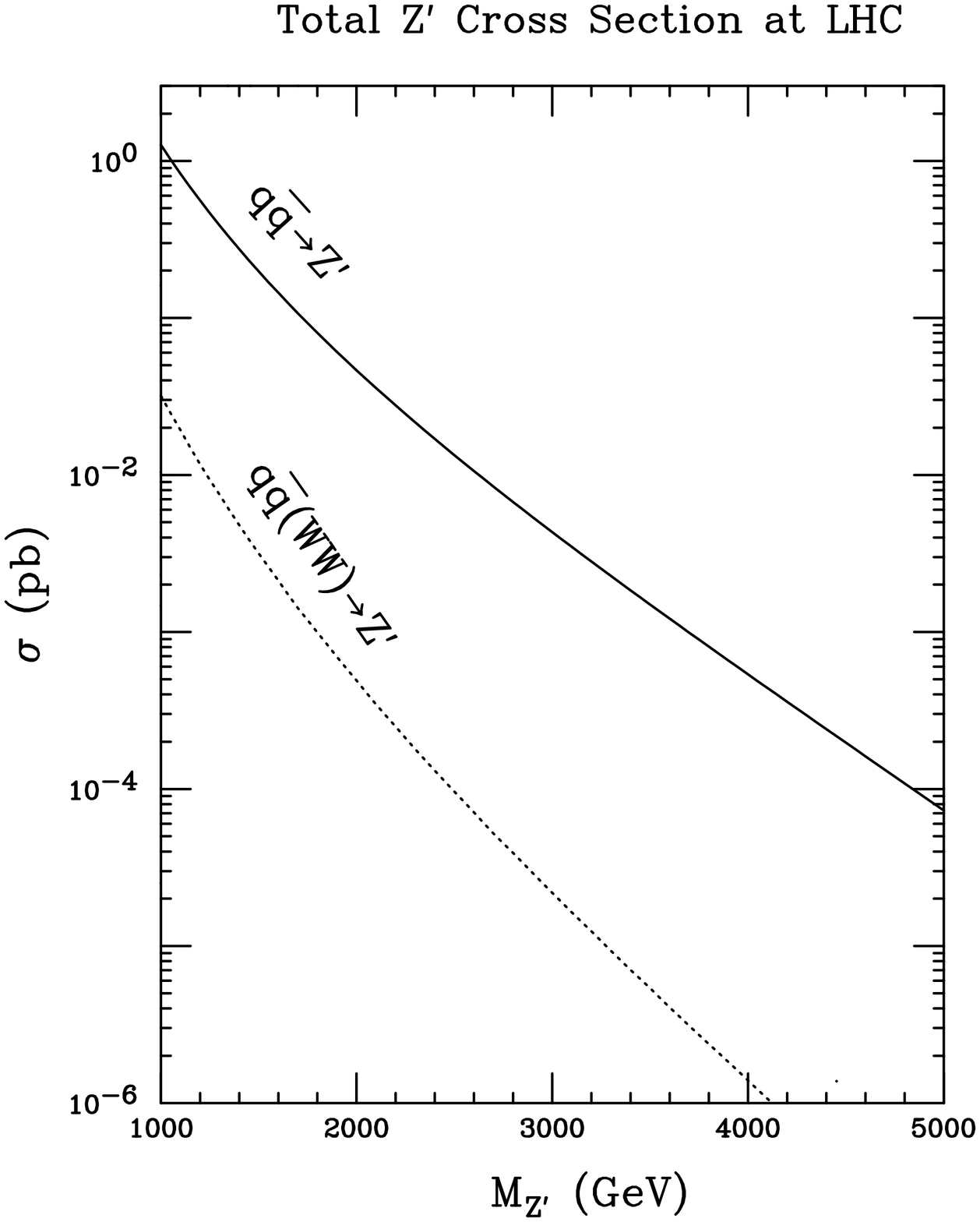}}
\caption{Total cross section for $\zpri$ production versus its mass,
(a) with the coupling constant squared ($\lambda^2$) factored out as in Table~\ref{lambda}
in Appendix B (for states in the KK eigenbasis, 
where $Z_{KK}$ includes $\ap$, $\zp$ and $\zx$, and the 
$q\bar q \zx$ coupling is vanishingly small), 
and (b) with the absolute normalization for the couplings (for states in the mass eigenbasis). 
\label{xsmassz.FIG}}
\end{center}
\end{figure}

\subsection{$\ap,\ \zxt  \rightarrow W^+W^-$}
\label{zpri2WW.SEC}
As seen from the discussion for the $\zpri$ decay in Sec.~\ref{zpridec.SEC}, 
$\ap$ and $\zxt$ decay to   $W^+W^-$ with substantial branching fraction
of $30-40\%$, 
while for $\zpt$ it is down by more than one order of magnitude. 

\begin{figure}
\begin{center}
\scalebox{0.5}{\includegraphics[angle=270]{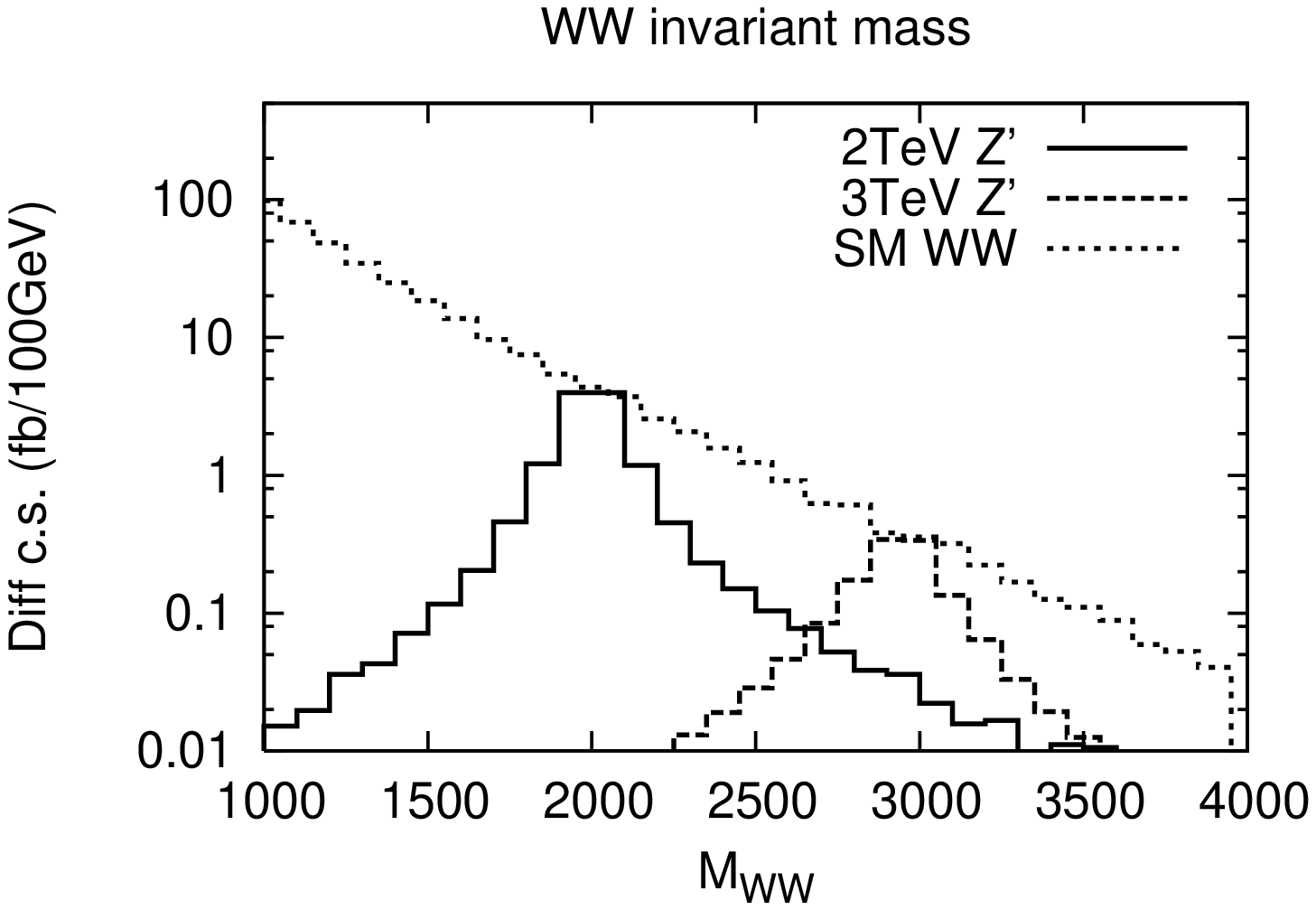}}
\scalebox{0.5}{\includegraphics[angle=270]{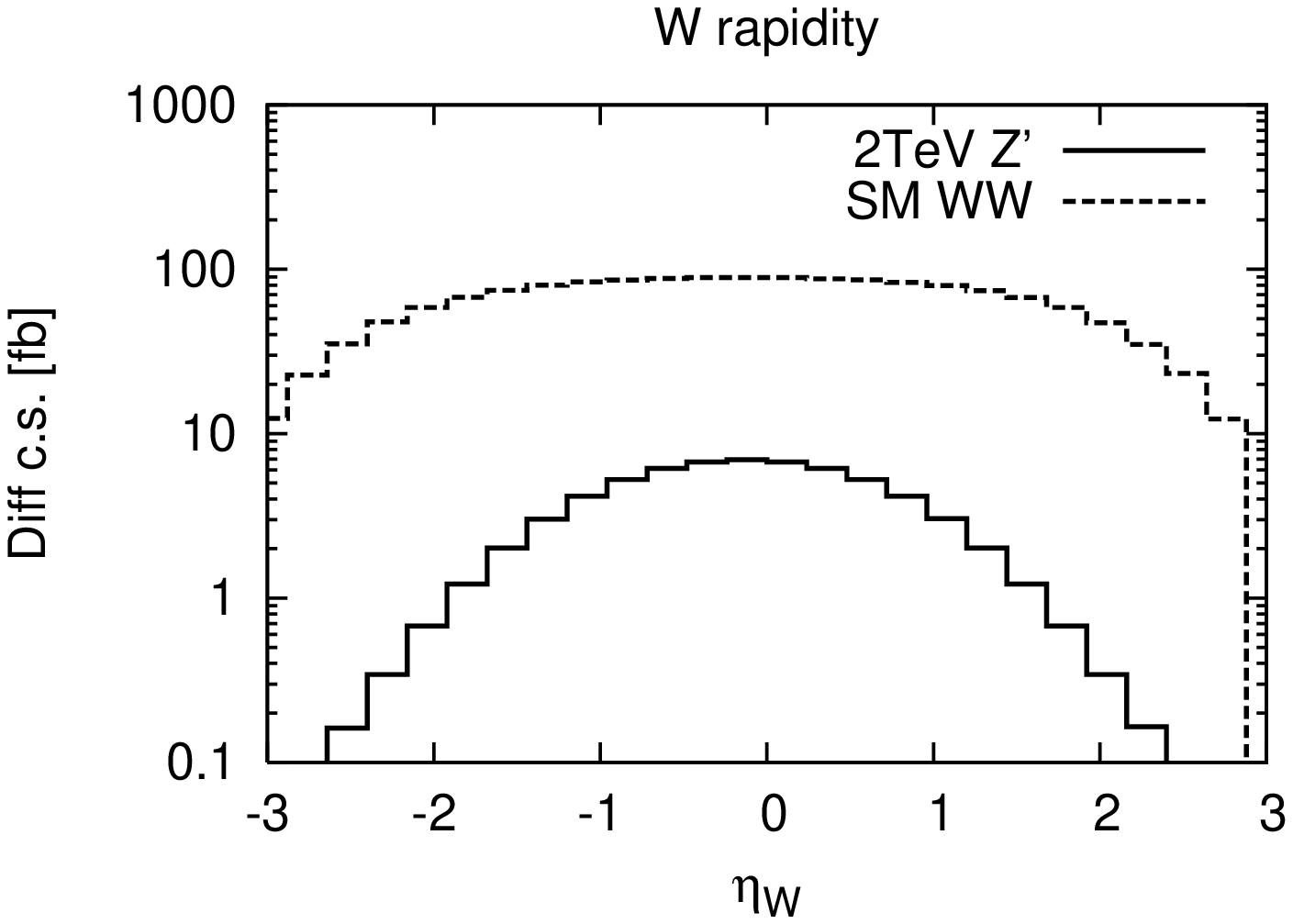}}
\caption{Distributions of the $WW$ final state  (a) for $W^+ W^-$ invariant mass variable
(in GeV) of a 2 and 3 TeV $\zpri$ along with the SM $W^+W^-$ background, and 
(b) for rapidity of a $W$. These are after a ${p_T}_W > 250~$GeV cut.
\label{dyzpriww.FIG}}
\end{center}
\end{figure}
To gain a qualitative sense first, we consider the differential cross section for the
signal with a  mass of 2 and 3 TeV and the irreducible SM background of $W^+W^-$ pair
production in Fig.~\ref{dyzpriww.FIG}, for (a) the invariant mass distributions $M_{WW}$,
and for (b) the rapidity distribution $\eta_W$. 
These are after a ${p_T}_W > 250$~GeV cut. 
The signal cross-section before any cuts is about 16~fb for a mass of 2~TeV, and
1.3~fb for 3~TeV mass.  
Based on the distributions, the signal can be enhanced relative to background 
by the application of suitable $M_{WW}$ and $\eta$ cuts.
We see clearly the good signal observability, and we consider in the following how
to realize these cuts using only the observable particles resulting from the
decay of the two $W$'s. Additional sources of background will have to be contended with when 
one considers specific decay modes.
%
%

For the observable final states, we will not consider the  fully hadronic mode 
for $WW$ decays due to the formidable QCD di-jet background.
We will propose to focus on the purely-leptonic and semi-leptonic channels.

\subsubsection{Purely leptonic channel:}

We first consider the purely leptonic mode, $\zpri \to WW\to \ell\nu \ell\nu\ (\ell=e,\mu)$, 
which provide the clean channels from the observational point of view. 
The price to pay is the rather small branching ratio $BR(WW) \approx (2/9)^2= 4/81$,
in addition to the inability to reconstruct the total invariant mass due to the 
presence of two neutrinos carrying away missing momentum.
We select the events with the basic acceptance cuts 
\bea
 p_{T\ell} > 50\ {\gev},\quad |\eta_\ell | < 3,\quad  \Delta R_{\ell\ell} > 0.4, \quad \etmiss > 50\ {\gev},
\eea
where $p_{T\ell},\ \eta_\ell$ are the transverse momentum and pseudo-rapidity of the
charged leptons,  $\Delta R_{\ell\ell}$ the separation of $\ell\ell$, and $\etmiss$ the
missing transverse energy due to the neutrinos.
The leading irreducible backgrounds include $W^+W^-  \to \ell^+ \ell^- \etmiss$ and 
$Z/\gamma^*\to \tau^+\tau^- \to \ell^+ \ell^- \etmiss$. 

\begin{figure}
\begin{center}
\scalebox{0.5}{\includegraphics[angle=270]{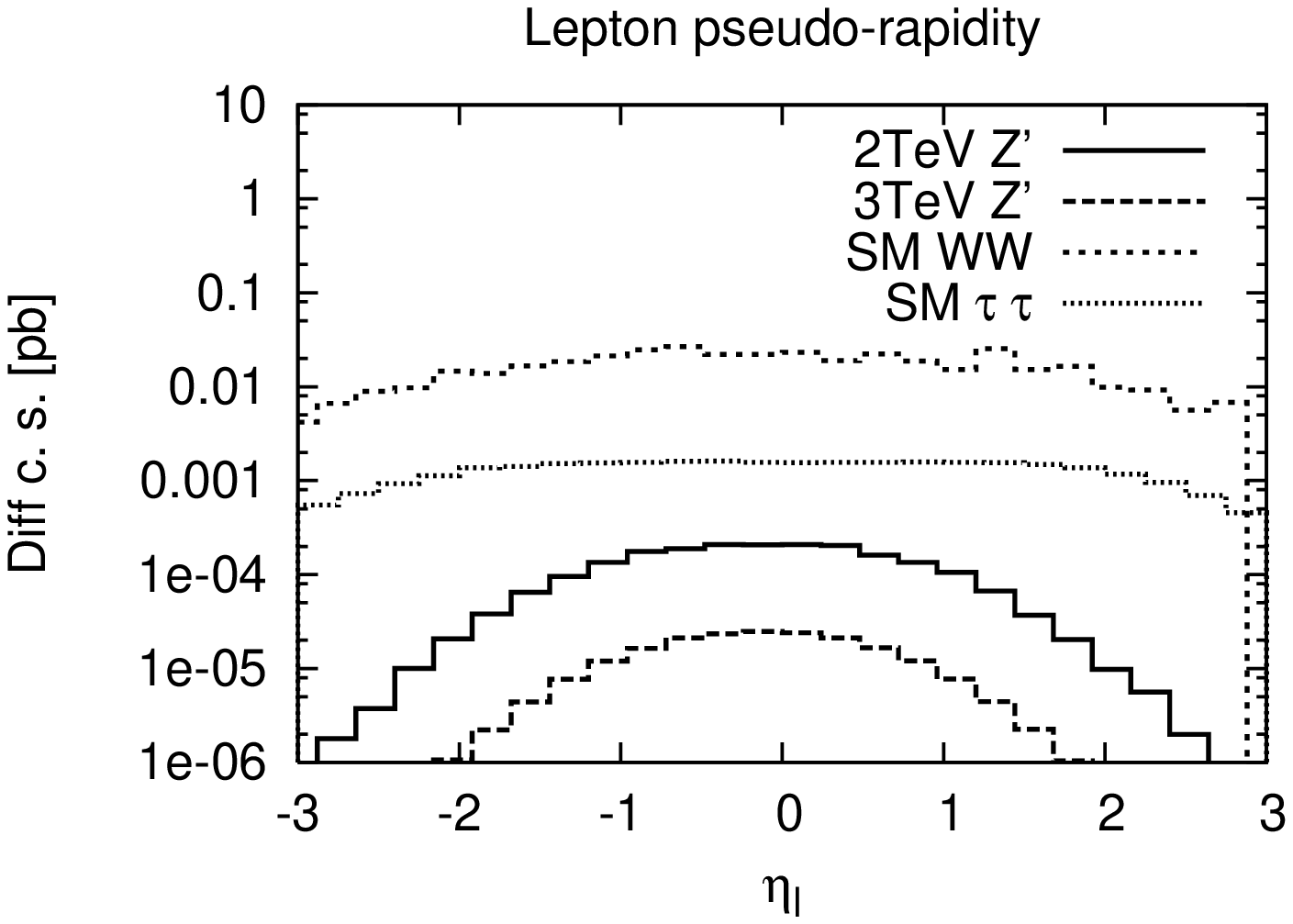}}
\scalebox{0.5}{\includegraphics[angle=270]{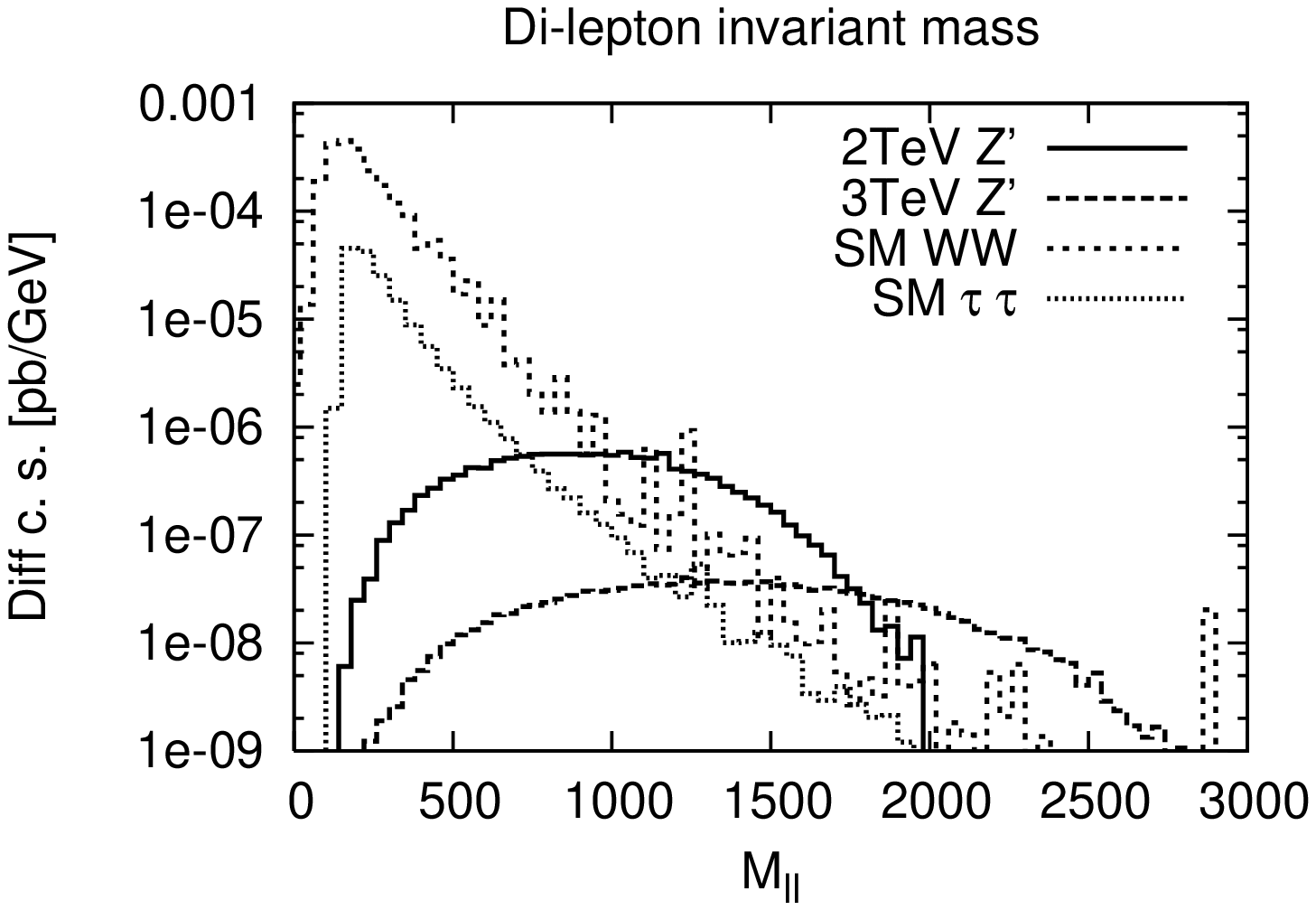}}
\scalebox{0.5}{\includegraphics[angle=270]{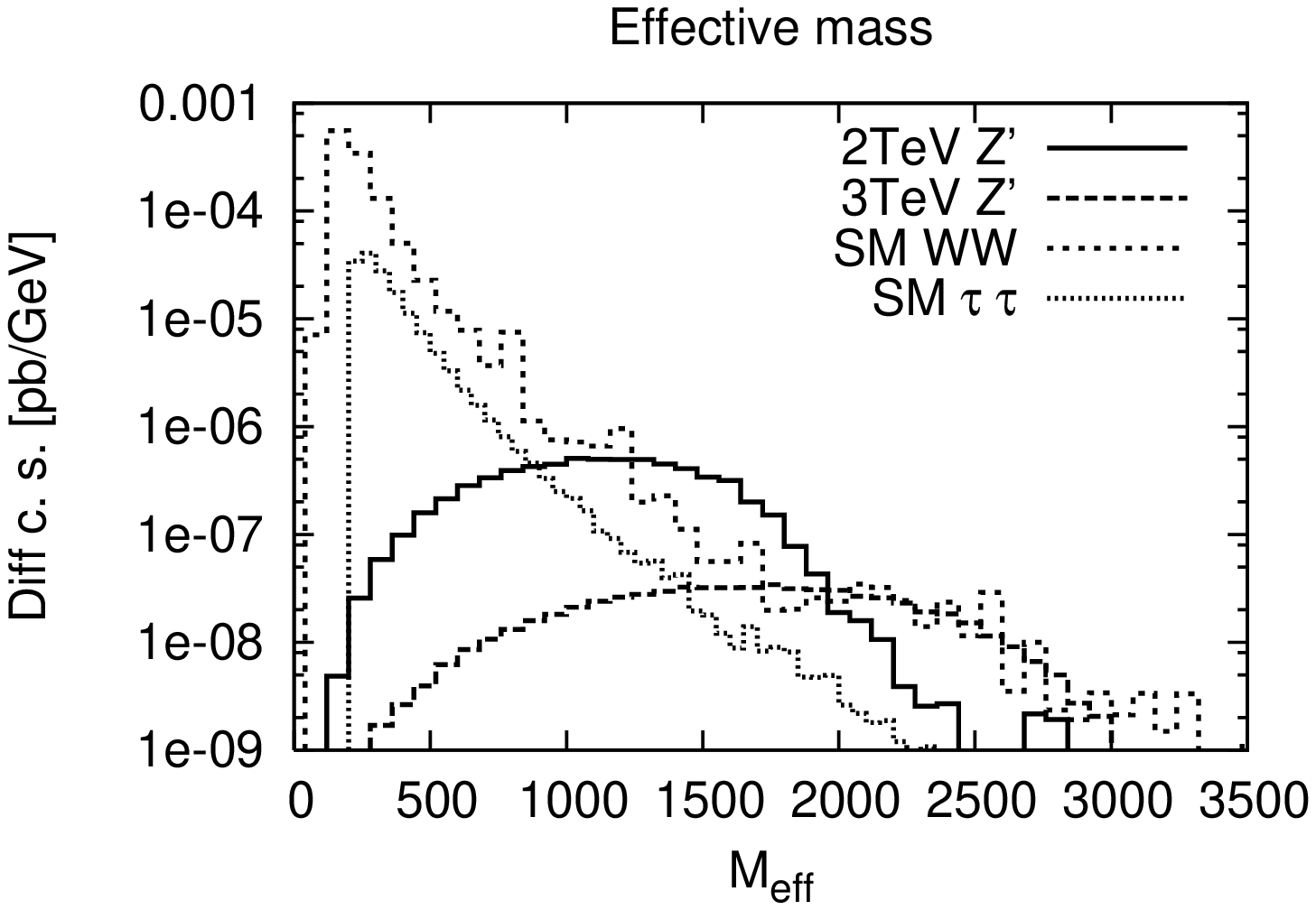}}
\scalebox{0.5}{\includegraphics[angle=270]{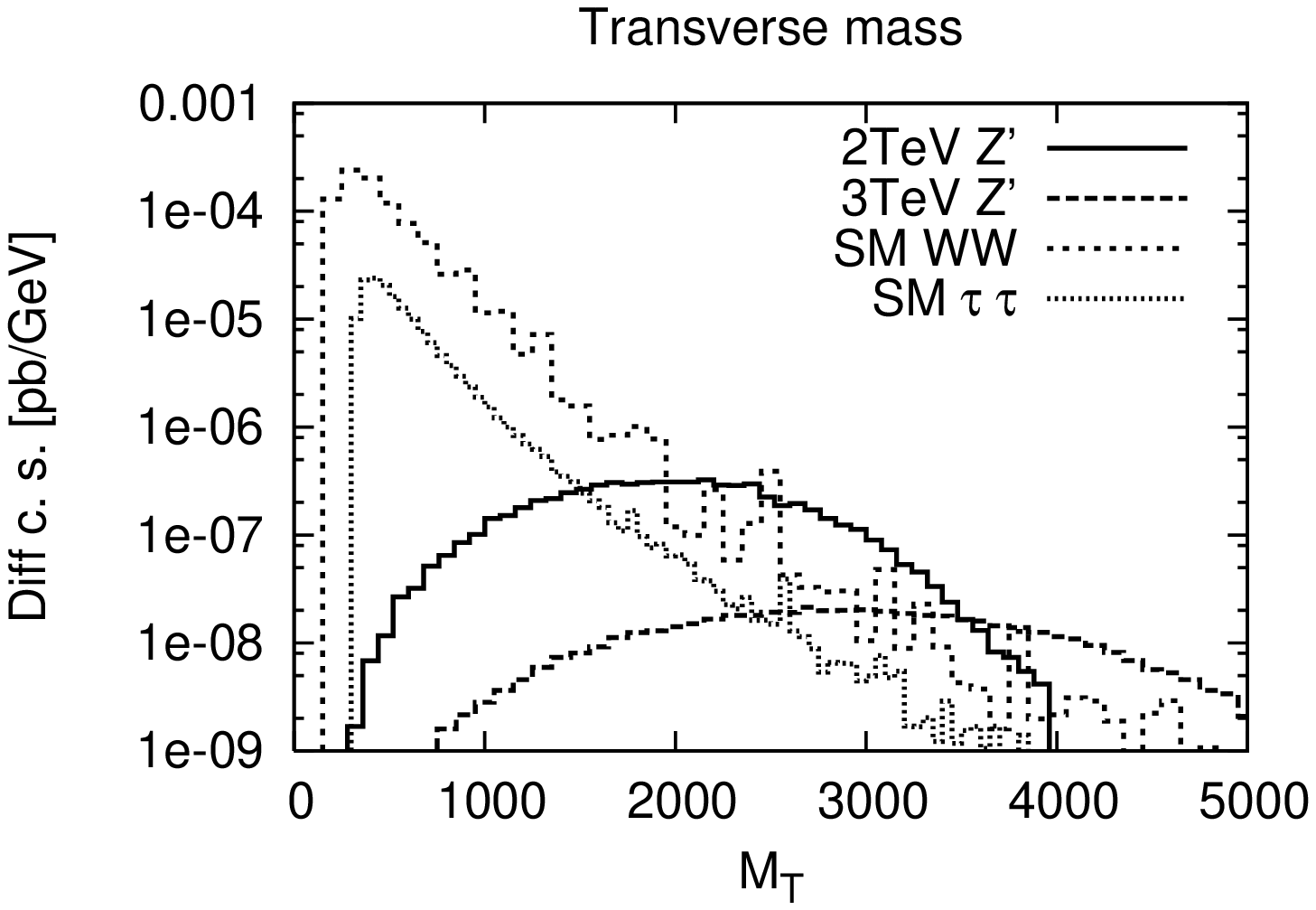}}
\caption{The differential cross-section from Drell-Yan production of 2~TeV $\zpri$ into
the $WW$ final state followed by $W^+\rightarrow \ell^+\nu$ and 
$W^-\rightarrow \ell\bar\nu$ at the LHC (all horizontal axis masses in GeV) for 
(a) pseudo-rapidity distribution of the charged lepton, 
(b) invariant mass of the charged lepton pair,
(c) the effective mass distribution, 
and (d) the cluster transverse mass distribution. 
These distributions are after the basic cuts.
\label{dyzpriEmNn.FIG}}
\end{center}
\end{figure}

Although we will not be able to fully reconstruct the resonant variable of the invariant
mass $M_{WW}$, 
we form the ``effective mass''  and cluster transverse mass defined by
\beq
M_{eff} \equiv {p_T}_{\ell_1} + {p_T}_{\ell_2} + \ptmiss,\quad
{M_T}_{WW} \equiv 2 \sqrt{{p^2_T}_{\ell\ell} + M_{\ell\ell}^2}.
\eeq
In Fig.~\ref{dyzpriEmNn.FIG} we show the different characteristics of the backgrounds
and the signal for a 2 and 3 TeV mass, for (a) the pseudo rapidity $\eta$ distribution, 
(b) the effective mass distribution, (c) invariant mass of the lepton pair,  and (d) the
cluster transverse mass distribution. The variable ${M_T}_{WW}$ ($M_{\ell\ell}$)
should be broadly peaked at the resonance mass (half of it), the $M_{eff}$ gives
the typical energy scale of the object produced. 
We are motivated to tighten up the kinematical cuts to further improve the signal
observability. The cuts and results are  shown in Table~\ref{WW2llNn.TAB}.  
We see that the backgrounds can be suppressed to the level of $S/B \sim 1$,
but the signal rate is rather low. For a 2 TeV $Z'$, it is conceivable to reach a
5$\sigma$ statistical sensitivity with an integrated luminosity of 100 ${\fbi}$, while
for a 3 TeV $Z'$ a higher luminosity would be needed to have a clear observation
of the signal.

\begin{table}[h]
\caption{$pp\rightarrow \ell^+ \ell^- \etmiss$ cross-section (in fb) for 
the signal with
$\mzpri=2$, 3 TeV and the $WW$ and $\tau\tau$ backgrounds, with cuts 
applied
successively ($M_{eff}$ and $M_T$ are in TeV).
The number of events and statistical significance are shown
for  100 ${\rm fb}^{-1}$ ($\mzpri = 2$ TeV)
and  1000 ${\rm fb}^{-1}$ (3 TeV), respectively.
\label{WW2llNn.TAB}}
\begin{center}
\begin{tabular}{|c||c|c|c|c|c|c|c|}
\hline
$2$ TeV
& Basic cuts &
$|\eta_\ell | < 2$&
$M_{eff}>1$ TeV&
$M_{T}>$1.75 TeV&
\# Evts &
$S/B$&
$S/\sqrt{B}$\tabularnewline
\hline
\hline
Signal &
$0.48$&
$0.44$&
0$.31$&
$0.26$&
$26$&
$0.9$&
$4.9$
\tabularnewline
\hline
$WW$&
$82$&
$52$&
$0.4$&
$0.26$&
$26$&
$$&
$$\tabularnewline
\hline
$\tau\tau$&
$7.7 $&
$5.6 $&
$0.045 $&
$0.026$&
$ 2.6$&
&
\tabularnewline
\hline
\hline
$3$ TeV&
 Basic cuts&
$|\eta_{\ell}|<2$&
$1.5<M_{eff}<2.75$&
$2.5<M_{T}<5$&
\# Evts&
$S/B$&
$S/\sqrt{B}$\tabularnewline
\hline
\hline
Signal&
$0.05$&
$0.05$&
$0.03$&
$0.025$&
$25$&
&
\tabularnewline
\hline
$WW$&
$82$&
$52$&
$0.08$&
$0.04$&
$40$&
$0.6$&
$3.8$\tabularnewline
\hline
$\tau\tau$&
$7.7$&
$5.6$&
$0.015$&
$0.003$&
$3$&
&
\tabularnewline
\hline
\end{tabular}
\end{center}
\end{table}
%

\subsubsection{Semi-leptonic channel:}
\label{zpri2WW2SL.SEC}
To increase the statistics, we next consider the 
semi-leptonic mode when one $W$ decays as $W\rightarrow \ell\nu\ (\ell=e,\ \mu)$ 
while the other as $W\rightarrow j j^\prime$ ($j$ denotes a jet from a light quark). 
The branching ratio for this channel is $BR(WW) \approx 2/9 \times 6/9 \times 2= 8/27$, 
and the factor of 2 is due to including both $\ell^+$ and $\ell^-$.

\begin{figure}
\begin{center}
\scalebox{0.53}{\includegraphics[angle=270]{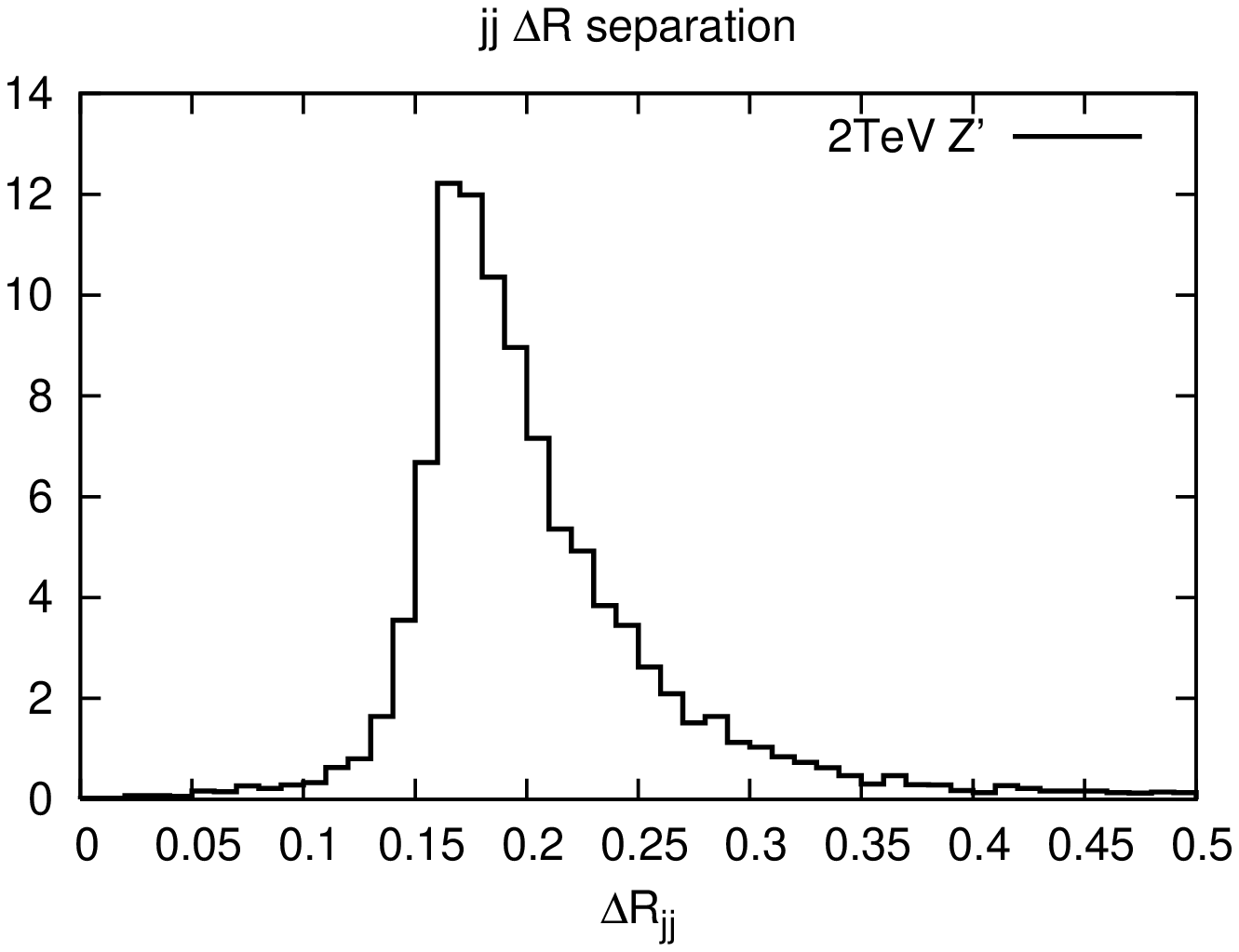}}
\scalebox{0.53}{\includegraphics[angle=270]{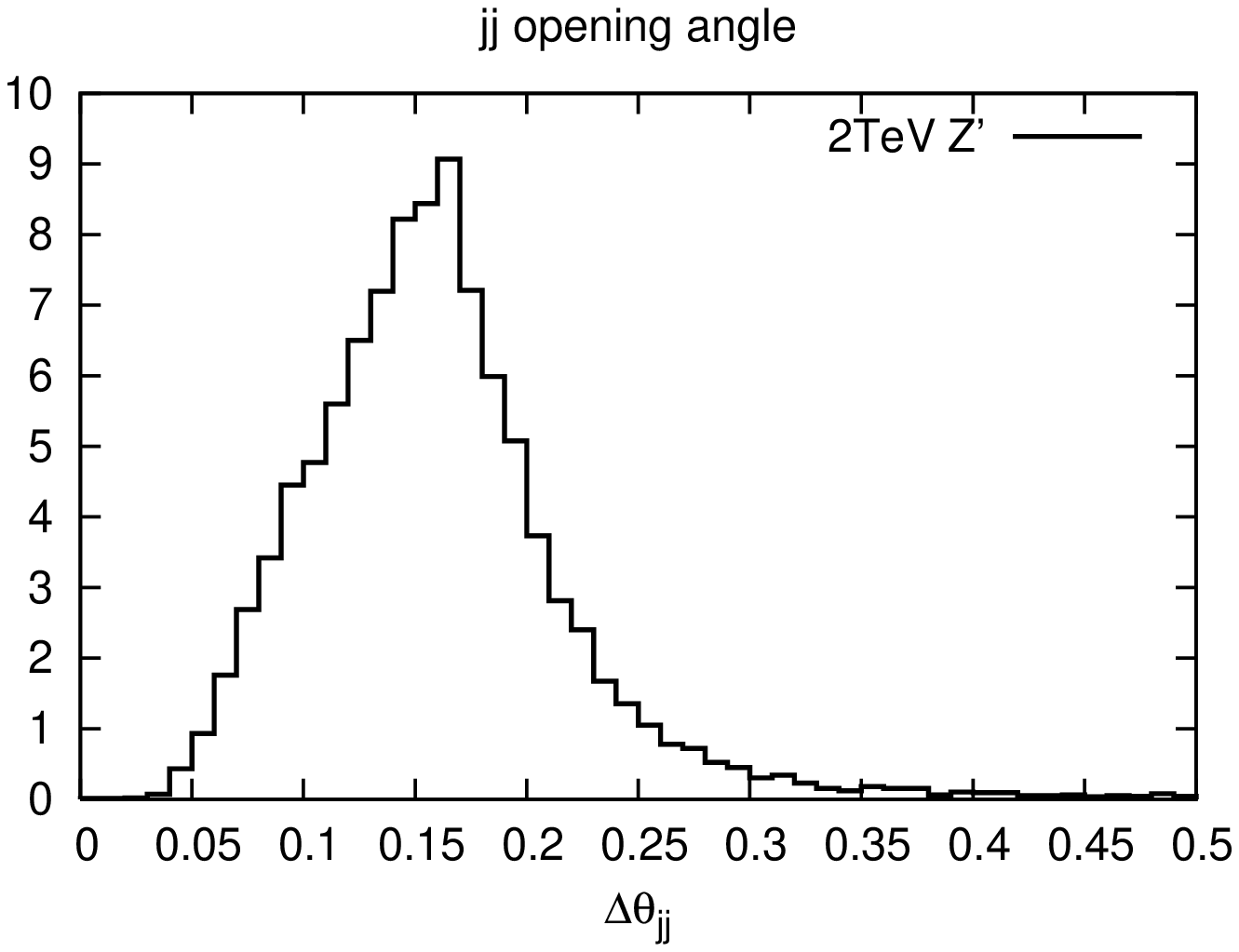}}
\caption{Distributions of the two jets resulting from $W\rightarrow jj$ 
from the Drell-Yan production of a 2~TeV $\zpri$ for (a) the
separation $\Delta R$ 
and (b) the lab-frame opening angle.
\label{dRzprijj.FIG}}
\end{center}
\end{figure}
%
Owing to the large mass of the $\zpri$, the two $W$'s are significantly
boosted resulting in their decay products highly collimated in the lab-frame. 
To illustrate this we show in Fig.~\ref{dRzprijj.FIG} the distribution of (a) the
separation $\Delta R$ 
and (b) the lab-frame opening angle of the decay products of two fermions 
of the $W$ for $\mzpri = 2$~TeV. It  can be seen from the 
figures that  the separation is strongly peaked around 0.16, consistent with $2 M_W/p_T$ 
for the opening angle. 
This kinematical feature has significant impact on the searches.
The presently typical jet reconstruction cone size of $\Delta R=0.4$ 
will cause these two jets from the $W$ decay to be reconstructed likely as a single jet 
(albeit a fat jet), 
This means that we would pick up the SM single jet as a background
for each $W$ decaying hadronically. 
For the leptonically decaying $W$, the charged lepton and the missing neutrino
will be approximately collinear as well, rendering the accurate determination of
the missing transverse energy difficult, although making the kinematics simpler.

As explained above, the two jets may not be resolvable due to collimation,
and merged as a single jet, and thus the process $W(\rightarrow \ell\nu) + 1$ QCD jet
turns out to be the leading background, aside from the semileptonic decay from 
$WW$ production.\footnote{$t\bar{t}$ production can be a source of 
(reducible) background, but a jet-veto on the leptonic side can be used to 
suppress this.}
We adopt the event selection criteria  with the basic cuts
\bea
&& p_{T\ell} > 50\ {\gev},\quad |\eta_\ell | < 1,\quad  \Delta R_\ell > 0.4, \quad \etmiss > 50\ {\gev},\\ 
&& E_{Tj } > 100\ {\gev},\quad |\eta_j | < 1, \quad \Delta R_j > 0.4.
\eea

In order to capture the feature of the production of a very massive object, and to
reconstruct the $\zpri$ mass, we once again consider the effective mass and 
the transverse mass defined as
\beq
M_{eff} \equiv {p_T}_{jj} + p_{T\ell} +  \ptmiss,\quad 
{M_T}_{WW} = 2 \sqrt{E_{Tjj}^2 + M^2_W}, 
\label{mTww}
\eeq
where $E_{Tjj}$ is the transverse energy for the jet pair which presumably 
reconstructs to the hadronic $W$. 
Alternatively, one can design a more sophisticated 
variable in the hope to reconstruct the invariant mass for the semi-leptonic system.
This makes use of the fact that the missing neutrino is collimated with the charged lepton 
and we thus can expect to approximate the unknown longitudinal component of the $\nu$
momentum by
\beq
p_{L\nu} = {\etmiss\over p_{T\ell} }\  p_{L\ell}.
\label{pln}
\eeq
%
The momentum of the leptonic $W$ is thus reconstructed and  we can evaluate the 
invariant mass of the semi-leptonic system by $\bar M_{WW}^2 = (p_{\ell \nu}+ p_{jj})^2$.
In Fig.~\ref{jetMwj.FIG}(a), we show the distributions for the two variables
$M_{TWW}$ (solid curve) and $\bar M_{WW}$ (dashed curve) along with the continuum 
background. These variables reflect the resonant  feature rather well. 
We find the following cuts effective in reducing the QCD background for two representative
values of $M_{Z'}$
\bea
&& M_{eff} >1000\ {\gev},\quad 1800 < M_{TWW} < 2200\ {\gev}\ {\rm for\ 2~TeV,}\\  
&& M_{eff} >1250\ {\gev},\quad 2800 < M_{TWW} < 3200\ {\gev}\ {\rm for\ 3~TeV.}
\label{MWWcuts_SL.EQ}
\eea
Another approach could be to constrain $(p_\ell + p_\nu)^2 = M_W^2$ which allows us to
infer the z-component of $p_\nu$ also, up to a quadratic ambiguity. Although 
the collimation of the $\ell\nu$ makes the mass determination
inaccurate, it can be treated in a manner that maximizes signal 
over background. We do not pursue this method here.

\begin{figure}
\begin{center}
\scalebox{0.5}{\includegraphics[angle=270]{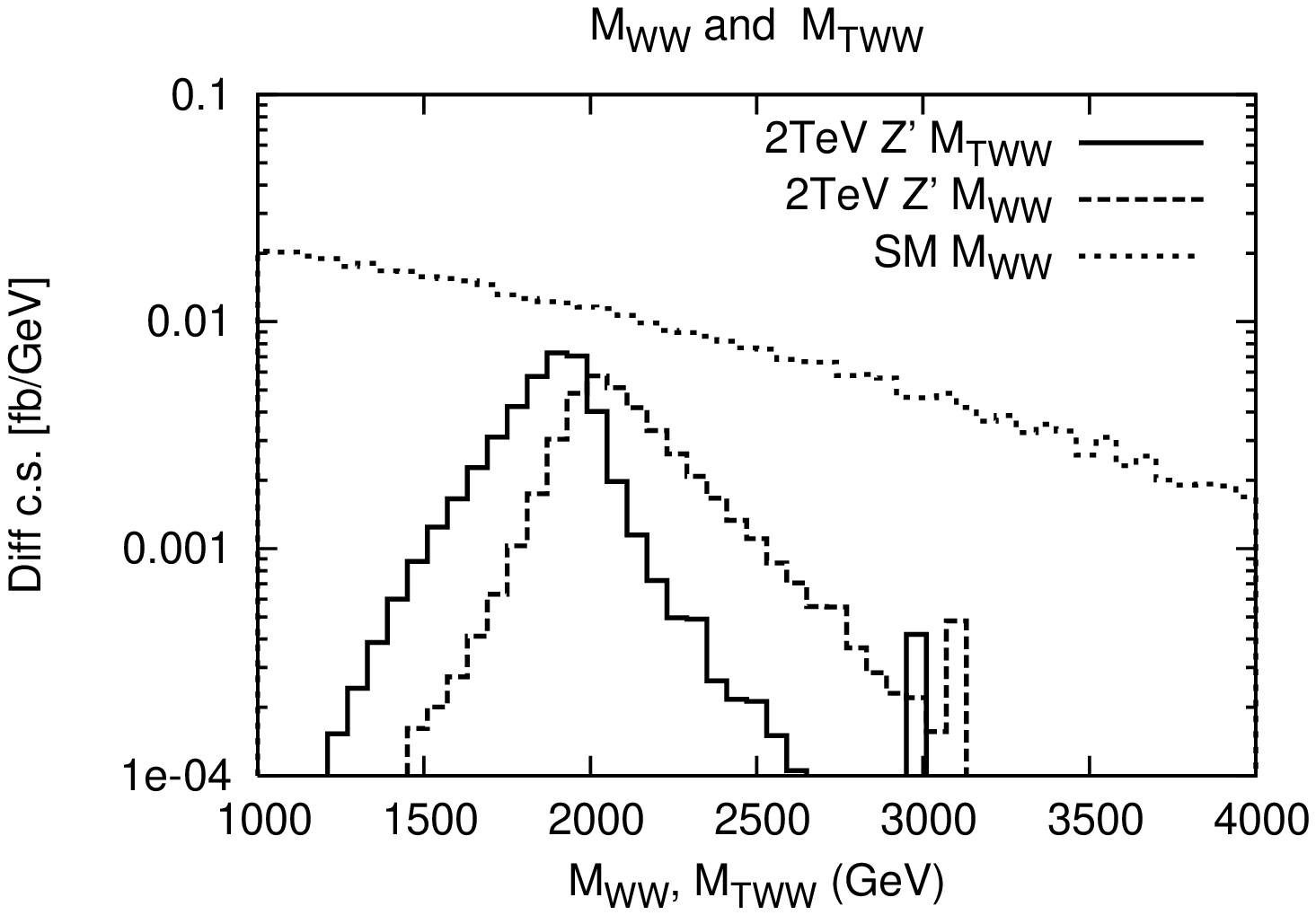}}
\scalebox{0.5}{\includegraphics[angle=270]{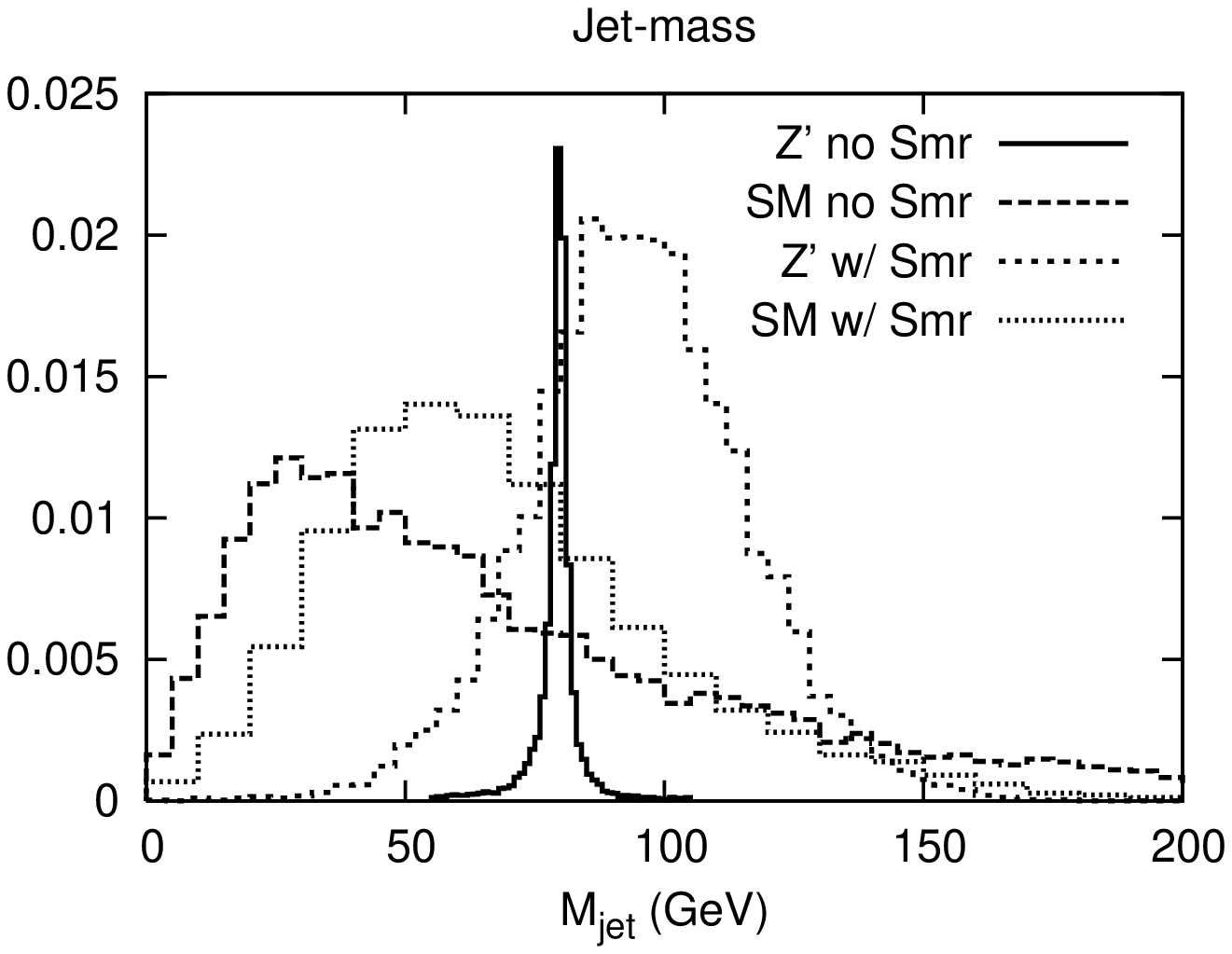}}
\caption{Distributions for the signal and background 
(a) the reconstructed invariant mass $\bar M_{WW}$ 
(after $p_T$ and $\eta$ cuts), and (b) jet-mass distribution 
for a cone-size of 0.4 with and without $E$, $\eta$ and $\phi$ smearing. 
\label{jetMwj.FIG}}
\end{center}
\end{figure} 

%

In order to improve the rejection of QCD background, we may be able to exploit more
differences between the signal and the QCD background (W+1jet).
One such quantity that may have discriminating power is the jet-mass, which is 
the combined invariant mass of the vector sum of 4-momenta of all hadrons 
making up the jet.
The jet-mass resolution is limited by our ability to reconstruct the angular separation 
of the constituents of the jet.  
For $\mzpri$ in the few TeV range, due to collimation, forming the jet-mass
becomes more challenging, since the cell size of the 
(ATLAS) hadronic calorimeter is of the order 
$\Delta\eta\times \Delta\phi\sim 0.1\times 0.1$.
Although the two jets from $W\rightarrow jj$ are severely overlapping in the hadronic
calorimeter, 
it may be possible to combine~\cite{Stras_ph07} it with the 
electromagnetic calorimeter and the tracker in order to obtain a reasonable discriminating
power using the jet-mass. 
Since the EM calorimeter has better granularity, the two jets from the signal $W$ events 
are expected to have two separated EM cores, and the finer segmentation of the 
EM calorimeter helps in improving the jet-mass resolution~\cite{FrankPaige}.
For the signal, we expect the jet mass to peak at $\mw$, and, 
although a QCD jet will develop a  mass due to the color radiation or showering, 
the jet-mass is limited when the jet size is fixed  by $\Delta R$ and only single
jet events are retained. 
We can thus use a jet-mass cut to suppress the QCD background. 
To obtain a rough estimate of how much background can be rejected, we have performed
a study with the leading-order $W + 1$ jet matrix element followed by showering in
Pythia 6.4~\cite{Sjostrand:2006za}. In Fig.~\ref{jetMwj.FIG}(b), 
we show the resulting jet-mass distributions for the signal and background where we have
smeared the 
energy by $80$\,\%/$\sqrt{E}$ and the $\eta$ and $\phi$ by $0.05$ to account for possible 
experimental uncertainties.\footnote{Although the hadronic calorimeter cell-size in 
$\eta$ and $\phi$ is $0.1$ one may be able to do better by combining the tracker and
electromagnetic calorimeter information as already mentioned; we therefore choose an 
angular uncertainty of $0.05$.}
We find that for a jet-mass cut 
\beq
75 < M_{jet} < 125\ {\gev},
\label{jetM}
\eeq
we obtain an acceptance fraction
of $0.78$ for the signal, while it is $0.3$ for the background. 
%
It is important to emphasize that this 
level of study gives us a rough estimate, and a more realistic determination 
would  require a study beyond leading order and including a detector simulation. 
A study along these lines albeit in a different context and cuts has been performed
in Refs.~\cite{jmassRefs}.

\begin{table}[h]
\caption{$pp\rightarrow \ell^\pm \etmiss +1$ jet cross-section (in fb) 
for $\mzpri=$2 and 3~TeV, and background, with cuts applied 
successively. The number of events is shown for
${\cal L} = 100~{\rm fb}^{-1}$ for 2~TeV, and $1000~{\rm fb}^{-1}$ for 
3~TeV.
\label{WW2lnjj.TAB}}
\begin{center}
\begin{tabular}{|c||c|c|c|c|c|c|c|c|}
\hline
$\mzpri=2$ TeV&
 $p_{T}$&
$\eta_{\ell,j}$&
$M_{eff}$&
$M_{T_{WW}}$&
$M_{jet}$&
\# Evts&
$S/B$&
$S/\sqrt{B}$\tabularnewline
\hline
\hline
Signal&
$4.5$&
$2.40$&
$2.37$&
$1.6$&
$1.25$&
$125$&
$0.39$&
$6.9$\tabularnewline
\hline
W+1j&
$1.5\times10^{5}$&
$3.1\times10^{4}$&
$223.6$&
$10.5$&
$3.15$&
$315$&
&
\tabularnewline
\hline
WW&
$1.2\times10^{3}$&
$226$&
$2.9$&
$0.13$&
$0.1$&
$10$&
&
\tabularnewline
\hline
\hline
$\mzpri=3$ TeV&
&
&
&
&
&
&
&
\tabularnewline
\hline
\hline
Signal&
$0.37$&
$0.24$&
$0.24$&
$0.12$&
-&
$120$&
$0.17$&
$4.6$\tabularnewline
\hline
W+1j&
$1.5\times10^{5}$&
$3.1\times10^{4}$&
$88.5$&
$0.68$&
-&
$680$&
&
\tabularnewline
\hline
WW&
$1.2\times10^{3}$&
$226$&
$1.3$&
$0.01$&
-&
$10$&
&
\tabularnewline
\hline
\end{tabular}
\end{center}
\end{table}

In Table~\ref{WW2lnjj.TAB} we show the cross-section (in fb) for the 
$pp\rightarrow \ell^\pm \etmiss +1$ jet process for $\mzpri=2$ and 3 TeV and 
the SM backgrounds of QCD and $W^+W^-$. The cuts as discussed  
in the text are applied successively and the improvement in $S/B$ is evident. 
For $\mzpri=3$~TeV the increased collimation of the $W$ decay products makes it
more challenging to use the jet-mass cut, and therefore we
have not applied the jet-mass cut in this case. 
We find that the ${M_T}_{WW}$ cut results in a slightly better efficiency compared
with the ${\bar M}_{WW}$ cut, and we therefore do not show the latter cut in the table.
%
We thus infer for $\mzpri=2$~TeV that 
a signal of about $7 \sigma$ significance may be reached with an
integrated luminosity of $100~\fbi$, with a $S/B=40\%$. A heavier
$Z'$ would need significantly more luminosity to see a clear signal.
For instance, 1000 $\fbi$ may be needed to reach about a 5$\sigma$ sensitivity
for $M_{Z'}=3$ TeV.

\subsection{$\tilde Z_1 \rightarrow Z h$}

\begin{figure}
\begin{center}
\scalebox{0.5}{\includegraphics[angle=270]{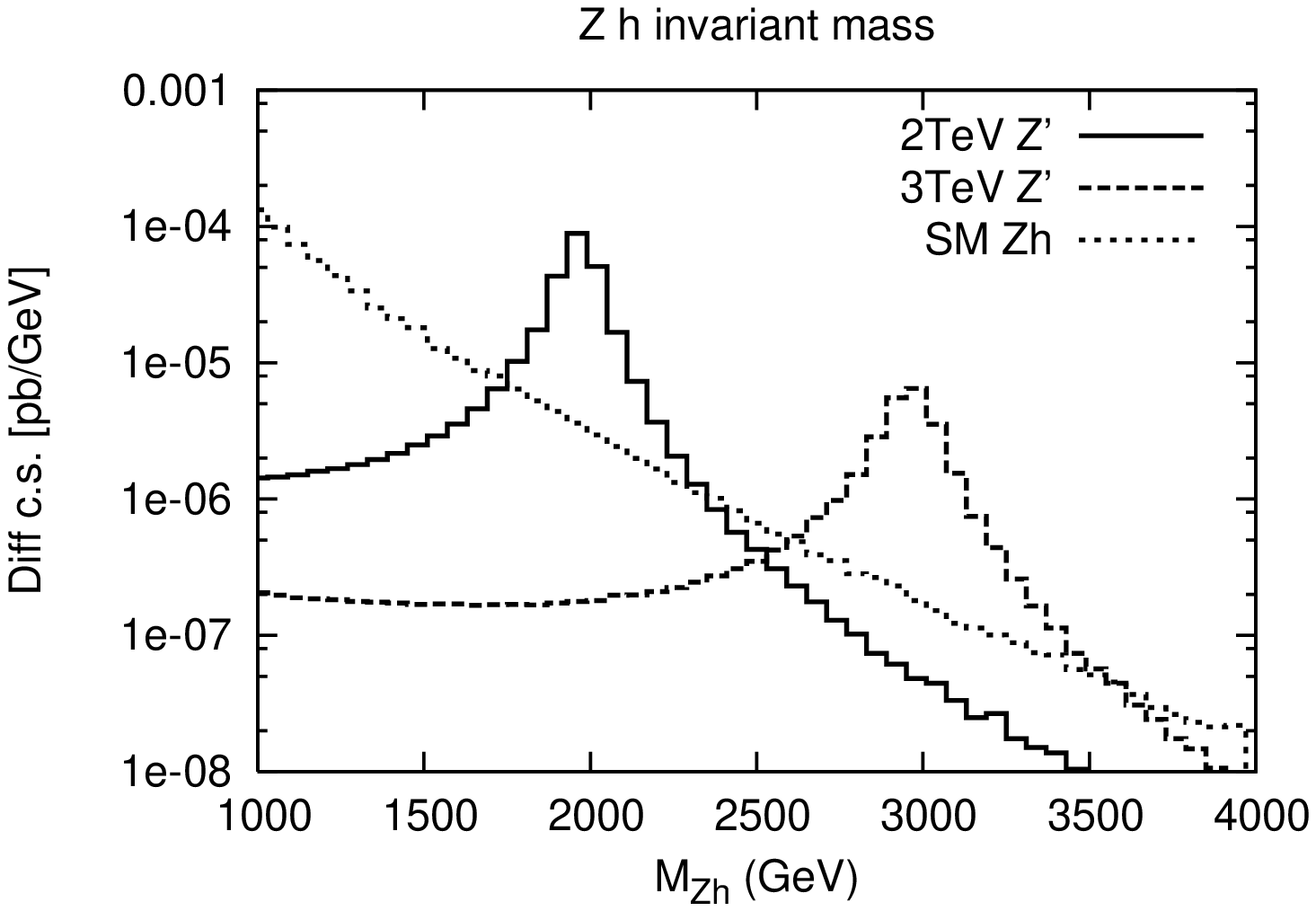}}
\scalebox{0.5}{\includegraphics[angle=270]{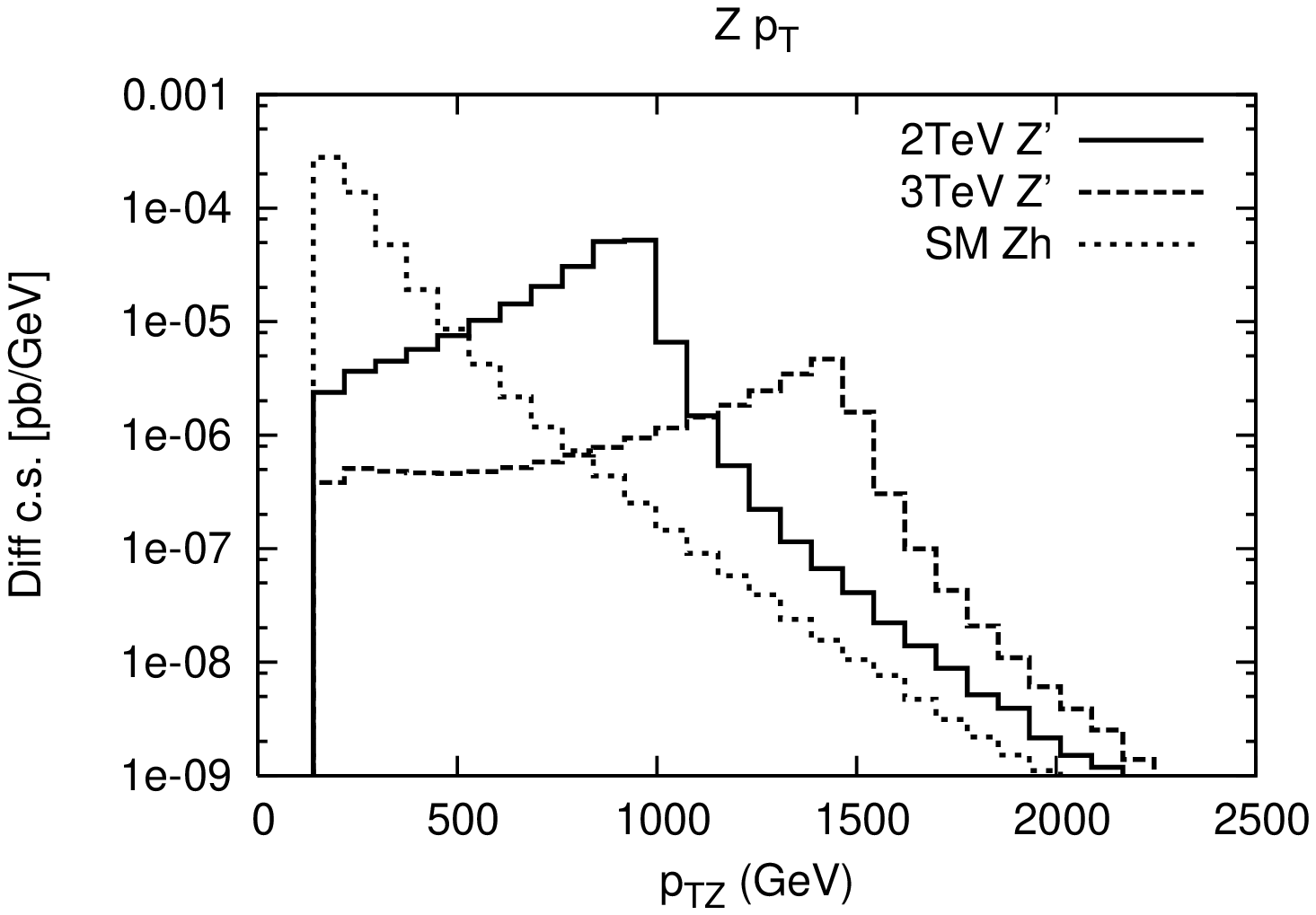}}
\caption{The differential cross-section as a function of (a) the $Z\, h$ invariant mass,
and (b) the ${p_T}_Z$, 
from Drell-Yan production of $\zpri$ (with mass 2~TeV and 3~TeV) at the LHC including the 
SM background. This is after the cuts ${p_T}_{Z,h} > 200$GeV, and 
$-3 < \eta_{Z,h} < 3$.
\label{dyZprismZh.FIG}}
\end{center}
\end{figure}

As discussed in the last section, the $W^+W^-$ mode from $\ap,\ \tilde Z_{X_1}$ decays will
lead to significant signals; while their decay to $Zh$ will be small. On the other hand,
the decay channel $\tilde Z_1 \to Zh$ is overall dominant as seen from Fig.~\ref{BR.vs.M.FIG}.
We impose the basic acceptance cuts for the event selection
\beq
{p_T}_Z, \ {p_T}_h > 200~{\rm GeV} \ \ ; \ \ -3 < \eta_{Z}, \ \eta_{h} < 3 \ .
\label{etazh}
\eeq
After the basic cuts, the cross-section for $\mzpri=2$ TeV into this final state is 
16.7~fb for 2 TeV mass, and 1.8~fb for 3~TeV mass.
Figure \ref{dyZprismZh.FIG} shows the differential cross-section as a function of
the $Z h$ invariant mass, and ${p_T}_Z$, along with the SM background 
arising from the $Zh$ production. 
%
%
Although the SM irreducible background from $Zh$ production is small, 
when a particular decay mode is 
considered we will pick up additional sources of background. Again due to the
large boost of fast moving $Z$ and $h$, the decay products are collimated, making
signal reconstruction more challenging. We will discuss these issues in greater detail
in the following when we consider particular decay modes.
%
%

For our purposes of illustration here, most important 
features can be highlighted by considering two cases: $m_h = 120~$GeV and 
$m_h = 150~$GeV.\footnote{In models where the Higgs is the $A_5$, $m_h$ is naturally
about 150~GeV \cite{Agashe:2004rs, Agashe:2005dk, Contino:2006qr, Medina:2007hz}}.

\subsubsection{$m_h = 120~$GeV}
The $h$ decay modes in this mass range 
(with branching fractions in parenthesis) are:  $b\bar b$ (0.7),
$\tau^+ \tau^-$ (0.07), $WW^*$ ($0.15$) and $ZZ^*$ (0.02). 
The leading Higgs decay is $h\to b\bar b$, we thus consider the leptonic modes
of the $Z$ decay as
\begin{enumerate}
\item $h\rightarrow b\bar b$, $Z\rightarrow \ell^+\ell^-$: BR $\approx 0.7\times 2/30 = 4.6\%$. 
If the two $b$-jets get merged, we demand to tag only one $b$ to be conservative. 
The background with one tagged $b$ thus is $Z+1~b\to \ell^+\ell^- +1$ tagged $b$.
\item $h\rightarrow b\bar b$, $Z\rightarrow \nu\bar\nu$: BR $\approx 0.7\times 
0.21 = 15\%$.
Here we can demand a large missing $E_T$ (of the order $\mzpri/2$). 
The background is mainly from $Z+1~b\to \etmiss +1$ tagged $b$.
%
%
\end{enumerate}
Obviously, the decay $Z\to \ell^+\ell^-$ yields a clean mode and we thus concentrate on
the channel $pp\rightarrow b\bar b\ \ell^+\ell^-$.
We start with the basic cuts as in Eq.~(\ref{etazh}), and 
we show the distributions after the  cuts in Fig.~\ref{pp2Zh2llbb.FIG}. 
It should be noted that the signal distributions and cross-sections are obtained
by multiplying the corresponding $Z h$ quantities by BR($h\rightarrow b\bar b$)
and BR($Z\rightarrow \ell^+\ell^-$). 
\begin{figure}
\begin{center}
\scalebox{0.5}{\includegraphics[angle=270]{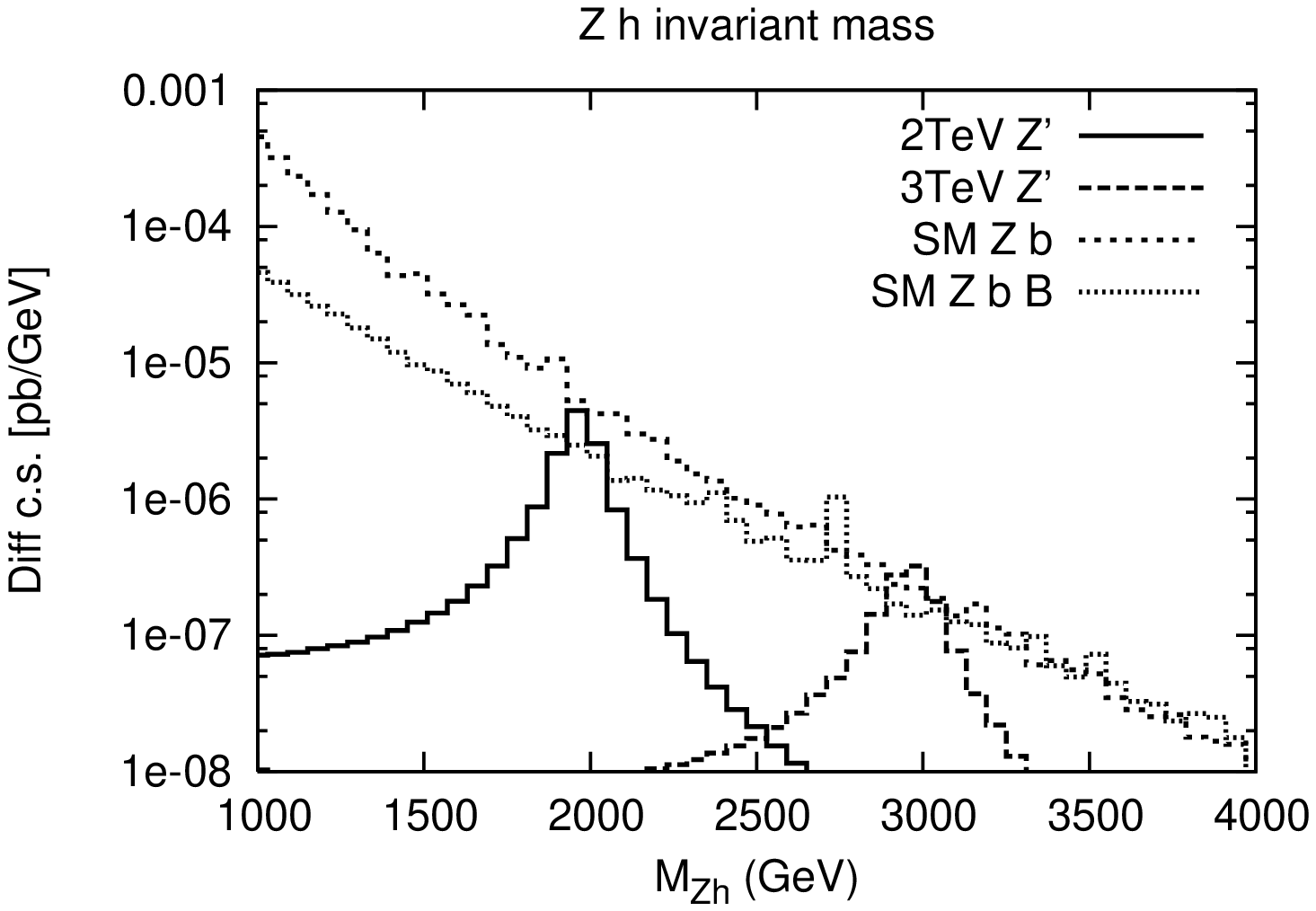}}
\scalebox{0.5}{\includegraphics[angle=270]{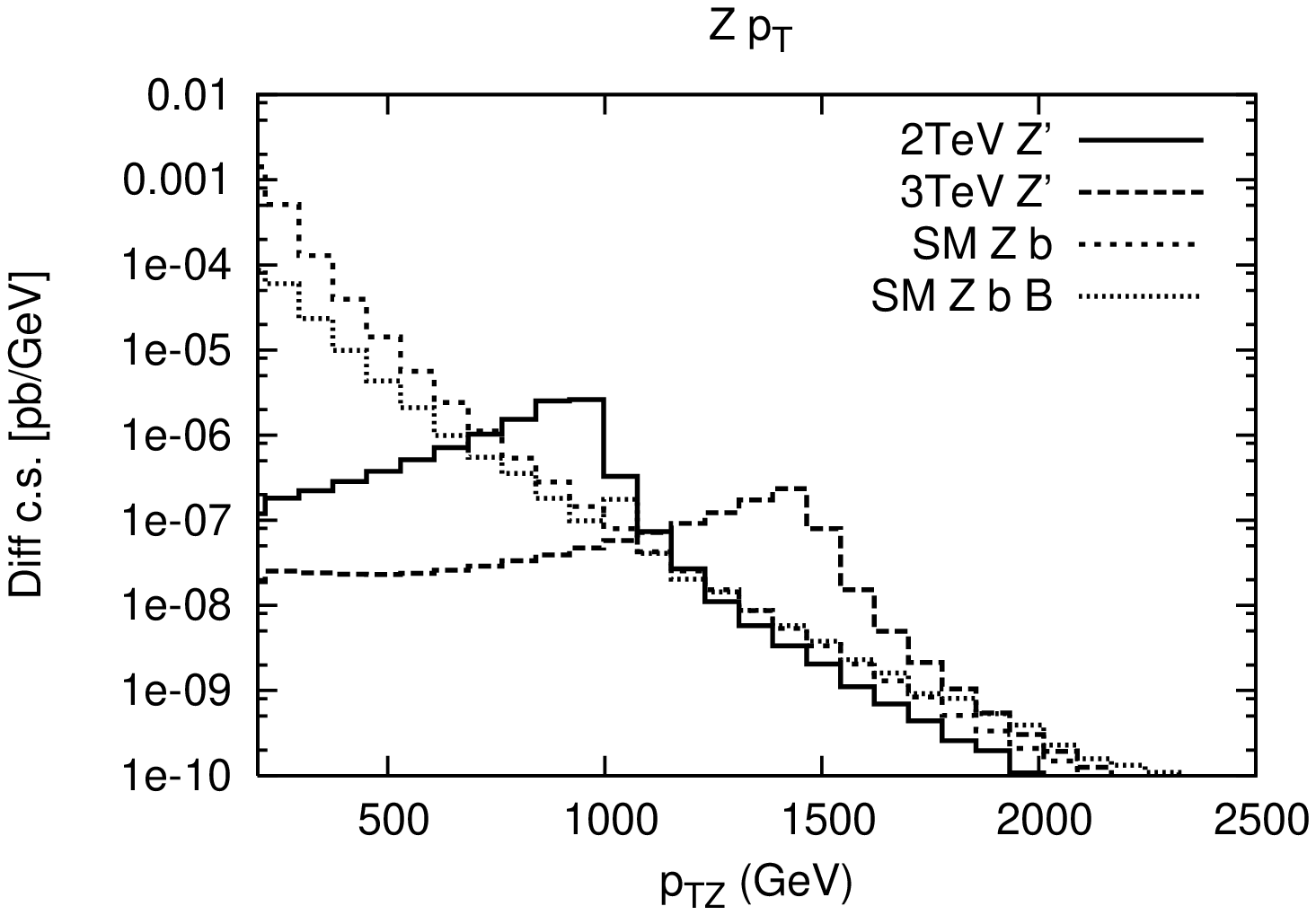}}
\scalebox{0.5}{\includegraphics[angle=270]{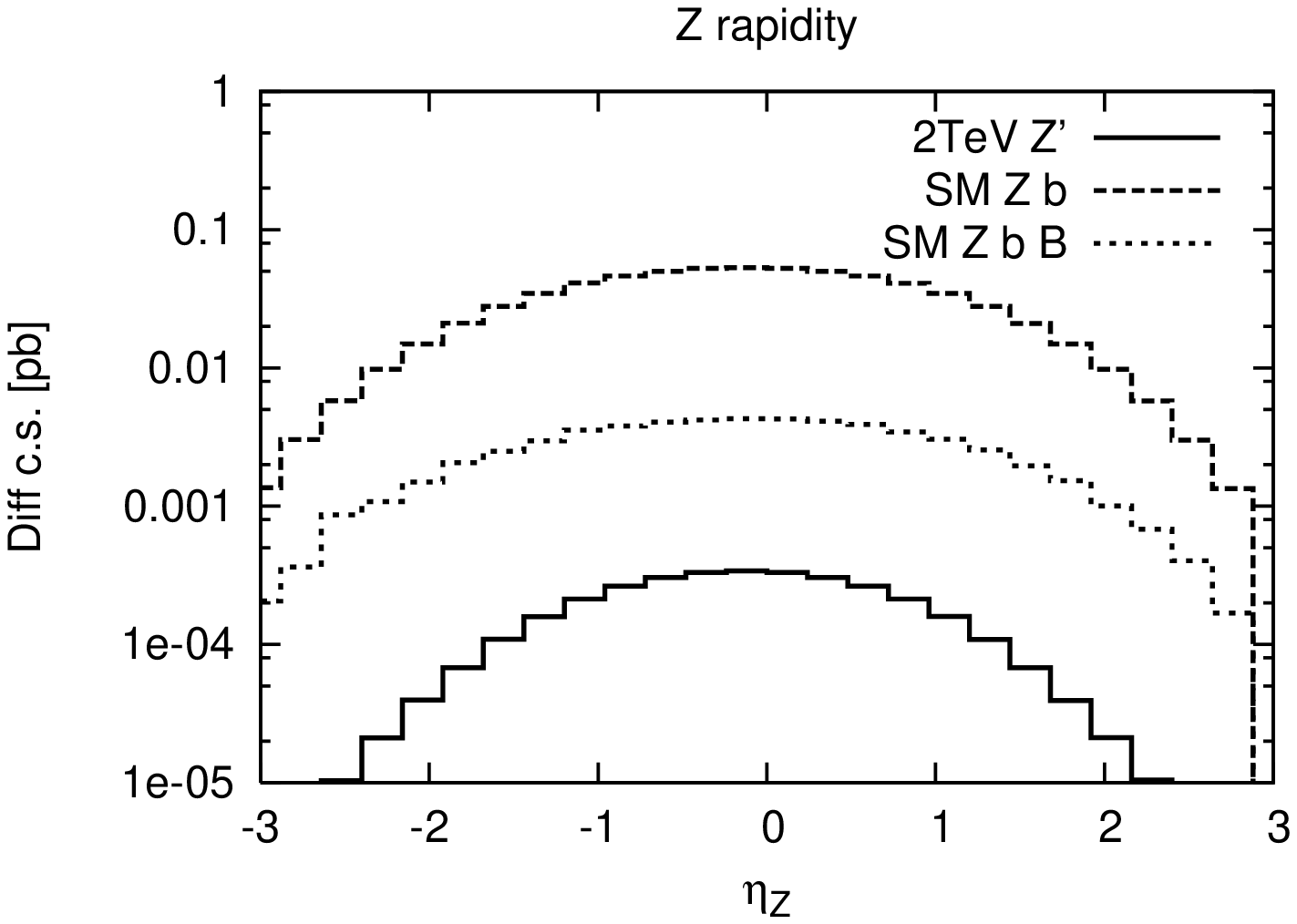}}
\scalebox{0.5}{\includegraphics[angle=270]{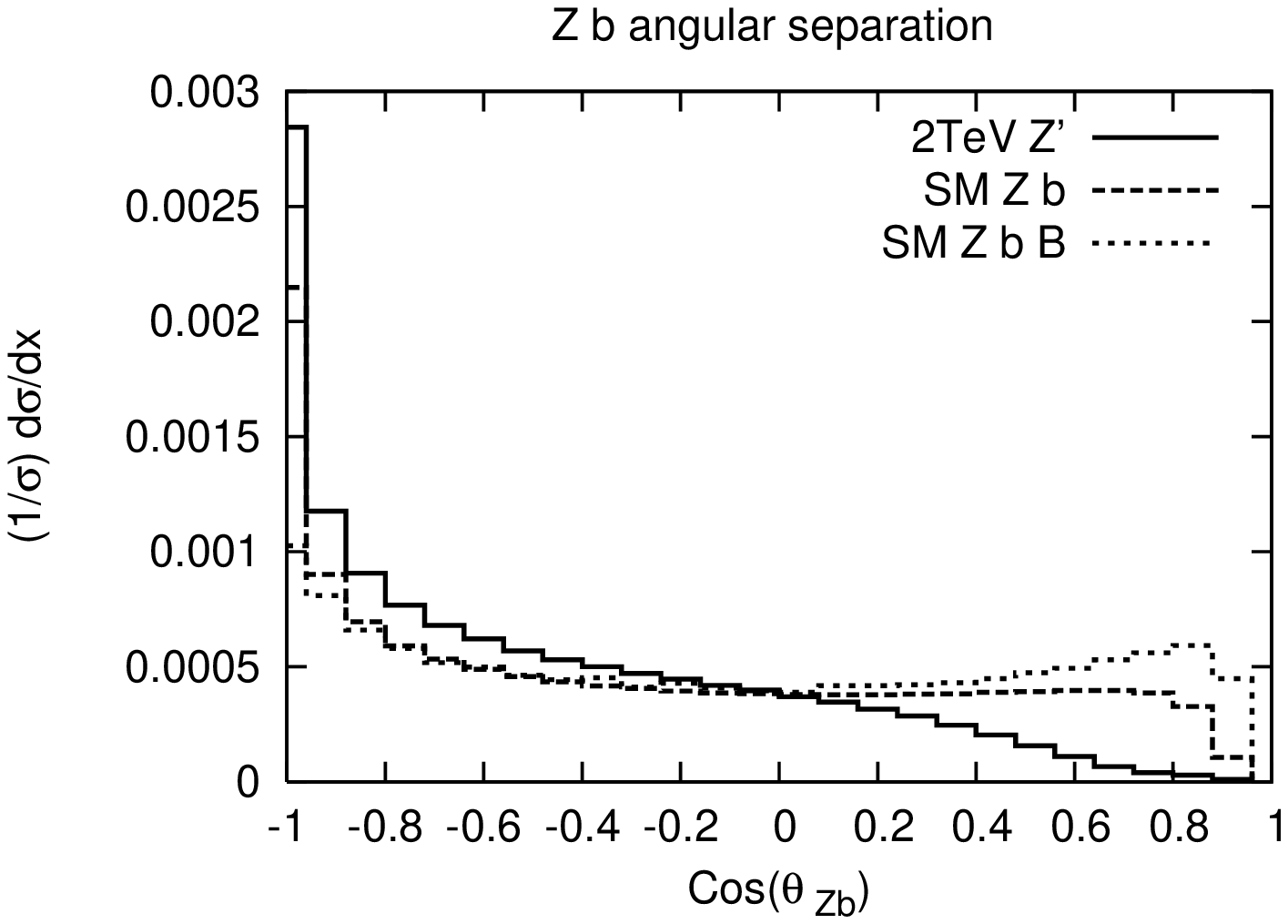}}
\caption{Signal and background distributions in $pp\rightarrow b\bar b\ell^+\ell^-$ for 
(a) $b\bar b\ell\ell$ invariant mass, (b) ${p_T}_Z$, 
(c)  $\eta_Z$, and (d) $\cos{\theta_{Zb}}$. These are after the basic cuts. 
\label{pp2Zh2llbb.FIG}}
\end{center}
\end{figure}
For the $\zpri \rightarrow h Z \rightarrow b\bar b ~ \ell \ell$ mode we show the 
significance in Table~\ref{ZH2llbb.TAB} as we tighten up the cuts 
successively as follows
\bea
\nonumber
&& {p_T}_\ell > 250\ {\gev},\   {p_T}_{b\bar b} > 0.5\ {\tev},\  \eta_{\ell,b} < 2,\ 
\cos\theta_{\ell b} < -0.5,\  1850 < M_{Zh} < 2150\ {\gev} \ \  {\rm for\ 2~TeV,}~~~ \\  
\nonumber
&& {p_T}_\ell > 500\ {\gev},\   {p_T}_{b\bar b} > 1\ {\tev},\  \eta_{\ell,b} < 2,\ 
\cos\theta_{\ell b} < -0.5,\  2800 < M_{Zh} < 3200\ {\gev} \ \  {\rm for\ 3~TeV}.
\eea
We use a b-tagging efficiency of $0.4$ with a rejection factor for light-jets (from 
$u,d,s,g$) $R_u = 20$~\cite{btagRef}.
It is noted that this efficiency/rejection is with b-tagging parameters optimized for
low ${p_T}_b$, and the rejection is expected to improve with tagging techniques 
optimized for high ${p_T}_b$. We use a charm quark rejection factor $R_c = 5$. 
We find a clear signal above the background. With an integrated luminosity
of 200 {$\fbi$} (1000 {$\fbi$}), we obtain 
$S/\sqrt{B} \approx 5.3$ (5.7) for $M_{Z'}=2$~TeV ($3$~TeV).\footnote{For the 
$3$~TeV case, since the number of background events is low, using Poisson statistics
leads to about 99.95\;\% CL.}
Improvements in the b-tagging light-jet rejection factor can improve the significance
even further.

\begin{table}[h]
\caption{$pp\rightarrow Zh\rightarrow b\bar b ~ \ell\ell$ cross-section (in fb) 
for the signal with $\mzpri=2$ TeV, and $\mzpri=3$ TeV,
and the corresponding backgrounds, with cuts applied successively.  
The statistical significance is  shown 
for  200~${\rm fb}^{-1}$ (for 2~TeV) and for 1000~${\rm fb}^{-1}$ (for 3~TeV).
\label{ZH2llbb.TAB}}
\begin{center}
\begin{tabular}{|c|c|c|c|c|c|c|c|c|}
\hline 
$\mzpri=2$ TeV&
Basic&
$p_{T},\eta$&
$\cos{\theta_{Zh}}$&
$M_{inv}$&
b-tag&
\# Evts&
$S/B$&
$S/\sqrt{B}$\tabularnewline
\hline
\hline 
$Z^{\prime}\rightarrow hZ\rightarrow b\bar{b}\,\ell\ell$ &
$0.81$&
$0.73$&
$0.43$&
$0.34$&
$0.14$&
$27$&
$1.1$&
$5.3$\tabularnewline
\hline 
SM $Z+b$&
$157$&
$1.6$&
$0.9$&
$0.04$&
$0.016$&
$3$&
&
\tabularnewline
\hline 
SM $Z+b\bar{b}$&
$13.5$&
$0.15$&
$0.05$&
$0.01$&
$0.004$&
$0.8$&
&
\tabularnewline
\hline
SM $Z+q_{l}$&
$2720$&
$48$&
$22.4$&
$1.5$&
$0.08$&
$15$&
&
\tabularnewline
\hline
SM $Z+g$&
$505.4$&
$11.2$&
$5.8$&
$0.5$&
$0.025$&
$5$&
&
\tabularnewline
\hline
SM $Z+c$&
$184$&
$1.9$&
$1.1$&
$0.05$&
$0.01$&
$2$&
&
\tabularnewline
\hline
\hline 
$\mzpri=3$ TeV&
&
&
&
&
&
&
&
\tabularnewline
\hline
\hline 
$Z^{\prime}\rightarrow hZ\rightarrow b\bar{b}\,\ell\ell$ &
$0.81$&
$0.12$&
$0.05$&
$0.04$&
$0.016$&
$16$&
$2$&
$5.7$\tabularnewline
\hline
SM $Z+b$&
$157$&
$0.002$&
$0.001$&
$3\times10^{-4}$&
$1.2\times10^{-4}$&
$0.12$&
&
\tabularnewline
\hline
SM $Z+b\bar{b}$&
$13.5$&
$0.018$&
$0.014$&
$0.002$&
$0.001$&
$1$&
&
\tabularnewline
\hline
SM $Z+q_{l}$&
$2720$&
$1.1$&
$0.7$&
$0.1$&
$0.005$&
$5$&
&
\tabularnewline
\hline
SM $Z+g$&
$505.4$&
$0.3$&
$0.2$&
$0.03$&
$0.0015$&
$1.5$&
&
\tabularnewline
\hline
SM $Z+c$&
$183.5$&
$0.03$&
$0.02$&
$0.002$&
$4\times10^{-4}$&
$0.4$&
&
\tabularnewline
\hline
\end{tabular}
\end{center}
\end{table}
%

%

\subsubsection{$m_h = 150~$GeV}
The $h$ decay modes in this mass range 
(with B.R.'s in parenthesis) are:  $b\bar b$ (0.2),
$\tau^+ \tau^-$ (0.001), $WW^*$ (0.7) and $ZZ^*$ (0.1), we will thus consider
the leading mode of $WW^*$. 
After including the $Z$ decay BR's we find the following modes to be significant:
\begin{enumerate}
\item $h\rightarrow WW\to$jets, $Z\rightarrow \ell^+\ell^-$: BR $\approx 0.7\times
(2/3)^2\times 1/15 = 2.1\%$.
The corresponding background is from $Z+2$ jets $\to$ 2 jets $+\ell^+\ell^-$.
\item $h\rightarrow WW\to$ jets, $Z\rightarrow \nu\bar\nu$: BR $\approx 0.7\times 
(2/3)^2\times 0.2 = 6.2\%$.
The background is from $Z+2$ jet $\to$ 2 jets $+\etmiss$.
\item $h\rightarrow WW\to \ell\nu\ jj$, $Z\rightarrow jj$: BR$\approx 0.7\times (2\times 2/9
\times 2/3) \times 0.7 = 15\%$.
The background is from $W+2$ jets $\to$ 2 jets $+\ell^\pm + \etmiss$.
\end{enumerate}
 
We consider the last channel that yields the largest branching fraction with
good experimental signatures. 
The $\Delta R$ separation of the two jets from the $W$ is about 0.3 and those from the 
$Z$ about 0.16. In our analysis, to be conservative, we will not require that the two jets 
be resolved and treat them as a single jet (one from the $W$ and another from the $Z$). 
We will refer to the jet(s) from the hadronic decay of the $W$ as the ``near-jet'' (since it is 
near to the leptonic $W$) and the jet(s) from the $Z$ as the ``far-jet''. We will denote them
as $j_N$ and $j_F$ respectively. 
The jet merging issues discussed in Sec.~\ref{zpri2WW2SL.SEC} are applicable identically 
to $j_F$, and much less severe for $j_N$ (due to larger $\Delta R \approx 0.3$). 

The irreducible SM background will be due to $Z W W$. However, owing to the above 
jet merging issues, we will additionally pick up $W Z j $, $W W j$ and $W j j$. 
The former two are smaller  than the last because of the electroweak
versus the strong coupling.

Since the final state has a neutrino carrying away (missing) momentum we will not be able to
reconstruct the full invariant mass of the system\footnote{However, similar to the case
explained below Eq.~(\ref{MWWcuts_SL.EQ}) it may be possible to use the $M_W$ constraint 
to some advantage, although we do not pursue this here.}. 
We therefore use the transverse-mass
of the $\ell j_N j_F$ system to enhance the signal resonance. 
The  transverse mass is defined by
\beq
{M_T}_{Zh} = \sqrt{{p_T}_Z^2 + M_Z^2} + \sqrt{{p_T}_h^2 + M_h^2}.
\eeq
In Fig.~\ref{MTWjjjj.FIG} we compare the $M_T$ and the true invariant mass ($M_{inv}$). 
We see that the $M_T$ distribution reflects the resonant structure rather well, although it
is broader.
\begin{figure}
\begin{center}
\scalebox{0.54}{\includegraphics[angle=270]{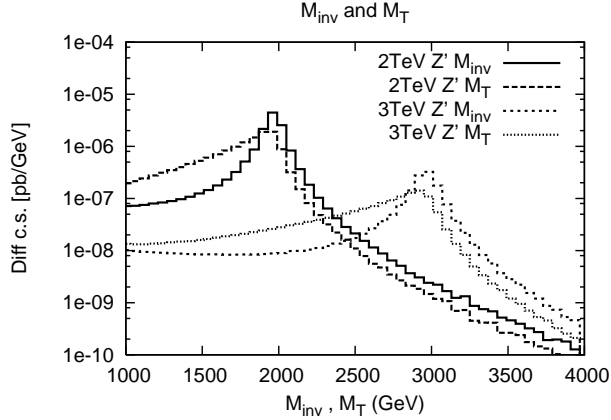}}
\caption{
The transverse mass distribution for the signal compared with the true invariant mass. 
\label{MTWjjjj.FIG}}
\end{center}
\end{figure}

We select $\ell$ + 2-jet events with the following basic cut: 
\beq
{p_T}_j > 100\ {\gev},\  {p_T}_\ell > 50\ {\gev},\ \ptmiss > 50\ {\gev},\ |\eta_{\ell, j}| < 3.
\eeq
We further apply various cuts to improve the significance, which is 
shown in Table~\ref{ZHlnujjjj.TAB}.
For $\mzpri=2$ TeV we apply successively the cuts:
\bea
&& {p_T}_{j_N} > 400\ {\gev},\quad {p_T}_{j_F} > 800\ {\gev},\\
&& 0.8 < \cos{\theta_{W j_N}} < 1,\quad -1 < \cos{\theta_{j_N j_F}} < -0.5,\\
&& 
1750\ {\gev} < {M_T}_{Wj_Nj_F} < 2150\ {\gev},\ \ 
100 < {M_T}_{Wj_N} < 175\ {\gev}\ \ ({\rm near}\ m_h), \\
&& 70\ {\gev}  < M_{j_F} < 110\ {\gev}\ \ ({\rm near}\ M_Z).
\eea
Similarly for $\mzpri=3$ TeV,
\bea
&& {p_T}_{j_N} > 500\ {\gev},\quad {p_T}_{j_F} > 1000\ {\gev},\\
&& 0.8 < \cos{\theta_{W j_N}} < 1,\quad -1 < \cos{\theta_{j_N j_F}} < -0.5,\\
&& 
2800\ {\gev} < {M_T}_{Wj_Nj_F} < 3100\ {\gev},\ \ 
100 < {M_T}_{Wj_N} < 175\ {\gev}\ \ ({\rm near}\ m_h).
\eea
Due to increased collimation we do not apply the jet-mass cut on $j_F$. Although we do not 
pursue it here we can apply a jet-mass cut on $j_N$ to further improve the significance.
For the 3~TeV case, we only show the SM $Wjj$ background in Table~\ref{ZHlnujjjj.TAB} 
but not the $W Z j$ and $W W j$ since they are much smaller as in the earlier case.
Once again, we obtain substantial statistical significance for the signals.

\begin{table}[h]
\caption{$pp\rightarrow Zh\rightarrow (jj)~(jj)~\ell \etmiss$ cross-section (in fb) 
for the signal with $\mzpri=2$ TeV and 3~TeV
and the corresponding backgrounds, with cuts applied 
successively.  The number of events and statistical significance are shown 
for  100 ${\rm fb}^{-1}$ (for 2~TeV) and 300 ${\rm fb}^{-1}$ (for 3~TeV).
\label{ZHlnujjjj.TAB}}
\begin{center}
\begin{tabular}{|c|c|c|c|c|c|c|c|c|}
\hline 
$\mzpri=2$ TeV~~~$m_{h}=150$ GeV&
Basic&
$p_{T},\eta$&
$\cos{\theta}$&
$M_{T}$&
$M_{jet}$&
\# Evts&
$S/B$&
$S/\sqrt{B}$\tabularnewline
\hline
\hline 
$Z^{\prime}\rightarrow hZ\rightarrow\ell\,\etmiss\,(jj)\,\,(jj)$ &
$2.4$&
$1.6$&
$0.88$&
$0.7$&
$0.54$&
$54$&
$2.5$&
$11.5$\tabularnewline
\hline 
SM $W\, j\, j$&
$3\times10^{4}$&
$35.5$&
$12.7$&
$0.62$&
$0.19$&
$19$&
$ $&
$ $\tabularnewline
\hline
SM $W\, Z\, j$&
$184$&
$0.45$&
$0.15$&
$0.02$&
$0.02$&
$2$&
&
\tabularnewline
\hline
SM $W\, W\, j$&
$712$&
$0.54$&
$0.2$&
$0.02$&
$0.01$&
$1$&
&
\tabularnewline
\hline
\hline
$\mzpri=3$ TeV~~~$m_{h}=150$ GeV&
&
&
&
&
&
&
&
\tabularnewline
\hline
\hline
$Z^{\prime}\rightarrow hZ\rightarrow\ell\,\etmiss\,(jj)\,\,(jj)$ &
$0.26$&
$0.2$&
$0.14$&
$0.06$&
$-$&
$18$&
$1.2$&
$4.7$\tabularnewline
\hline
SM $W\, j\, j$&
$3\times10^{4}$&
$ $&
$4.1$&
$0.05$&
$-$&
$15$&
&
\tabularnewline
\hline
\end{tabular}
\end{center}
\end{table}



\subsection{$\zpri \rightarrow \ell^+ \ell^-$}
The cleanest channel of all should be the di-lepton mode from the DY production.
However, due to the highly suppressed coupling of $Z'$ to the light fermions, this
channel requires a large integrated luminosity for observation. 
The cross-section into the $\ell^+ \ell^-$ final state (with $\ell = e,\mu$) for 
$\mzpri=2~$ TeV is 0.12~fb without any cuts. We select events with the basic cuts
\beq
{p_T}_\ell > 50\ {\gev},\quad |\eta_\ell | < 3.
\eeq
In Table~\ref{pp2ll.TAB} we show the improvement in S/B for $\mzpri=2$~TeV 
as we apply the following cuts:
\beq
{p_T}_\ell > 500\ {\gev},\quad 1900\ {\gev} < M_{\ell\ell} < 2100\ {\gev}. 
\eeq
One would need much larger integrated luminosity to reach a significant signal.
Although the event rate is rather low and high luminosity would be needed to 
reach a significant observation, it is noted that this clean channel is mainly 
statistically dominated and does not suffer from systematic effects 
present in some of the other channels.

\begin{table}[h]
\caption{
$pp\rightarrow \ell^+ \ell^-$ cross-section (in fb) 
for $\mzpri=2$ TeV signal and the corresponding backgrounds, with cuts applied 
successively.  The number of events and statistical significance are shown 
for  1000 ${\rm fb}^{-1}$.
\label{pp2ll.TAB}}
\begin{center}
\begin{tabular}{|c|c|c|c|c|c|c|}
\hline 
$\mzpri=2$ TeV&
Basic&
$p_{T\ell}$&
$M_{\ell\ell}$&
\# Evts&
$S/B$&
$S/\sqrt{B}$\tabularnewline
\hline
\hline 
Signal&
$0.1$&
$0.09$&
$0.06$&
$60$&
$0.3$&
$4.2$\tabularnewline
\hline 
SM $\ell\ell$&
$3\times10^{4}$&
$5.4$&
$0.2$&
$200$&
&
\tabularnewline
\hline 
SM $WW$&
$295$&
$0.03$&
$0.002$&
$2$&
&
\tabularnewline
\hline
\end{tabular}
\end{center}
\end{table}

\subsection{$\zpri \rightarrow t\bar{t},\ b \bar b$}
Because of the large coupling to the heavy fermions, the decay modes of $Z'$ to
$t\bar t,\ b\bar b$ are substantial. 
We start with the basic cut 
\beq
{p_T}_t > 100\ {\gev},\quad |\eta_t| < 3 .
\eeq
\begin{figure}
\begin{center}
\scalebox{0.53}{\includegraphics[angle=270]{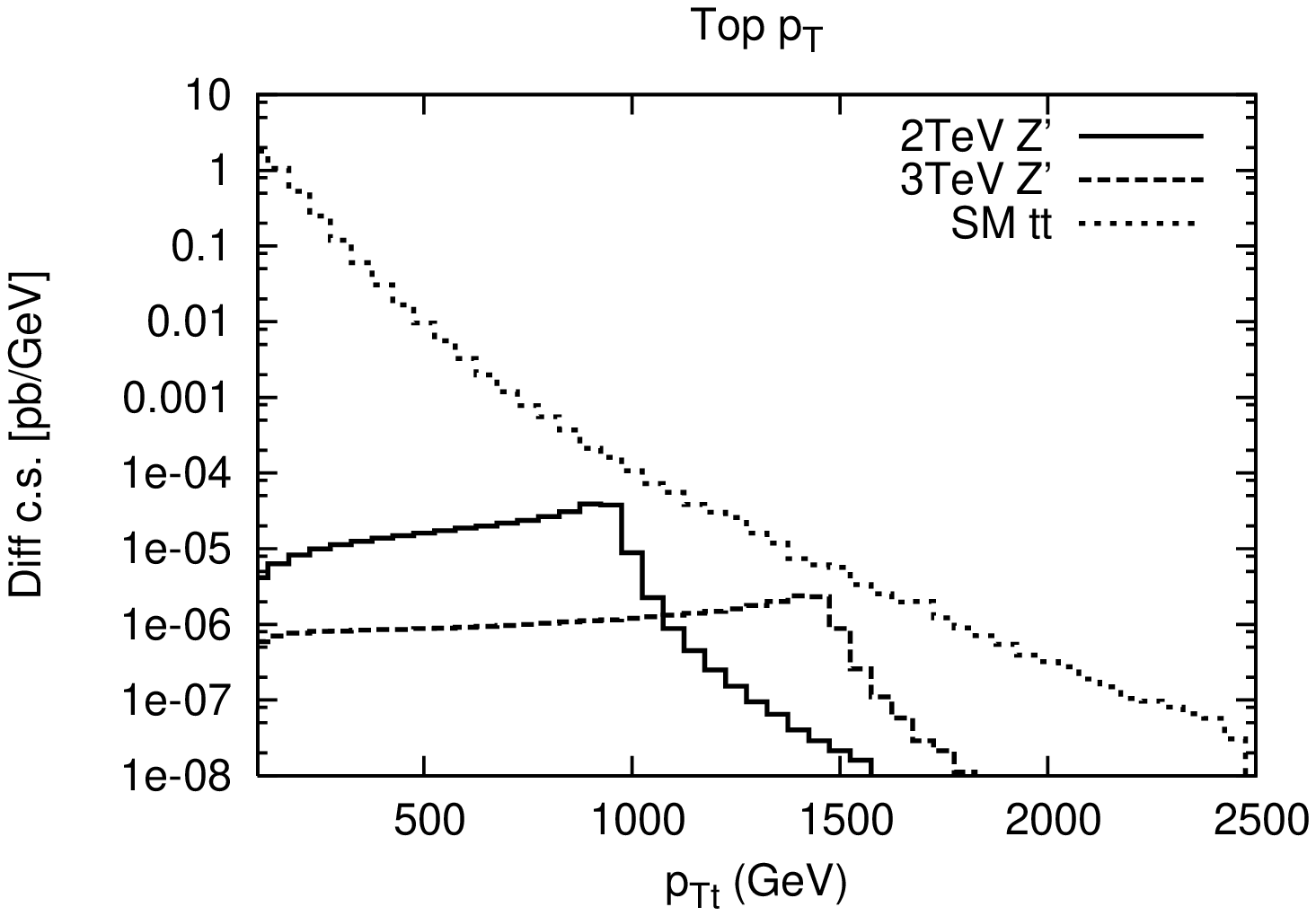}}
\scalebox{0.53}{\includegraphics[angle=270]{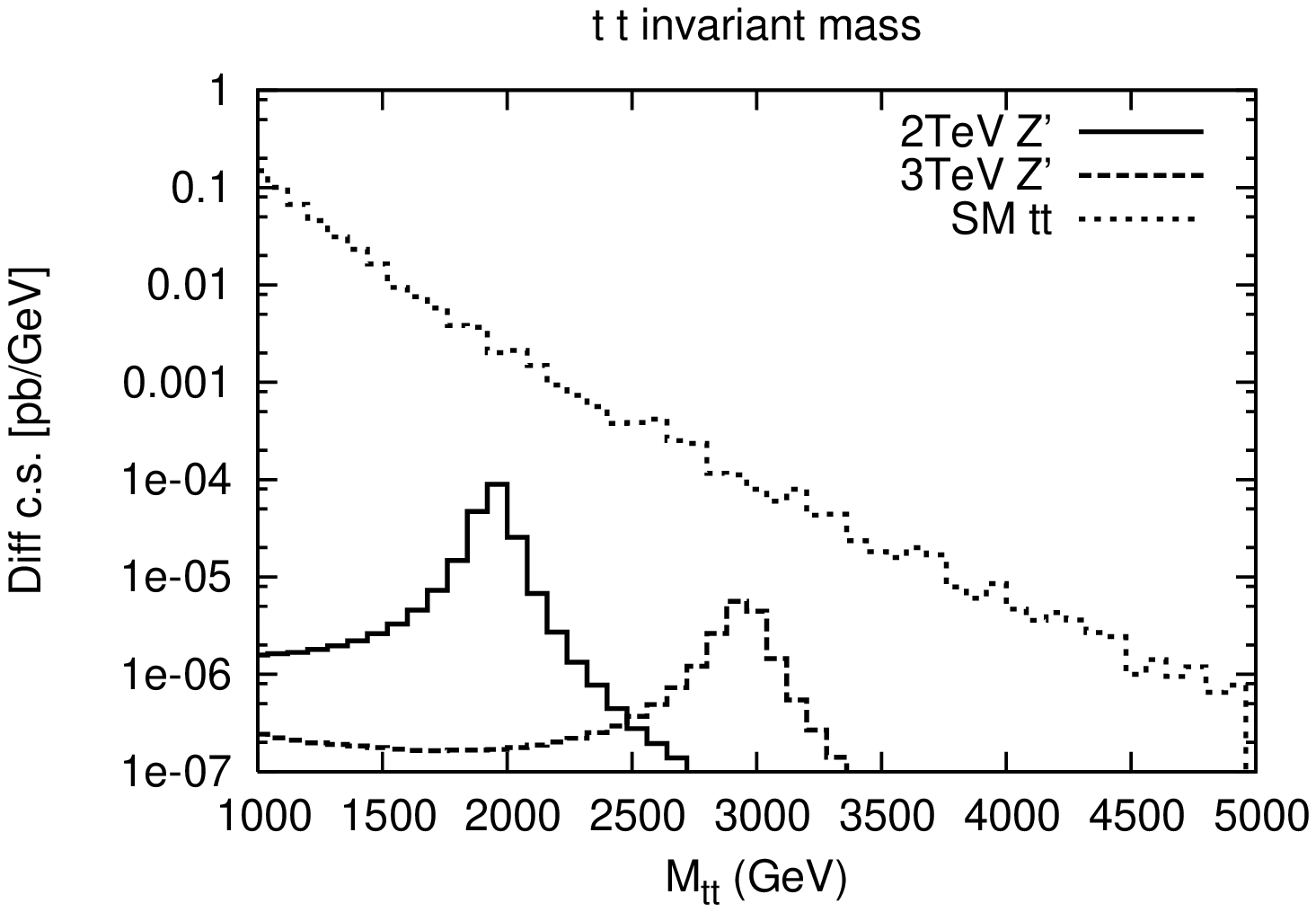}}
\caption{The differential cross-section of Drell-Yan production of 2 and 3 TeV signal
and SM background, as a function of ${p_T}_t$ (left), and $M_{tt}$ (right), after basic cuts. 
The KK gluon contribution has not been included in these plots.
\label{dyaptt.FIG}} 
\end{center}
\end{figure}
Fig.~\ref{dyaptt.FIG} shows the Drell-Yan cross-section of $\zpri$ with mass 2~and~3~TeV 
into the $t\bar{t}$ final state after basic cuts. 
\begin{table}[h]
\caption{
$pp\rightarrow t \bar t$ cross-section (in fb) 
for $\mzpri=2$ and $3$ TeV signal and the corresponding SM backgrounds, with cuts applied 
successively ($p_T$ and $M_{\ell\ell}$ are in GeV).  
\label{pp2tt.TAB}}
\begin{center}
\begin{tabular}{|c|c|c|c|}
\hline 
$\mzpri=2$ TeV&
Basic&
$p_{T}>800$&
$1900<M_{tt}<2100$\tabularnewline
\hline
\hline 
Signal&
$17$&
$7.2$&
$5.6$\tabularnewline
\hline 
SM $t\bar{t}$&
$1.9\times10^{5}$&
$31.1$&
$19.1$\tabularnewline
\hline
\hline 
$\mzpri=3$ TeV&
Basic&
$p_{T}>1250$&
$2850<M_{tt}<310$\tabularnewline
\hline
\hline 
Signal&
$1.7$&
$0.56$&
$0.45$\tabularnewline
\hline
SM $t\bar{t}$&
$1.9\times10^{5}$&
$4.1$&
$1.1$\tabularnewline
\hline
\end{tabular}
\end{center}
\end{table}
In Table~\ref{pp2tt.TAB} we show the cross-section as we tighten the cuts
without including any top decay branching ratios.
We see that the signal observability over the SM background is promising at this
level.

The fully hadronic mode from $t\bar t$ decays to $b \bar b + 4 j$ 
has a branching fraction about BR$\approx 0.65^2 = 43\%$.
In the hadronic mode, we expect for a 2~TeV $\zpri$ that the 
$jj$ opening angle of $jj$ from the $W$ is $2 M_W/p_T \sim 0.32$~rad.
The multiple-jet QCD background Will be difficult to overcome making this decay 
channel difficult to observe.
The semi-leptonic mode  for $\ell = e,\mu$
has with BR$=0.65 \times 0.12 \times 2 \times 2 =31\%$ and the event reconstruction
has been discussed in Refs.~\cite{Barger:2006hm}, 
and the signal significance is found to be
encouraging consistent with Table~\ref{pp2tt.TAB}.

Now we turn to {$\zpri \rightarrow b\bar{b}$}.
The $\zpri$ cross-section for $\mzpri=2~$TeV into this final state is 8.4~fb without any cuts.
With the cuts 
\beq
|\eta_{b,\bar b}| < 1,\quad 1900\ {\gev} < M_{bb} < 2100\ {\gev}\ ,
\label{bbarcuts.EQ}
\eeq
the cross-section
is $\sigma_S = 0.7$~fb, while the SM background with the same cuts is $\sigma_B = 14.9$~fb. 
The significance is thus marginal for this channel.

\begin{figure}
\begin{center}
\scalebox{0.75}{\includegraphics{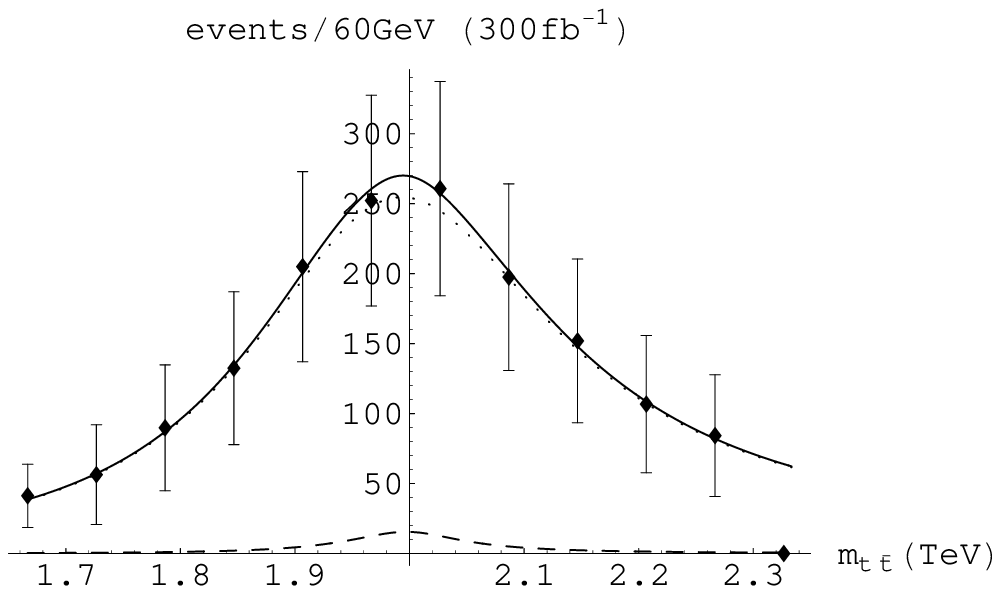}}\scalebox{0.75}{\includegraphics{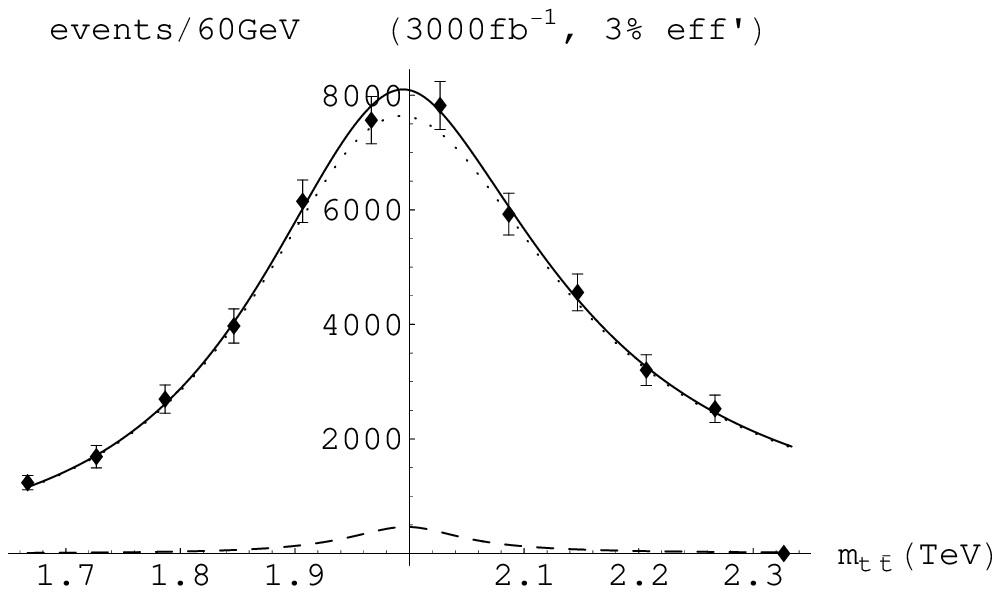}}
\scalebox{0.75}{\includegraphics{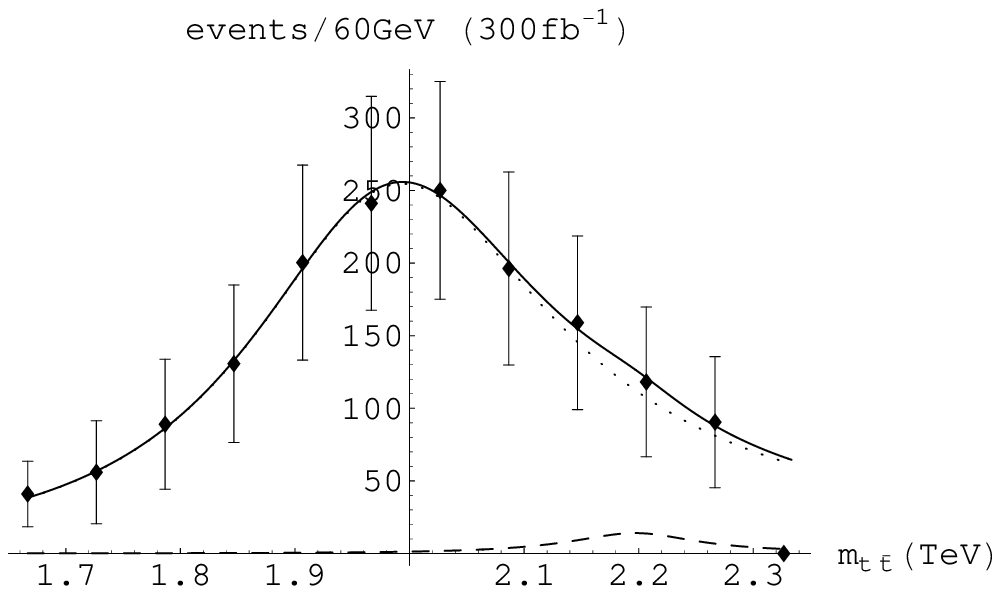}}\scalebox{0.75}{\includegraphics{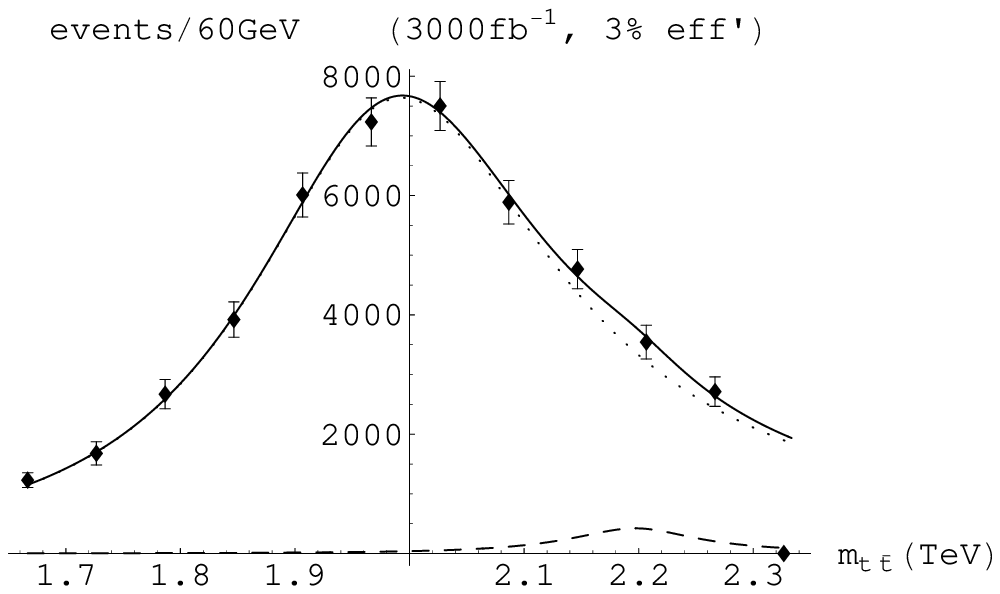}}
\caption{
The KK gluon and $\zpri$ line-shapes at the peak. The dotted line is the BW shape 
due to the KK gluon. The dashed line is due to the 3 neutral modes. 
The solid line is the sum. The error-bars shown are statistical only for the indicated 
integrated luminosity.
\label{kkGZtt.FIG}}
\end{center}
\end{figure}

However, as discussed in recent literature~\cite{kkgluon}, 
the KK gluon ($G_1$) contributes dominantly to the $t\bar t$ mode 
with a cross-section of 
about $938$~fb without any cuts for the $\mzpri = 2$~TeV case, 
and  is $108$~fb after the cuts similar to Eq.~(\ref{bbarcuts.EQ}).
This large production rate may prohibit the observation of the $Z'$ in this channel.
This is illustrated in Fig.~\ref{kkGZtt.FIG}, where we see that a $Z'$ peak may be
totally buried under the $G_1$ signal.

Note that, like in the case of the SM $Z$ boson, the $\zpri$ will
induce a tree-level forward-backward asymmetry (or charge asymmetry)
that can be observed via the
the distribution of the $t\bar t$ and $b\bar b$ final states.
The SM predicts a very small asymmetry which is dominantly due to next
to leading order
QCD processes (NLO)\cite{Afb} 
(in $q \bar{q}$ annihilation)
which are further diluted at the LHC due
to fact that the production is dominated by $gg$ fusion.
Interestingly enough, since the KK gluon is dominantly  produced via
$q\bar q$ annihilation
then the asymmetry due to the NLO processes will be enhanced and
expected to be of $\cal O$ (10\%).
Furthermore, near the peak of $\zpri$ the ratio between the KK gluon
background and the signal
is roughly about ten (depends on how degenerate they are). 
The nondegeneracy between the KK gluon and $\zpri$ masses can be 
generated for example from loop corrections to the 
brane kinetic terms~\cite{BrKin.REF},
and we illustrate its effect in the lower panel of Fig.~\ref{kkGZtt.FIG}.
The $\zpri$ would yield an additional source of
forward backward asymmetry roughly at the same size (after taking into
account the ratio of cross sections).
If measured, this pattern, in the asymmetries associated with the
location of the KK gluon and the location of the $\zpri$ resonance
would yield an intriguing hints that the signal indeed originated
from the above set-up.
Since the new boson would generically decay dominantly either to RH or
LH tops we expect
that the sign of the  resulting left-right polarization asymmetry
would be the same as the one
induced by the KK gluon~(see the first Ref. in \cite{kkgluon}).

\section{Conclusions}
\label{conclude}
The Randall-Sundrum I (RS1) framework of a warped extra dimension 
provides a novel and very interesting resolution to the Planck-weak 
{\em and} flavor hierarchy problem of the SM. 
As we enter the LHC era, however, it is of crucial importance to 
know the prospects for experimentally verifying this framework.
This amounts to observing the Kaluza-Klein (KK) 
excitations
of bulk fields in RS1
and 
measuring their couplings.

Considerations related to flavor and electroweak  (EW) physics, 
as well as the requirement of UV insensitivity, 
suggest that the SM fields may propagate in the bulk, 
with the light fermions being localized near the UV brane (Planck brane) 
and heavier fermions closer to the IR brane (TeV brane).
The resulting setup leads to two serious challenges for the LHC phenomenology:
{\it (i)} KK couplings to light fermions, and in particular to proton's constituents, are suppressed
since the KK states are localized near the TeV brane. 
{\it (ii)} The dominant decay channels of the new states is to TeV-brane localized fields, namely  
longitudinal gauge bosons, the Higgs and third generation quarks.  
These features make most of the new  states rather elusive.    

Nonetheless, it was shown in Ref.~\cite{kkgluon} 
that a KK gluon with a mass up to $\cal O$(4 TeV) is within the reach of the LHC.
However, observing a single KK state would not suffice to verify the above class of models.
The aforementioned challenging features were shown to make the discovery of a 
KK graviton questionable, unless it is unexpectedly light~\cite{KKGravitonRefs}.

In this work, we considered the corresponding neutral KK states of the EW sector. 
We focused on a class of models with  custodial symmetry for $Z\to b\bar b$ (and the $\rho$ parameter). 
In these models in addition to the SM fields, there are three neutral KK modes
present, denoted collectively as $Z'$, with masses of order of a few TeV, 
in compliance with precision tests.

In accordance with the above discussion, we find 
that discovering these states is a non-trivial task.
The leading production channel is the Drell-Yan process.
We investigated various decay modes and analysis strategies.  We showed that, 
unlike the often studied $Z^{ \prime}$ cases, 
the LHC $Z'$ mass reach for the above RS1 models is more limited, where  KK states of mass  
$\sim 2 \; (3)$ TeV can be discovered with a 
$\sim 100$ fb$^{-1}$ ($\sim 1$ ab$^{-1}$) of integrated luminosity.
Since the electroweak and flavor precision tests favor
KK mass $\gtrsim 3$~TeV in the simplest existing models, our results 
clearly motivate luminosity upgrade for the LHC.
The best discovery mode is via a $Zh$ (a $Z$ and a Higgs) final state 
which works both for a light and heavy Higgs.  
The assumption here is that the decays of the 
Higgs are dominated by SM final states and the corresponding branching fractions 
are approximately like in the SM.

However, only one of the three neutral eigenstates dominantly decays into the $Zh$ 
final state. We demonstrated that the two other modes 
can be discovered via    
longitudinal $WW$ final state, although it will require a higher luminosity. 
The $WW$ semi-leptonic mode 
(in general modes in which the $W$ and $Z$ decay hadronically) will benefit from  
unconventional jet-mass reconstruction 
techniques that may be devised in the future. 
It is worth noting, in addition, that with enough statistics one 
can look at the $W/Z$ polarization, 
associated with the signal in the differential cross section, and observe that 
they are dominantly longitudinally polarized as predicted by our framework.

The three neutral states have a sizable branching ratio  
into top pairs. However, they tend to be degenerate in mass with the KK gluon so that
the signal is completely swamped by KK gluon decay into tops. 
Precision measurements of
the top final state, such as forward backward asymmetry  
can, nevertheless, allow for indirectly observing 
the presence of the $Z'$.

Finally, we emphasize that, via the AdS/CFT duality \cite{Maldacena:1997re}, 
the RS framework can be viewed as a tool to study $4D$ strong dynamics. 
%
%
%
In fact, the idea of a composite pseudo-Goldstone boson (PGB) Higgs, in $4D$, 
has been studied in the RS framework (called ``holographic'' PGB Higgs) \cite{Contino:2003ve, Agashe:2004rs}.  
It is therefore likely that our results apply (in general) to $4D$ 
TeV-scale strong dynamics responsible for EWSB. 
In particular, our analysis with regard to the RS1 LHC signals 
suggests that little hierarchy models with
UV completion via strong dynamics\footnote{In fact, see reference 
\cite{Thaler:2005en} for UV completion of the 
Little{\em st} Higgs model using RS framework.}
({\it i.e.}, little Higgs 
and some flat extra dimensional models)
would
be characterized by LHC signals which are quite different from those 
usually emphasized in the literature. 
The reason is that the couplings between the extended electroweak sector and the 
light (heavy) {\em SM} particles may be actually highly suppressed (enhanced), 
unlike what is typically assumed in other 
LHC studies.\footnote{Ref.~\cite{Perelstein:2003wd} does mention, in the context of LHC signals, 
that suppressed couplings of light fermions to $Z^{ \prime }$, $W^{ \prime }$ are 
motivated in order to satisfy electroweak precision tests. 
However, most of these studies still assume {\em universal} 
fermionic couplings so that couplings to top quark are 
also suppressed in this case.
Whereas, we emphasize that top quark couplings to 
the new states are likely to be enhanced, 
leading to difficulties in detection of the new states.}
Generically, the new particles will be broader,
with small production rates and non-leptonic decay channels. 
As such, these models may face similar challenges regarding the 
detection of new states.

\section*{Acknowledgments}
We are very grateful to F.~Paige for many discussions,
particularly on jet-mass and b-tagging issues.
For help with Monte Carlo tools, we thank A.~Belayev (CalcHEP), R.~Frederix (MadGraph) 
and M.~Reece (Bridge). 
We would also like to thank U.~Baur, G.~Brooijmans, B.~Kilgore and G.~Sterman
for discussions.
KA is supported in part by the U.~S.~DOE under
Contract no. DE-FG-02-85ER 40231.
HD, SG and AS are supported in part by the DOE grant
DE-AC02-98CH10886 (BNL).
TH and G.-Y.H are supported in part by a DOE grant
No. DE-FG02-95ER40896 and in part by the Wisconsin Alumni Research Foundation.
KA, TH and GP  thank the Aspen Center for Physics for hospitality.

\appendix


\section{Model details and Heavy Electroweak Gauge Bosons}
\label{couplings}

This section describes the couplings of $1$ heavy $W$ or $Z$ state
to $2$ SM 
states
%
%
in a model
with the SM gauge fields propagating in the bulk of a warped extra dimension
and the Higgs being localized close to the TeV brane. 
In subsection \ref{simple}, we begin with 
couplings of heavy
$W$ or $Z$ to (i) $2$ SM $W$ or $Z$ and (ii) Higgs and SM $W$ or $Z$, considering  first
the simplified case of a {\em single} $SU(2)$ and giving a detailed derivation
of the couplings {\em in unitary gauge}. 
Since
these unitary gauge couplings have not been explicitly derived in the literature before
to our knowledge, 
we feel that such a pedagogical treatment will be useful.
If the reader wishes, 
she/he can skip the derivation and go directly to the couplings
in Eq.~(\ref{massbasis}) for the simplified case. We give a check
against equivalence theorem in subsection \ref{equivalence}, followed
by an outline only of the derivation of the same couplings
for the realistic case in subsection \ref{realistic}.
Finally, 
in subsection \ref{fermion}, we discuss the couplings of heavy $W$ or $Z$ to fermions
with the charge assignments of reference \cite{Agashe:2003zs}
and in subsection \ref{Zbb}, we consider the charge assignments
of reference \cite{Agashe:2006at} with the custodial symmetry for $Z b \bar{b}$.
Various expressions directly relevant to our numerical study are summarized in
App.~\ref{coupl.APP}.

\subsection{Simplified case of single $SU(2)$}
\label{simple}

\subsubsection{Couplings to two SM $W$ or $Z$ in unitary gauge} 
\label{twosmwz.sec}
The basic idea is that 
there are {\em no} couplings of $2$ gauge
%
%
zero-modes to $1$ gauge KK mode
at {\em tree}-level
due to flatness of zero-mode profile and orthogonality of 
profiles.\footnote{This also follows from $4D$ gauge invariance.}
However, Higgs vev mixes zero and KK modes of $W$ so that mass eigenstates
--
``heavy'' $W$ and SM $W$ -- are admixtures of the two, with former being
mostly KK $W$ and the latter being mostly zero-mode $W$. So,
we can start with a coupling of $3$ $W$ zero-modes
or a coupling of $2$ $W$ KK modes 
and $1$ $W$ zero-mode 
and use
the above mixing to obtain a coupling
of $1$ heavy $W$ to $2$ SM $W$. There are also couplings with $3$ KK modes
which requires mixing {\em twice} to obtain coupling of $1$ heavy $W$ to $2$ SM
$W$
and hence will be a higher order effect.

The mass terms (restricting to only zero
and 1st KK modes)
are
\begin{eqnarray}
\frac{1}{4} g^{ (0) \; 2 } v^2 \Big[ W^{ + \; (0) } W^{ - \; (0) }
+ \sqrt{ k \pi r_c }  W^{ + \; (0) } W^{ - \; (1) } \; (+ \; h.c.)
+ k \pi r_c  W^{ + \; (1) } W^{ - \; (1) } + \nonumber \\
\frac{1}{2}  W^{3  \; (0) } W^{ 3 \; (0) }
+ \sqrt{ k \pi r_c }  W^{ 3 \; (0) } W^{ 3 \; (1) } \; 
+ \frac{1}{2} k \pi r_c  W^{ 3 \; (1) } W^{ 3 \; (1) } \Big] 
+ \nonumber \\
m^2_{ W^{ (1) } } 
\Big[
W^{ + \; (1) } W^{ - \; (1) }
 + \frac{1}{2} 
W^{ 3 \; (1) } W^{ 3 \; (1) }
\Big]
\label{massterms}
\end{eqnarray}
where
$g^{ (0) } = g_{ 5D } / \sqrt{ \pi r_c }$ is the zero-mode 
(or $4D$) gauge
coupling.
The factor of $\xi \equiv \sqrt{ k \pi r_c }$ comes from the enhanced coupling
of the Higgs\footnote{We assume
Higgs as $A_5$ here \cite{Contino:2003ve, Agashe:2004rs}.} 
that is peaked near the TeV brane to gauge KK modes, in turn, 
due to 
enhanced wavefunction
of KK modes compared to zero-mode at the TeV brane. Also, the KK mass
is
\begin{eqnarray}
m_{ W^{ (1) } } \equiv m_{ KK } & \approx & 2.45 \; k \; e^{ - k \pi r_c }
\end{eqnarray}
where (as usual) $k \pi r_c \sim \log \left( M_{ Pl }/ \hbox{TeV} 
\right) \sim 34$ and $k \sim M_{ Pl }$
so that $m_{ W^{ (1) } }
%
%
\sim$ (a few) TeV. 
%
We define
\beq
\xi \equiv \sqrt{ k \pi r_c } \ ,
\label{xidefn.EQ}
\eeq
and we take $\xi = \sqrt{34} = 5.83$ for our numerical study.

The mass eigenstates, denoted by $W$ (``SM'') and $W^{ \prime }$
(heavy $W$), are
\begin{eqnarray}
W^{ (1) } & \approx & \cos \theta  
W^{ \prime } + \sin \theta  W   
\nonumber \\
W^{ (0) } & \approx & \cos \theta W - 
\sin \theta W^{ \prime }
\label{basischange} 
\end{eqnarray}
where 
\begin{eqnarray}
\tan 2 \theta & = & \frac{ \frac{1}{2}
g^{ (0) \; 2 } v^2 \sqrt{ k \pi r_c } }{ m_{ W^{ (1) } }^2 + \frac{1}{4}
g^{ (0) \; 2 } v^2 \left( k \pi r_c - 1 \right) }
\end{eqnarray}
valid for both charged and neutral $W$.
Clearly, the mass and couplings of {\em SM} $W$ are shifted relative to those of the
zero-mode due to the above mixing with the KK mode,
but this effect can be neglected for our purposes since it will be 
higher order
in $v / m_{ KK }$.
So, we set $1/2 \; g^{ (0) } v \approx m_W$ and $g^{ (0) } \approx g$,
{\it i.e.}, the SM $SU(2)_L$ gauge coupling, also denoted by $g_L$. 
Assuming $m_W^2 \; k \pi r_c 
\ll m_{ KK }^2$ (which holds for $m_{ KK } \stackrel{>}{\sim}$ a few TeV),
we get
\begin{eqnarray}
\sin \theta & \approx & \frac{ m_W^2 \sqrt{ k \pi r_c } }{ m_{ KK }^2 }
\label{mixingangle}
\end{eqnarray}

The Feynman rules in KK basis are 
(1) $3$ zero-mode couplings: 
\begin{eqnarray}
W^{ + \; (0) }_{ \nu }  \left( k_2 \right) W^{ - \; (0) }
_{ \lambda } \left( k_3 \right)
W^{ 3 \; (0) }_{ \mu }  \left( k_1 \right) & : & - i g 
\Big[ \left( k_1 - k_2 \right)_{ \lambda } g_{ \mu \nu }  +
\left( k_2 - k_3 \right)_{ \mu } g_{ \nu \lambda } + 
\left( k_3 - k_1 \right)_{ \nu } g_{ \lambda \mu }
\Big]
\label{allzero}
\end{eqnarray}
and (2) KK $W^3$, KK $W^-$ and $W^+$ zero-mode and (3)
KK $W^3$, KK $W^+$ and $W^-$ zero-mode
coupling which are identical to that in Eq.(\ref{allzero}). 
As mentioned above, the $3$ KK $W$ coupling
will give a higher order effect.

We now go from the KK basis to the
mass eigenstate basis using Eq. (\ref{basischange}). 
Schematically,
we use the above mixing to ``convert''
$W^3$ zero-mode to heavy $W^3$ in coupling (1) in Eq.~(\ref{allzero})
(which gives a factor of $\sin \theta$) and convert KK $W^{ \pm }$ to SM $W^{ \pm }$ in 
couplings (2)
and (3)
(which gives a factor of $-\sin \theta$). Thus, we obtain a coupling
of $1$ heavy $W_3$ to SM $W^-$ and SM $W^+$ (setting
$\cos \theta \approx 1$):
\begin{eqnarray}
W^+_{ \nu }  \left( k_2 \right) W^-_{ \lambda } \left( k_3 \right)
W^{ 3 \; \prime}_{ \mu }  \left( k_1 \right) & : & - i g \sin \theta
\Big[ \left( k_1 - k_2 \right)_{ \lambda } g_{ \mu \nu }  +
\left( k_2 - k_3 \right)_{ \mu } g_{ \nu \lambda } + 
\left( k_3 - k_1 \right)_{ \nu } g_{ \lambda \mu }
\Big]
\label{massbasis}
\end{eqnarray}
with $\sin \theta$ given in Eq.~(\ref{mixingangle}).

Similarly couplings of heavy $W^+$ to SM $W_3$ and SM $W^-$ can be
obtained.

\subsubsection{Couplings to Higgs and SM $W$ or $Z$}

>From Eq.~(\ref{massterms}), {\it i.e.}, replacing single $v$
by physical Higgs ($h$), and going from KK to mass basis, we get (setting
$g^{ (0) } \approx g$,
$g^{ (0) } v / 2 \approx m_W$ and
$\cos \theta \approx 1$)
\begin{eqnarray}
{\cal L}_{ Higgs } & \approx & 
m_W
%
%
g \sqrt{ k \pi r_c }
h \Big[ W^{ 3 \; \prime } W^3 + 
W^{ + \; \prime } W^- ( + h.c. ) \Big]
\label{higgs}
\end{eqnarray}

\subsubsection{Check against Equivalence Theorem}
\label{equivalence}

From Eq.~(\ref{massterms}), 
we can see that the couplings of complex Higgs
doublet ($H$) to a single gauge KK mode only are given by
$\sim \partial_{ \mu } H^{ \dagger } H W^{ (1) } g \sqrt{ k \pi r_c }$
-- to be explicit, replace $g^{ (0) } W^{ (0) }$
by $\partial_{ \mu }$ and $v / \sqrt{2}$ by $H$ in 2nd term of 
Eq.~(\ref{massterms}).
By equivalence theorem,
longitudinal $W$ and $Z$ are (approximately)
the unphysical Higgs and hence the coupling
of heavy $W$ to (i) 2 longitudinal SM $W$'s and
also to (ii) physical Higgs and longitudinal SM $W$ 
is expected to be of the above size, 
i.e., $W_{ long. } W_{ long } W^{ \prime} $ 
and $W_{ long. } h W^{ \prime}$ couplings 
$\sim g \sqrt{ k \pi r_c }$ (up to the factor of derivative/momentum).
Using the longitudinal polarization vector ($\sim E / m_W$) with $E 
\sim m_{ KK }$
(which is valid for production/decay of heavy $W$), 
we do indeed get same result in unitary gauge from
Eq.~(\ref{massbasis}) and similarly from Eq.~(\ref{higgs}).

\subsection{Realistic case}
\label{realistic}

We have the gauge group
$SU(2)_L \times SU(2)_R \times U(1)_X $ in the bulk
with the SM Higgs doublet being promoted to a bi-doublet of
$SU(2)_L \times SU(2)_R$, {\em i.e.}, $\left( H, i \sigma_2 H^{ \ast }
\right)$ transform as a doublet of $SU(2)_R$,
where $H$ and $i \sigma_2 H^{ \ast }$ 
are each doublets of $SU(2)_L$ as usual and does not carry any $U(1)_X$ charge.
The charged boson matrix is
\begin{eqnarray}
\left(
\begin{array}{ccc}
W_L^{ + \; (0) } & W_{ L_1 }^+  & 
W_{ R_1 }^+  
\end{array}
\right)
{\cal M}_{ charged }^2
\left(
\begin{array}{c}
W_L^{ - \; (0) } \\ 
W_{ L_1 }^-  \\
W_{ R_1 }^-
\end{array}
\right)
\label{LwpriM.EQ}
\end{eqnarray}
with
\begin{eqnarray}
{\cal M}_{ charged }^2 & = & 
\left(
\begin{array}{ccc} 
m_W^2 & m_W^2 \sqrt{ k \pi r_c } &
- m_W^2 \sqrt{ k \pi r_c } \frac{ g_R }{ g } \\
m_W^2 \sqrt{ k \pi r_c } & m_{ KK }^2 + m_W^2 
k \pi r_c & - m_W^2 
k \pi r_c 
\frac{ g_R }{ g } \\
- m_W^2 \sqrt{ k \pi r_c } \frac{ g_R }{ g_L } & - m_W^2 
k \pi r_c 
\frac{ g_R }{ g } & 0.963 \; m_{ KK }^2 + m_W^2 k \pi r_c 
\left( \frac{ g_R }{ g } \right)^2
\end{array}
\right)
\end{eqnarray}
where we have restricted to 
only the first KK modes, denoted by $W^{ \pm }_{ L_1, \; R_1 }$.
Note that 
there is no zero-mode for $W_{R}^+$ due to choice of Dirichlet 
boundary condition (BC) on Planck brane
so that 
$g_R$ is the ``would-be'' zero-mode (or $4D$) $SU(2)_R$ gauge coupling.
Due to the different
BC on Planck brane relative to $W^+_L$, the KK mass for $W^+_{ R_1 }$
is also slightly smaller:
\begin{eqnarray}
M_{ W^+_{ R_1 } } & \approx & 0.981 \; m_{ KK }
\end{eqnarray}

The 
mass eigenstates -- $W$ (SM) and $\wpLt$,
$\wpRt$ ({\em two} heavy $W$'s) -- 
are mixtures of these $3$ modes.
Note that (EW preserving) KK masses for $W^+_{ R_1 }$
and $W_{ L_1 }^+$ are quite degenerate such that 
the EWSB mixing (mass)$^2$
term is larger than KK (mass)$^2$ 
splitting for $m_{ KK } \stackrel{<}{\sim} 3.5$ 
TeV.
Hence, for the interesting range of KK masses, we expect large mixing
between $W_{ R_1}^+$
and $W_{ L_1 }^+$, {\it i.e.}, $\wpLt$ and
$\wpRt$ will be roughly $50-50$ admixtures
of $W_{ R_1 }^+$ and $W_{ L_1 }^+$, of course
with a small component of $W_L^{ + (0) }$.

In the neutral gauge boson sector, it is convenient to define
the KK $Z$ (denoted by $Z_{ \mu }^{ ( 1, \; 2...) }$) and KK photon 
(denoted by $A_{ \mu }^{ ( 1, \;  2,...) }$) to be linear combinations of 
KK $W^3_L$, i.e., $W^{ 3 \; (1, \;  2,...) }_L$, and KK hypercharge, i.e.,
$B^{ (1, \;  2,...) }$, with mixing 
identical to that for zero-modes, i.e.,
\begin{eqnarray}
A_{ \mu }^{ (n) } & = & \sin \theta_W W_{ \mu \; L }^{ 3 \; (n) }
+  \cos \theta_W B^{ (n) }_{ \mu } \nonumber \\
Z_{ \mu } ^{ (n) } & = & \cos \theta_W W_{ \mu \; L }^{ 3 \; (n) }
- \sin \theta_W B^{ (n) }_{ \mu }
\end{eqnarray}
where $n = 0, 1...$ and
with $\sin \theta_W$ being the ratio of $4D$ (or zero-mode) hypercharge
and $Z$ gauge couplings\footnote{The weak
mixing angle defined in this manner differs from
the observed $\sin^2 \theta_W$
by higher order (in $v / m_{ KK }$) corrections
coming from the zero-KK mode
mixing.
Such effects are 
important
in the EW fit, but they can be neglected for our purpose
and hence we can set $\sin^2 \theta_W$ defined {\em as above}
to be the observed one.}
or equivalently the $5D$ hypercharge and $5D$ $Z$ gauge couplings,
where $g_{ 5D \; Z } \equiv \sqrt{ g_{ 5D \; L }^2 + g_{ 5D \; Y }^2 }$.
In turn, the hypercharge gauge boson KK modes are
linear combinations of KK modes of $U(1)_R$ and $U(1)_{ X }$
gauge boson (denoted by $X_{ \mu }^{ ( 1, \;  2,...) }$), with the 
combination 
orthogonal to hypercharge gauge boson
being denoted by 
$Z_X$, i.e., 
\begin{eqnarray}
B_{ \mu }^{ (n) } 
& = & \sin \theta^{ \prime } W_{ \mu \; R }^{ 3 \; (n) } 
+ \cos\theta^{ \prime } 
X_{ \mu }^{ (n) } \nonumber \\
Z_{ \mu \; X }^{ (n) }  & = & \cos \theta^{ \prime } 
W_{ \mu \; R }^{ 3 \; (n) } - \sin \theta^{ \prime } X_{ \mu }^{(n)}
\end{eqnarray}
In analogy with $B-W^3_L$ mixing, we have 
$\sin \theta^{ \prime } = g_{ X } / g_{ Z^{ \prime } }$, where
$g_{Z ^{ \prime } } = \sqrt{ g_R^2 + g_{ X }^2 }$ 
and $ g_{ X }$ are the
``would-be'' zero-mode ($4D$) couplings for
$Z_X$ and $U(1)_{ X }$, respectively. 
Note that since
the hypercharge gauge coupling, $g^{ \prime }
=  g_R  \; g_{ X } / \sqrt{ g_R^2 + g_{ X }^2 }$\footnote{Equivalently,
$1 / g_{ 5D \; Y }^2 \equiv 1 / g_{ 5D \; R }^2 +  1 / g_{ 5D \; X }^2$ 
with $g_{ 5D \; X }$ being the $5D$ $X$ gauge coupling.},
there is only $1$ gauge
coupling (say, either $g_R$ or $g_{ Z^{ \prime } }$)
which is a free parameter. See reference \cite{Agashe:2003zs} for more details.

The 
photon and $Z$
KK masses are given by
\begin{eqnarray}
M_{ A_1, \; Z_1 } & = & m_{ KK }
\end{eqnarray}
since both have
Neumann BC's on both branes.
Whereas $Z_X$ does not
have a zero-mode due
to (effectively) Dirichlet BC on Planck brane
so that
\begin{eqnarray}
\mzx & \approx & 0.981 \; m_{ KK }
\end{eqnarray}
Here, 
we have denoted the first KK 
excitation of the three neutral gauge bosons as $A_1$, $\zp$ and $\zx$.

The advantage of
this definition of KK $Z$ and KK photon is that the photon 
(zero and KK) modes do not couple
to Higgs at leading order and hence do not mix with each other or with $\zp$ or
$\zx$ modes even {\em after} EWSB.
Hence, the SM photon {\em is} the zero-mode photon, i.e., EM
coupling is not modified with respect to that of the zero-mode
(this is guaranteed by $4D$ gauge invariance), unlike
for the case of $W$ and $Z$.

Similar to the case of charged gauge bosons, the neutral gauge boson mass matrix is
\begin{eqnarray}
\left(
\begin{array}{ccc}
Z^{ (0) } & \zp & 
\zx  
\end{array}
\right)
\frac{1}{2} {\cal M}_{ neutral }^2
\left(
\begin{array}{c}
Z^{ (0) } \\ 
\zp \\
\zx
\end{array}
\right)
\label{LzpriM.EQ}
\end{eqnarray}
%
\begin{eqnarray}
{\cal M}_{ neutral }^2 & = & 
\left(
\begin{array}{ccc} 
m_Z^2 & m_Z^2 \sqrt{ k \pi r_c } &
- m_Z^2 \sqrt{ k \pi r_c } \frac{ g_{ Z^{ \prime } } }{ g_Z } 
{c^\prime}^2 \\
m_Z^2 \sqrt{ k \pi r_c } & m_{ KK }^2 + m_Z^2 
k \pi r_c & - m_Z^2 \;
k \pi r_c 
\frac{ g_{ Z^{ \prime } } }{ g_Z } {c^\prime}^2 \\
- m_Z^2 \sqrt{ k \pi r_c } \frac{ g_{ Z^{ \prime } } }{ g_Z } 
{c^\prime}^2 & - m_Z^2 \;
k \pi r_c 
\frac{ g_{ Z^{ \prime } } }{ g_Z } {c^\prime}^2 
& 0.963 \; m_{ KK }^2 + m_Z^2 k \pi r_c 
\left( \frac{ g_{ Z^{ \prime } } }{ g_Z } {c^\prime}^2 \right)^2
\end{array}
\right)
\nonumber \\
\end{eqnarray}
where ${c^\prime}\equiv \cos{\theta^\prime}$.

As before, we start with the gauge couplings in KK basis -- of
$3$ zero-modes or $1$ zero-mode and $2$ KK modes -- and 
use zero-KK mode mixing (i.e., go to mass eigenstate
basis) to obtain couplings of $1$ heavy $W$ (or $Z$) to
$2$ SM $W$ or $Z$.

In addition to the trilinear gauge couplings from $SU(2)_L$
group,
we also need to take into account those from $SU(2)_R$. Note that 
$W_R^{ \pm }$ does not have zero-modes so that there are no $3$ zero-mode
couplings from $SU(2)_R$. However,
the $W_R^{ + \; (1) }$-$W_R^{ - \; (1) }$-$W_R^{ 3 \; (0) }$ coupling does
contribute 
to the coupling of heavy $W$ to SM $W$ and SM $Z$
via mixing of $W^{ + (1) }_R$ with $W_L^{ + (0) }$, i.e.,
SM $W$ has a (small) admixture of $W_R^{ + \; (1) }$. 

{\bf Coupling of KK photon to $WW$}

We also obtain a coupling of KK {\em photon} 
to 2 SM $W$'s via trilinear $SU(2)_L$ 
coupling between $W_L^{ 3 \; (1) }$, $W_L^{ \pm (1) }$
and $W_L^{ \mp (0) }$ followed by
$W_L^{ \pm (1) }$ mixing with $W_L^{ \pm (0) }$ via Higgs vev
-- the point is that the 
KK photon has an admixture of $W_L^{ 3 \; (1) }$. It is clear that 
we 
cannot obtain such a coupling of KK
photon from trilinear $SU(2)_R$ coupling
(at the same order in $v / m_{ KK }$).
We can also obtain this coupling from equivalence theorem,
i.e., coupling of KK photon to 
(unphysical) charged Higgs.

\subsection{Couplings of fermions to heavy $W/Z/\gamma$}
\label{fermion}
Here we take $U(1)_X$ to be $U(1)_{B-L}$ as usual.
Neglecting effects suppressed by SM Yukawa couplings,
couplings of (zero-modes of) light quarks (including $b_R$ and excluding 
$(t,b)_L$ and $t_R$) to electroweak 
gauge KK modes in weak/KK basis are suppressed by $\sim \xi
\equiv \sqrt{ k \pi r_c } \sim 5$
compared to the SM couplings:
\begin{eqnarray}
{\cal L} & \ni & \frac{ - 1.13 }{ \xi } \left( 
\frac{ g_Z }{2} \zp\,_{ \mu } 
\Big[
\bar{u} \gamma^{ \mu }
\left( \frac{1}{2} - \frac{4}{3} \sin^2 \theta_W 
- \frac{1}{2} \gamma_5 \right) u 
+ \bar{d} \gamma^{ \mu } \left( \frac{-1}{2} + \frac{2}{3} \sin^2 \theta_W 
+ \frac{1}{2} \gamma_5 \right) d \Big] + \right. \nonumber \\
 & & \left. 
 \frac{g}{ 2 \sqrt{2} } \bar{u} \gamma^{ \mu } 
\left( 1 - \gamma_5 \right) 
d W_{ \mu }^{ + \; (1) } \; (+ \; \hbox{h.c.})
+ e A_{ \mu }^{ (0) } 
\Big[ \frac{2}{3}
\bar{u} \gamma^{ \mu } u + \frac{-1}{3} \bar{d} \gamma^{ \mu } d \Big] \right)
\label{light}
\end{eqnarray}
%
%
The couplings of KK $W$ or $Z$ to {\em all} leptons can be similarly obtained.
In particular, 
there's no coupling of $\zx$ and
$W_{ R_1 }^{ \pm }$ in this approximation.

%
%
Next, we give the couplings
of $(t,b)_L$ and $t_R$ to electroweak gauge
KK modes. For this purpose, we 
choose $c_{ t_R } = 0$ and $c_{ (t,b)_L } 
= 0.4$, as favored by combination of large
$m_t$ and constraint from $Z \rightarrow b \bar{b}$. We find
\begin{eqnarray}
{\cal L} & \ni &  
\frac{ g_Z }{2} \zp\,_{ \mu }
\Big[ 
\left( \frac{ -1.13 }{ \xi }
+ 0.2 \xi \right)
\left( 
\left( \frac{1}{2} - \frac{2}{3} \sin^2 \theta
\right)  
\bar{t} \gamma^{ \mu } \left( 1 - \gamma_5 \right) t 
+
\left( - \frac{1}{2} + \frac{1}{3} \sin^2 \theta
\right)
\bar{b} \gamma^{ \mu } \left( 1 - \gamma_5 \right) b
\right)
+ \Big. \nonumber \\
 & &\hspace*{6cm} \Big. \left( \frac{ -1.13 }{ \xi }
+  0.7 \xi \right) \frac{-2}{3}
\sin^2 \theta_W
\bar{t} \gamma^{ \mu } \left( 1 + \gamma_5 \right) t 
\Big] 
+ \nonumber \\
& & \frac{ g }{ 2 \sqrt{2} } \left( \frac{ -1.13 }{ \xi }
+ 0.2 \xi \right) 
W_{ \mu \; L }^{ + \; (1) }
\bar{t} 
\gamma^{ \mu } \left( 1 - \gamma_5 \right) b \; (+ \; \hbox{h.c.})
+ \nonumber \\
 & & \frac{ g_{ Z^{ \prime } } }{2} \zx\,_{ \mu } 
\Big[ 
- \frac{1}{6} \sin^2 \theta^{ \prime }_W  0.2 \xi
\left( \bar{t} \gamma^{ \mu } \left( 1 - \gamma_5 \right) t + 
\bar{b}
\gamma^{ \mu } \left( 1 - \gamma_5 \right) b
\right) + \nonumber \\
& & \hspace*{6cm} 0.7 \xi \left( 
\frac{1}{2}
- \frac{2}{3} \sin^2 \theta^{ \prime }_W
\right) \bar{t} \gamma^{ \mu } \left( 1 + \gamma_5 \right) t
\Big] + \nonumber \\ & & 
\ap\,_{ \mu } \frac{e}{2} 
\Big[ \left( \frac{ -1.13 }{ \xi }
+ 0.2 \xi \right) 
\left( \frac{2}{3} \bar{t} \gamma^{ \mu }
\left( 1 - \gamma_5 \right) t 
+ \frac{-1}{3} \bar{b} \gamma^{ \mu }
\left( 1 - \gamma_5 \right) b 
\right) + \nonumber \\
& &\hspace*{6cm} \frac{2}{3} \left( \frac{ -1.13 }{ \xi } + 0.7 \xi \right) \bar{t} \gamma^{ \mu }
\left( 1 + \gamma_5 \right) t 
\Big] 
\label{topbottom}
\end{eqnarray}
using the ``charge'' under
$Z_X$ (which multiplies $g_{ Z^{ \prime } }$ and 
the factors from the profiles) 
given by 
\begin{eqnarray}
Q_{ Z^{ \prime } } & = & 
T_{ 3 R } - Y \sin^2 \theta^{ \prime }_W
\end{eqnarray}

The reason for 
writing the couplings in this way 
in both Eqs.~(\ref{light}) and
(\ref{topbottom}) is as follows. We can show that the  
$1 / \xi$ terms (in the prefactors) originate
from overlap ``near'' the Planck brane\footnote{Based on AdS/CFT
duality, this is
the dual of the
coupling of SM fermions to techni-$\rho$ induced 
via first coupling of SM fermions 
to ``$\gamma$'' followed by $\gamma - \rho$
mixing.} and hence these terms are absent for $\zx$
(which vanishes near Planck brane). Moreover,
this overlap of profiles near Planck brane
(for KK $Z/W^{ \pm }_L$ only)
is universal (i.e., {\em in}dependent of $c$) and hence
is the same for $(t,b)_L$ and
$t_R$ in Eq.~(\ref{topbottom}) as for light fermions
in Eq.~(\ref{light}). Whereas, the terms $\propto$ 
$\xi$ (in prefactors) can be shown to come 
from overlap near the TeV brane\footnote{This is the dual
of {\em direct} coupling of SM fermions
to techni-$\rho$ (cf. via $\gamma - \rho$ mixing).} 
(and is present
for $\zx$ as well) and hence is
suppressed by Yukawas for light fermions (and was not therefore shown in
Eq.~(\ref{light})).
We can also show that the coefficients of the $\xi$-terms, i.e.,
$0.2$ for $(t,b)_L$
and $0.7$ for $t_R$ are
(roughly) proportional to $\left( 1/2 - c \right)$, at least for $c$ close to $+1/2$.

In reality, all we require is for
$t_R$ to have a profile highly peaked near TeV brane, {\it i.e.},  
$c_{ t_R }$ can vary (roughly) from $0$ to
$-1/2$ and
also that $(t,b)_L$ has
close to a flat profile, {\it i.e.}, 
$c_{ (t,b)_L }$ can vary (roughly) from $0.4$ to
$0.3$. However, based on the above discussion, 
the effect of these variations in $c$'s on the
couplings of top and bottom to gauge KK modes will be at most a factor of
$2$.

The coupling of $W^+_{ R_1 } - t^{ (0) }_R - \tilde{b}^{ (1) }_R$,
where $\tilde{b}^{ (1) }_R$ is the $SU(2)_R$ partner of the $t_R$ as
explained in Ref.~\cite{Agashe:2003zs}, 
does induce
a coupling to SM
$b$ via mass mixing of $\tilde{b}^{ (1) }_R$ with $b^{ (0) }_L$. However,
this coupling requires electroweak symmetry breaking
(EWSB),  {\it i.e.}, it 
will be suppressed by $v / m_{ KK }$ and hence
is sub-leading to the above couplings (which appear even at $0^{ \hbox{th} }$
order in $v$).

Finally, 
as usual, the couplings 
of SM fermions to {\em heavy} gauge bosons
can be obtained
using the transformation from KK (or weak) basis to
mass basis for the gauge bosons which is derived above.

Note that there is also 
a transformation from KK (or weak) to mass basis for {\em fermions}
due to mixing
between zero and KK fermion modes (and also among KK modes) induced
by EWSB. Therefore,
SM fermions are mostly zero-modes, but with an admixture of
KK modes. However, this zero-KK 
mode mixing is
%
%
proportional 
to (roughly) SM Yukawa couplings so that it's effect
on couplings of heavy $W$ or $Z$ to SM fermions
is important only for top and bottom quarks. Even for top and bottom quarks, 
this effect is higher order in $v$
and hence can be neglected.
Whereas, the effect of the transformation
from KK to mass basis in {\em gauge} sector on couplings
of heavy $W$ or $Z$ to SM fermions is 
possibly large. The reason is that, even though mass mixing terms
among gauge modes
are 
%
%
suppressed
by $v$ (just like for fermion modes), mixing {\em angles} between
$2$ KK modes ({\em not} between zero and KK modes)
can be large (i.e., {\em not} suppressed by $v$)
due to the degeneracy between gauge KK states which was mentioned above.

This argument also
indicates that we can {\em neglect} mixing between {\em zero}
and KK gauge boson modes  
(but not the KK-KK mixing)
in obtaining couplings 
of heavy $W/Z/\gamma$ to 
SM {\em fermions} since the effect of this mixing is indeed 
higher-order. This approximation (which we use) is useful because 
it is easier to diagonalize $2 \times 2$ mass matrix
(for KK modes only)
instead of $3 \times 3$ mass matrix (including zero-modes).
Of course, for determining the coupling of heavy gauge boson to 
2 SM gauge bosons
(in unitary gauge) we must include the mixing of zero
and KK gauge modes, i.e., diagonalize the {\em full}
$3 \times 3$ matrix.

\subsection{Other possibilities for top/bottom couplings}
\label{Zbb}

In general, the $U(1)_X$ factor
multiplying $SU(2)_L \times SU(2)_R$ 
does not have to be $U(1)_{ B - L }$.
So, there is a freedom in the choice for charges under $SU(2)_R$ and $U(1)_X$
for the SM fermions:
SM LH fermions can transform under $SU(2)_R$ and
RH fermions might not transform under $SU(2)_R$.
The only requirement is that the correct hypercharge is reproduced
\begin{eqnarray}
Y & = & T_{ 3 R } + X
\end{eqnarray}
and that the SM Yukawa couplings are $SU(2)_R \times U(1)_X$
invariant - they are automatically invariant if we identify $X = B - L$.

%

In particular, it was shown in reference \cite{Agashe:2006at}
that for the choice
\begin{eqnarray}
T_{ 3 R } & = & 
%
%
\begin{array}{c}
- \frac{1}{2} \; \hbox{for} \; (t,b)_L \nonumber \\
0 \; \hbox{for} \; t_R
\end{array}
%
%
\; \hbox{so that} \\
X 
& = & 
\frac{2}{3} \; \hbox{for} \; (t,b)_L \; \hbox{and} \; t_R 
\end{eqnarray}
and
\begin{eqnarray}
g_{ 5D \; L } & = & g_{ 5D \; R } 
\label{LR}
\end{eqnarray}
with Higgs having $X = 0$
there is a ``custodial symmetry'' which suppresses $Zb \bar{b}$.
Without this symmetry,
the KK mass $\stackrel{>}{\sim} 5$ TeV
based on the conservative limit that shift in
$Z b \bar{b} \stackrel{<}{\sim} 0.25 \%$.

In this case, we can have the other extreme profiles for $(t,b)_L$
and $t_R$, for example $c_ { (t,b)_L } = 0$ (near TeV brane)
and $c_{ t_R } = 0.4$ (close to flat profile)
giving the following couplings\footnote{obtained
from Eq. (\ref{topbottom}) by exchanging the profiles of $t_R$ and
$(t,b)_L$, i.e., $0.2 \leftrightarrow 0.7$ for the coefficient of
the $\xi$ terms and also the new $T_{ 3 R }$'s.}
\begin{eqnarray}
{\cal L} & \ni &  
\frac{ g_Z }{2} \zp\,_{ \mu }
\Big[ 
\left( \frac{ -1.13 }{ \xi }
+ 0.7 \xi \right)
\left( 
\left( \frac{1}{2} - \frac{2}{3} \sin^2 \theta
\right)  
\bar{t} \gamma^{ \mu } \left( 1 - \gamma_5 \right) t 
+
\left( - \frac{1}{2} + \frac{1}{3} \sin^2 \theta
\right)
\bar{b} \gamma^{ \mu } \left( 1 - \gamma_5 \right) b
\right)
+ \Big. \nonumber \\
& & \hspace*{6cm} \Big. \left( \frac{ -1.13 }{ \xi }
+  0.2 \xi \right) \frac{-2}{3}
\sin^2 \theta_W
\bar{t} \gamma^{ \mu } \left( 1 + \gamma_5 \right) t 
\Big] 
+ \nonumber \\
& & \frac{ g }{ 2 \sqrt{2} } \left( \frac{ -1.13 }{ \xi }
+ 0.7 \xi \right) 
W_{ \mu \; L }^{ + \; (1) }
\bar{t} 
\gamma^{ \mu } \left( 1 - \gamma_5 \right) b \; (+ \; \hbox{h.c.})
+ \nonumber \\
 & & \frac{ g_{ Z^{ \prime } } }{2} \zx\,_{ \mu } 
\Big[ \left( - \frac{1}{2} 
- \frac{1}{6} \sin^2 \theta^{ \prime }_W  \right) 0.7 \xi
\left( \bar{t} \gamma^{ \mu } \left( 1 - \gamma_5 \right) t + 
\bar{b}
\gamma^{ \mu } \left( 1 - \gamma_5 \right) b
\right)  + \nonumber \\
& &\hspace*{6cm} 0.2 \xi \left( 
- \frac{2}{3} \sin^2 \theta^{ \prime }_W
\right) \bar{t} \gamma^{ \mu } \left( 1 + \gamma_5 \right) t
\Big] + 
\nonumber \\ 
& & \ap\,_{ \mu } \frac{e}{2} 
\Big[ \left( \frac{ -1.13 }{ \xi }
+ 0.7 \xi \right) 
\left( \frac{2}{3} \bar{t} \gamma^{ \mu }
\left( 1 - \gamma_5 \right) t 
+ \frac{-1}{3} \bar{b} \gamma^{ \mu }
\left( 1 - \gamma_5 \right) b 
\right) \nonumber \\
& & \hspace*{6cm} + \frac{2}{3} \left( \frac{ -1.13 }{ \xi }
+ 0.2 \xi \right) \bar{t} \gamma^{ \mu }
\left( 1 + \gamma_5 \right) t 
\Big]
\label{tRflat}
\end{eqnarray}

Note that 
\begin{eqnarray}
\sin^2 \theta^{ \prime } & = & \tan^2 \theta_W \\
g_{ Z^{ \prime } }^2 & = & g_Z^2 \frac{ \cos^2 \theta_W }
{ 1 - \tan^2 \theta_W } 
\end{eqnarray}
due to Eq. (\ref{LR}).

However, constraints from flavor violation might still prefer $(t,b)_L$ to have
close to a
flat profile instead of close to TeV brane since there is 
no symmetry to suppress couplings of $b_L$ to KK
{\em gluon} (which give 
the dominant contribution to FCNC's). So, if we choose
$c_{ (t,b)_L } = 0.4$ (which 
can be consistent with FCNC for KK mass scale as low as $\sim 3$ TeV)
and $c_{ t_R } = 0$ as before, but with
the custodial symmetry for protecting $Z b \bar{b}$,
the couplings in Eq. (\ref{topbottom}) are modified to
\begin{eqnarray}
{\cal L} & \ni &
\frac{ g_{ Z^{ \prime } } }{2} \zx\,_{ \mu } 
\Big[ 
\left( - \frac{1}{2} - \frac{1}{6} \sin^2 \theta^{ \prime }_W  \right) 0.2 \xi
\left( \bar{t} \gamma^{ \mu } \left( 1 - \gamma_5 \right) t + 
\bar{b}
\gamma^{ \mu } \left( 1 - \gamma_5 \right) b
\right)
+ \Big. \nonumber \\ 
& & \Big. 
0.7 \xi \left( - \frac{2}{3} \sin^2 \theta^{ \prime }_W
\right) \bar{t} \gamma^{ \mu } \left( 1 + \gamma_5 \right) t
\Big] 
\label{LcustZbb.EQ}
\end{eqnarray}
The couplings of 
KK $Z$ and KK photon are unchanged.

Finally, there is of course the intermediate case
where {\em both} $c_{ t_R }$ and $c_{ (t,b)_L }$ 
are in-between $\sim 0$ and $\sim +1/2$.

{\bf Preferences for profiles from EW fit:}
Note that references \cite{Carena:2006bn, Carena:2007ua}
argued that if $t_R$ is {\em singlet} of $SU(2)_R$,
then $t_R$ having a close to flat profile 
is preferred by the EW fit (specifically, the requirement
of $T >0$ at one-loop level)
in models with custodial symmetries for
both the $T$ parameter and $Z b \bar{b}$. Whereas, 
for $t_R$ being {\em triplet} of $SU(2)_R$ instead, it is possible to obtain
$T > 0$ even with $t_R$ close to TeV brane \cite{Carena:2006bn}.
In this latter case, $(t,b)_L$ can then have close to flat profile,
as favored by {\em flavor} tests. 
Of course, 
for both these representations of $t_R$, the group theory factors in 
the couplings of $t_R$ to neutral gauge KK modes are identical since
$T_{ 3 R } = 0$ for $t_R$ in both these cases.
We mainly focus on the choice in Eq.~(\ref{LcustZbb.EQ}) in this work since 
this does the best in evading {\em both}
precision electroweak and flavor constraints. 
As discussed in section \ref{summary}, the specific choice of
representations of top/bottom will not affect
our results for $WW$, $Zh$ and $l^+ l^-$ final states  
by more than an $O(1)$ factor.

\section{Couplings}
\label{coupl.APP}
In this section we collect from the previous section, expressions for couplings 
and mixing angles. We focus mainly on the fermion representation given in 
Eq.~(\ref{LcustZbb.EQ}) with the custodial symmetry protecting $Zb\bar b$.
For our numerical study, we assume $g_L = g_R$ throughout.
The mixing angles and couplings are related through (with $s\equiv \sin()$ and $c\equiv cos()$)
\bea
g^\prime &=& \frac{g_X g_R}{\sqrt{g_R^2 + g_X^2}} \ \ , \ \ s^\prime = \frac{g_X}{\sqrt{g_R^2 + g_X^2}} \ \ , \ \ c^\prime = \sqrt{1-{s^\prime}^2} \ , \\
e &=& \frac{g_L g^\prime}{\sqrt{g^{\prime 2} + g_L^2}} \ \ , \ \ s_W = \frac{g^\prime}{\sqrt{g^{\prime 2} + g_L^2}} \ \ , \ \ c_W = \sqrt{1-s_W^2} \ , \\
g_Z &=& g_L/c_W \ \ , \ \ g_{Z^\prime} = g_R/c^\prime \ .
\eea
For the case $g_R = g_L$, we have $s^\prime = 0.55$, $c^\prime = 0.84$.

As explained in App.~\ref{couplings}, EWSB induces a mixing between 
$Z^{(0)} \leftrightarrow \zp$ (with mixing angle $\theta_{01}$) 
and $Z^{(0)} \leftrightarrow \zx$ (with mixing angle $\theta_{01X}$). 
To leading order in $\mz/\mzpri$ these mixing angles are given by
\bea
\sin{\theta_{01}} \approx \left(\frac{\mz}{\mzp} \right)^2 \sqrt{k \pi r_c} \ ,\label{sth01.EQ} \\
\sin{\theta_{01X}} \approx -\left(\frac{\mz}{\mzx} \right)^2 \left(\frac{g_{Z^\prime}}{g_Z}\right) c^{\prime\,2} \sqrt{k \pi r_c} \ .
\label{sth01X.EQ}
\eea
For example, for $\mzpri = 2$~TeV, $s_{01} = 0.013$ and $s_{01X} = -0.01$.

EWSB similarly induces mixing in the charged $W^\pm$ sector i.e. mixing between 
$W \leftrightarrow \wpri$, with mixing angle given by
\bea
\sin{\theta_{0L}} \approx \left(\frac{\mw}{\mwpL} \right)^2 \sqrt{k \pi r_c} \ , \label{sth0L.EQ} \\
\sin{\theta_{0R}} \approx -\left(\frac{\mw}{\mwpR} \right)^2 \left(\frac{g_R}{g_L}\right) \sqrt{k \pi r_c} \ . \label{sth0R.EQ}
\eea
For example, for $\mzpri = 2$~TeV, $s_{0L} \approx 0.01$ and $s_{0R} \approx - 0.01 $.

EWSB also induces $\zp\leftrightarrow \zx$ mixing, with mixing angle given by
\beq
\tan{2\theta_1} = \frac{-2 \mz^2 (g_{Z^\prime}/g_Z) c^{\prime 2} k\pi r_c }{(\mzx^2 - \mzp^2) + 
\mz^2\left( (g_{Z^\prime}/g_Z)^2 c^{\prime 4} - 1\right) k\pi r_c} \ .
\eeq
For example, for $\mzp = 2000; \mzx = 1962$~GeV, this implies that $s_1 = 0.48$, $c_1 = 0.88$.
After this mixing, we will refer to the mass eigenstates as $\zpt$ and $\zxt$.

EWSB similarly induces $\wpL \leftrightarrow \wpR$ with  mixing angle given by
\beq
\tan{2\theta_1^c} = \frac{-2 \mw^2 (g_R/g_L) k\pi r_c }{(\mwpR^2 - \mwpL^2) + 
\mw^2\left( (g_R/g_L)^2 - 1\right) k\pi r_c} \ .
\eeq
For example, for $\mwpL = 2000; \mwpR = 1962$~GeV, this implies that $s^c_1 = 0.6$, $c^c_1 = 0.8$.
After this mixing, we will refer to the mass eigenstates as $\wpLt$ and $\wpRt$.

The $\zpri$ coupling to a fermion as developed in 
Eqs.~(\ref{light})~(\ref{topbottom})~(\ref{tRflat})~and~(\ref{LcustZbb.EQ}) is given by
\beq
\bar{\psi} i \gamma^\mu D_\mu \supset \bar{\psi}_{L,R} \gamma^\mu \left[ e Q {\cal I} \ap\, _\mu + g_Z \left(T^3_L - s_W^2 T_Q\right) {\cal I} \zp\, _\mu + g_{Z^\prime} \left(T^3_R - s^{\prime 2} T_Y\right) {\cal I} \zx\, _\mu \right] \psi_{L,R} \ ,
\eeq
where ${\cal I}$ is the $\psi \psi \zpri$ overlap integral with profiles in the 
extra-dimension. They are given by
\beq
{\cal I}^{
+, - 
%
%
} = \int [dy] f_{\psi}^2 g^{(++),(-+)}  \ , 
\eeq
where $f_\chi$ is the fermion profile (specified by $c$), $g^{(+,+)}$ is the profile of
a gauge boson with $(+,+)$ boundary condition ($\ap$ and $\zp$), and $g^{(-.+)}$ is that
for $(-,+)$ boundary condition ($\zx$), and $[dy]$ includes an appropriate measure. 
As explained in App.~\ref{couplings}, we choose the fermion representation in
Eq.~(\ref{LcustZbb.EQ}) since it does the best in satisfying the combined FCNC and 
precision constraints.
We take the fermion $c$ values $c_{Q_L} = 0.4$, $c_{t_R} = 0$ and $c_\chi > 0.5$ with
$\chi$ denoting all other fields. 
The Higgs is taken to be localized close to the TeV brane so that
the values of the overlap integrals are as shown in Table~\ref{ovlap_ffG.TAB},
with $\xi = \sqrt{k\pi r_c} = 5.83$.
The $T^3_L$ charges of the fermions are as in the SM, and the $T^3_R$ charges are $-1/2$
for the $t_L,b_L$ and zero for 
$t_R$.
%
%
\begin{table}[h]
\begin{center}
\caption{Values of $\psi \psi \zpri$ overlap integrals
for $c_{ Q_L^3 } = 0.4$ and $c_{ t_R } = 0$
and all the other $c$'s $> 0.5$. We take $\xi = \sqrt{k\pi r_c} = 5.83$.
 \label{ovlap_ffG.TAB}}
\begin{tabular}{|c||c|c|c|}
\hline 
&
$Q_{L}^{3}$&
$t_{R}$& other fermions
%
%
\tabularnewline
\hline
\hline 
${\cal I}^{+}$
&
$- \frac{ 1.13 }{ \xi } + 0.2 \xi \approx 1$
& $- \frac{ 1.13 }{ \xi } + 0.7 \xi \approx 3.9$ 
& $- \frac{ 1.13 }{ \xi } \approx -0.2$
\tabularnewline
\hline 
${\cal I}^{-}$&
$ 0.2 \xi \approx 1.2 $
& $0.7 \xi \approx  4.1$ 
& 0\tabularnewline
\hline
\end{tabular}
\end{center}
\end{table}

We define the couplings of the $\zpri$ to SM fields relative to the SM coupling as $\kappa$.
These couplings (including the SM factors) are given in Table~\ref{fermKappa.TAB}. In order
to appreciate the bare-bone feature of the processes, we further separate the model-dependent
factors called $\lambda$ (leaving the SM couplings still in) as given in Table \ref{lambda}.

\begin{table}[tb]
\begin{center}
\caption{Couplings $\kappa_L,\ \kappa_R$ of $Z'$ to SM fields, with 
$\tilde Z_1$ and $\tilde Z_{ X_1 }$ denoting the mass eigenstates.  
For $\tilde Z_{X_1}$, $a = s_1$ and
$b = c_1$; $a\leftrightarrow b$
for $\tilde Z_1$.  The overlap
integrals, ${\cal I}$'s are given in Table~\ref{ovlap_ffG.TAB}.
Here, $u$ and $d$ denote quarks other than top and bottom quarks,
except for $d_R$ which includes $b_R$. \label{fermKappa.TAB}
}
\begin{tabular}{|c||c|c|}
\hline
& $A_1$ & $\tilde Z_{ X_1 }/\tilde Z_1$
\tabularnewline
\hline
\hline
$t_L \bar t_L$
& $\frac{2}{3} e {\cal I}^+$
&  $g_Z \left( \frac{1}{2} - \frac{2}{3} s^2_W \right)
a {\cal I}^+ \pm
g_{ Z^{ \prime } } \left( - \frac{1}{2} - \frac{1}{6} s^{ \prime \; 2 }_W
\right) b {\cal I}^-$
\tabularnewline
\hline
$b_L \bar b_L$
& $- \frac{1}{3} e {\cal I}^+$
& $g_Z \left( - \frac{1}{2} + \frac{1}{3} s^2_W \right)
a {\cal I}^+ \pm g_{ Z^{ \prime } }
\left( - \frac{1}{2} - \frac{1}{6} s^{ \prime \; 2 }_W
\right) b {\cal I}^-$
\tabularnewline
\hline
$t_R \bar t_R$
&  $\frac{2}{3} e {\cal I}^+$
& $g_Z \left(  - \frac{2}{3} s^2_W \right) a {\cal I}^+ \pm g_{ Z^{ \prime } }
\left( - \frac{2}{3} s_W^{ \prime \; 2 } \right) b {\cal I}^-$
\tabularnewline
\hline
$u_L \bar u_L$
& $\frac{2}{3} e {\cal I}^+$
&  $g_Z \left( \frac{1}{2} - \frac{2}{3} s^2_W \right)
a {\cal I}^+$
\tabularnewline
\hline
$d_L \bar d_L$
& $- \frac{1}{3} e {\cal I}^+$
& $g_Z \left( - \frac{1}{2} + \frac{1}{3} s^2_W \right) a {\cal I}^+$
\tabularnewline
\hline
$u_R \bar u_R$
&  $\frac{2}{3} e {\cal I}^+$
& $g_Z \left(  - \frac{2}{3} s^2_W \right) a {\cal I}^+$
\tabularnewline
\hline
$d_R \bar d_R$
& $-\frac{1}{3} e {\cal I}^+$
& $g_Z \left( \frac{1}{3} s^2_W \right) a {\cal I}^+$
\tabularnewline
\hline
$\ell^+_R \ell^-_L$
& $- e {\cal I}^+$
& $g_Z \left( - \frac{1}{2} + s^2_W \right) a {\cal I}^+$
\tabularnewline
\hline
$\ell^+_L \ell^-_R$
& $-e {\cal I}^+$
& $g_Z \left( s^2_W \right) a {\cal I}^+$
\tabularnewline
\hline
$\nu_L \bar \nu_L$
& 0
& $g_Z \left( \frac{1}{2} \right) a {\cal I}^+$
\tabularnewline
\hline
$W^+W^-$
& $-2 e s_{ 0 L } $
& $g_L  c_W \left( s_{ 0 1 } a \pm s_ { 0 1 X } b - 2 a s_{ 0 L } \right)$
\tabularnewline
\hline
$Zh$
& 0
& $g_Z \sqrt{ k \pi r_c } \left( a \mp \frac{ g_R }{ g_L } c_W c^{ \prime } b \right)$
\tabularnewline
\hline
\end{tabular}
\end{center}
\end{table}

\begin{table}[h]
\begin{center}
\caption{Scaling factors $\lambda$ as used in Fig.~\ref{xsmassz.FIG}(a). 
Here, $Z_1$ and $Z_{ X_1 }$ denote states in KK basis.
\label{lambda}
}
\begin{tabular}[t]{ |c | c | c|c|}
\hline
      $Z'$& $qq\to Z'$& $bb_L\to Z'$ 
&WBF  \\
      \hline \hline 
      $A_1$&$-1.13/\xi$& $(-1.13/\xi+0.2\xi)\dfrac{-(1/3)s_W c_W}{-1/2+s^2_{W}/3}$
& $-2s_{0L}$  \\
      \hline
       $Z_1$ & $-1.13/\xi$ & $-1.13/\xi+0.2\xi$ 
& $s_{01}-2s_{0L}$  \\
      \hline
       $Z_{X1}$ & & $0.2\xi\dfrac{c_W}{c'}\dfrac{-1/2-s'^2/6}{-1/2+s^2_{W}/3}$ 
& $s_{01X}$ \\
\hline
\end{tabular}
\end{center}
\end{table}



\end{document}